\documentclass[12pt]{article}

\usepackage{scicite}
\usepackage{times}

\usepackage{graphicx} 

\usepackage{color}

\newenvironment{sciabstract}{%
\begin{quote} \bf}
{\end{quote}}

\title{How does salinity shape ocean circulation and ice geometry on Enceladus and other icy satellites?}

\author
{Wanying Kang$^{1\ast}$, Tushar Mittal $^{1}$, Suyash Bire $^{1}$, Jean-Michel Campin $^{1}$, John Marshall $^{1}$\\
\normalsize{$^{1}$Earth, Atmospheric and Planetary Science Department, Massachusetts Institute of Technology}\\
\normalsize{77 Massachusetts Ave., MA02139, USA}
\\
\normalsize{$^\ast$ E-mail:  wanying@mit.edu}
}

\date{}

\begin{document} 
\baselineskip24pt
\maketitle 

\begin{sciabstract}
  Of profound astrobiological interest, Enceladus appears to have a global subsurface ocean that is salty, indicating water-rock reaction at present or in the past, important for its habitability. Here, we investigate how salinity and the partition of heat production between the silicate core and the ice shell affect ocean dynamics and the associated heat transport -- a key factor that determines the equilibrium ice shell geometry. Assuming steady state conditions, we show that the meridional overturning circulation of the ocean, driven by heat and salt exchange with the ice, has opposing signs at very low and very high salinities. Regardless of these differing circulations, heat and freshwater converge towards the equator, where the ice is thick, acting to homogenize thickness variations. In order to maintain the observed ice thickness variation, the polar-amplified ice dissipation needs to be strong enough and ocean heat convergence cannot overwhelm well-constrained heat loss rates through the thick equatorial ice sheet. This requirement is found violated if the main heat source is in the core rather than the ice shell, or if the ocean is very fresh or very salty. Instead, with a salinity of intermediate range, the temperature- and salinity-induced density gradient largely cancel one another, leading to much reduced overturning and equatorial heat convergence rates and consistent budgets in appearance of a significant ice dissipation. 
\end{sciabstract}

\section*{Introduction}

Since the Cassini and Galileo missions, Enceladus (a satellite of Saturn) and Europa (a satellite of Jupiter) have been revealed to have a high astrobiological potential, satisfying all three necessary conditions for life: 1) the presence of liquid water \cite{Postberg-Kempf-Schmidt-et-al-2009:sodium, Thomas-Tajeddine-Tiscareno-et-al-2016:enceladus}, 2) a source of energy \cite{Choblet-Tobie-Sotin-et-al-2017:powering, Beuthe-2019:enceladuss}, and 3) a suitable mix of chemical elements \cite{Waite-Combi-Ip-et-al-2006:cassini, Postberg-Kempf-Schmidt-et-al-2009:sodium, Hsu-Postberg-Sekine-et-al-2015:ongoing, Waite-Glein-Perryman-et-al-2017:cassini, Postberg-Khawaja-Abel-et-al-2018:macromolecular, Taubner-Pappenreiter-Zwicker-et-al-2018:biological,  Glein-Waite-2020:carbonate}. In particular, the geyser-like sprays ejected from the fissures over Enceladus's south pole \cite{Hansen-Esposito-Stewart-et-al-2006:enceladus, Howett-Spencer-Pearl-et-al-2011:high, Spencer-Howett-Verbiscer-et-al-2013:enceladus} provide a unique opportunity to understand the chemistry and dynamics of Enceladus' interior without landing on and drilling through a typically 20km-thick ice shell. Within the geyser samples collected by Cassini, CO$_2$, methane \cite{Waite-Combi-Ip-et-al-2006:cassini}, sodium salt \cite{Postberg-Kempf-Schmidt-et-al-2009:sodium}, hydrogen \cite{Waite-Glein-Perryman-et-al-2017:cassini}, and macromolecular organic compounds \cite{Postberg-Khawaja-Abel-et-al-2018:macromolecular} have been found. This suggests a chemically active environment able to sustain life \cite{Glein-Waite-2020:carbonate, Taubner-Pappenreiter-Zwicker-et-al-2018:biological}. However, to make optimum use of these samples to infer the chemical environment of the subsurface ocean, one needs to better understand tracer transport by the interior ocean and hence ocean circulation itself.

  Ocean circulation on Enceladus is driven by heat and salinity fluxes from the silicate core \cite{Choblet-Tobie-Sotin-et-al-2017:powering} and the ice shell \cite{McCarthy-Cooper-2016:tidal, Tobie-Mocquet-Sotin-2005:tidal, Beuthe-2019:enceladuss}, as well as mechanical forcing, such as tides and libration \cite{Rekier-Trinh-Triana-et-al-2019:internal, Soderlund-Kalousova-Buffo-et-al-2020:ice}. The partition of heat production between the ice and the silicate core has a direct control over ocean dynamics \footnote{Ocean dissipation is thought to be weak \cite{Hay-Matsuyama-2019:nonlinear, Rekier-Trinh-Triana-et-al-2019:internal}, but it can affect circulation through its effect on mixing.}. Moreover, ocean salinity plays a key role since it determines whether density decreases or increases with temperature \cite{McDougall-Barker-2011:getting} (see Fig.\ref{fig:EOS-Hice-FreezingRate}c). For example if the ocean is very fresh then heat released by hydrothermal vents will not trigger penetrative convection from below \footnote{With extremely low vertical mixing, the ocean will form two layers: the bottom layer is likely to be convectively active and well mixed with a temperature equal to the critical temperature; the upper layer wouuld likely be diffusive and stably stratified, with temperature decreasing upward.} \cite{Zeng-Jansen-2021:ocean}. Furthermore, the global scale circulation of a salty ocean could be completely different from that in a fresh ocean, as has been explored in Earth's ocean and terrestrial exoplanets \cite{Cullum-Stevens-Joshi-2016:importance, Cael-Ferrari-2017:oceans}.

Despite its importance, the heat partition is poorly constrained due to our limited understanding of the rheology of both the ice shell and the silicate core.
Hydrogen and nanometre-sized silica particles have been detected on Enceladus, providing clear geochemical evidence for active seafloor venting \cite{Hsu-Postberg-Sekine-et-al-2015:ongoing, Waite-Glein-Perryman-et-al-2017:cassini}. However, whether this submarine hydrothermalism is the dominant heat source preventing the ocean from freezing remains inconclusive due to our limited understanding of the core's rheology \cite{Travis-Schubert-2015:keeping, Choblet-Tobie-Sotin-et-al-2017:powering}.
Another potential heat source is tidal dissipation within the ice shell itself. While the ice geometry on top of the ocean is qualitatively consistent with heating primarily in the ice shell \cite{Hemingway-Mittal-2019:enceladuss}, present dynamical models of ice are unable to reproduce enough heat to maintain such a thin ice shell \cite{Beuthe-2019:enceladuss, Soucek-Behounkova-Cadek-et-al-2019:tidal}. Attempts to account for higher heat generation through use of more advanced models of ice rheology have thus far not been successful \cite{Robuchon-Choblet-Tobie-et-al-2010:coupling, Shoji-Hussmann-Kurita-et-al-2013:ice, Behounkova-Tobie-Choblet-et-al-2013:impact, McCarthy-Cooper-2016:tidal, Beuthe-2019:enceladuss, Soucek-Behounkova-Cadek-et-al-2019:tidal, Gevorgyan-Boue-Ragazzo-et-al-2020:andrade}.

  An additional complication is that the salinity of Enceladus' ocean remains uncertain. Calculations of thermochemical equilibria over a range of hydrothermal and freezing conditions for chondritic compositions, suggest a salinity ranging between $2$-$20$~psu (g/kg), with a higher likelihood of it being below 10~psu \cite{Zolotov-2007:oceanic, zolotov2014can, Glein-Postberg-Vance-2018:geochemistry}. However, at least $17$~psu is required to keep the geysers' liquid–gas interface convectively active to ensure that they do not freeze up \cite{Ingersoll-Nakajima-2016:controlled}. Sodium-enriched samples taken from south pole sprays by Cassini have a salinity of 5-20~psu. This can be considered a lower bound since the interaction of cold water vapor sprays with their environment is likely to lower the salinity of droplets through condensation \cite{Postberg-Kempf-Schmidt-et-al-2009:sodium}. This is also uncertain, however, since fractional crystallization and disequilibrium chemistry may partition components in such a way that geyser particles are not directly representative of the underlying ocean \cite{Fox-Powell-Cousins-2021:partitioning}. Furthermore, if particles originate from a hydrothermal vent, composition can also deviate far from that of the overall ocean \cite{Glein-Postberg-Vance-2018:geochemistry, Choblet-Tobie-Sotin-et-al-2017:powering}. In a separate line of argument, the size of silica nano-particles carried along in the sprays suggests a salinity $<40$~psu, but this is sensitive to assumptions about ocean pH and the dynamics of hydrothermal vents \cite{Hsu-Postberg-Sekine-et-al-2015:ongoing}. 


Given the above uncertainties, here we use our understanding of ocean dynamics to provide further constraints on ocean salinity and the partition of heating between the ice shell and the core. To do this we make use of observations of ice shell geometry to provide boundary conditions for ocean circulation and model the ocean circulation driven by them as a function of salinity and tidal heating partitions. Some lead to heat transports which are consistent with what is known about tidal heating rates, others do not, enabling us to discriminate between them.

\section{Boundary conditions on ocean circulation}

\begin{figure*}[htp!]
    \centering
    \includegraphics[width=\textwidth]{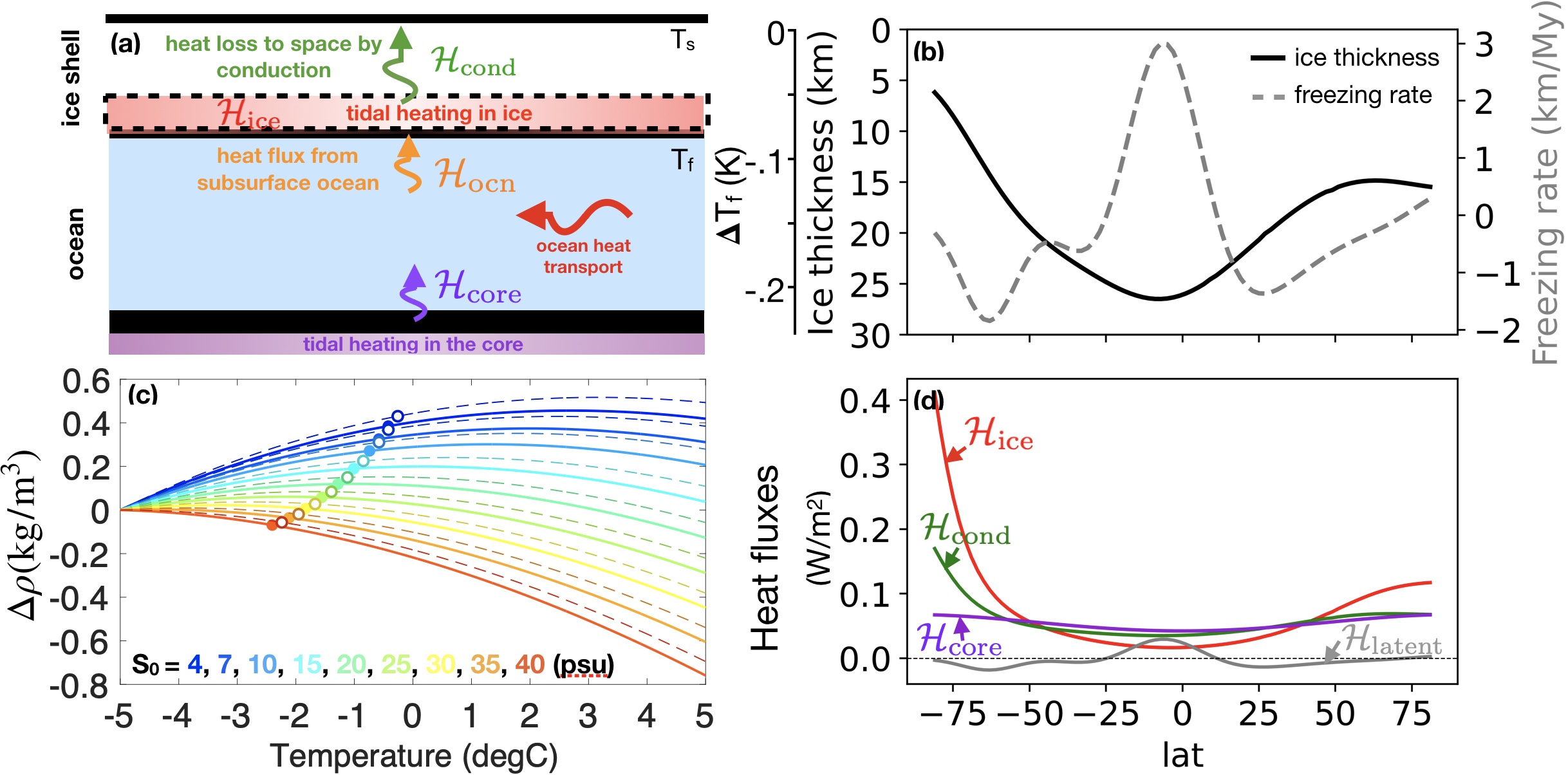}
    
    \caption{\small{ Panel (a) presents the primary sources of heat and heat fluxes in an icy moon which include: heating due to tidal dissipation in the ice $\mathcal{H}_{\mathrm{ice}}$ and the silicate core $\mathcal{H}_{\mathrm{core}}$, the heat flux from the ocean to the ice $\mathcal{H}_{\mathrm{ocn}}$ and the conductive heat loss to space $\mathcal{H}_{\mathrm{cond}}$. Ocean heat transport is shown by the horizontal arrow. Panel (b) shows the observed ice shell thickness of Enceladus based on shape and gravity measurements \cite{Hemingway-Mittal-2019:enceladuss} (black solid curve, left y-axis). The suppression of the freezing point of water by these thickness variations, relative to that at zero-pressure, is indicated by the outer left y-axis. The gray dashed curve shows the freezing (positive) and melting rate (negative) required to maintain a steady state based on an upside-down shallow ice flow model (y-axis on the right). Panel (c) shows how the density anomaly of water varies as its temperature varies around -5$^\circ$C as a function of salinity. Moving from cold to warm colors denotes increasing salinity, as indicated by the colored lettering. The solid (dashed) curves are computed assuming the pressure under the 26.5~km (5.6~km) of ice at the equator (south pole). The freezing points are marked by the circles. Panel (d) shown typical magnitudes and profiles of $\mathcal{H}_{\mathrm{ice}}$, $\mathcal{H}_{\mathrm{core}}$, $\mathcal{H}_{\mathrm{cond}}$ and $\mathcal{H}_{\mathrm{latent}}$. The models of heat fluxes and ice flow on which all these curves are based can be found in section~1D of the SM.}}
    
    \label{fig:EOS-Hice-FreezingRate}
  \end{figure*}

 Data provided by Cassini has enabled reconstructions to be made of Enceladus' ice thickness variations \cite{Iess-Stevenson-Parisi-et-al-2014:gravity, Beuthe-Rivoldini-Trinh-2016:enceladuss, Tajeddine-Soderlund-Thomas-et-al-2017:true, Cadek-Soucek-Behounkova-et-al-2019:long, Hemingway-Mittal-2019:enceladuss}. The solid curve in Fig.\ref{fig:EOS-Hice-FreezingRate}b show the zonal mean ice thickness deduced by \textit{Hemingway \& Mittal 2019}\cite{Hemingway-Mittal-2019:enceladuss}. Thick ice at the equator with a poleward thinning trend is notable. The thinnest ice shell over the south pole is only a fifth as thick as the equatorial ice shell. Such ice thickness variations have two effects. First, thick equatorial ice creates high pressure depressing the local freezing point leading to a roughly -0.1~K temperature anomaly just beneath the ice compared to the poles (solid curve in Fig.\ref{fig:EOS-Hice-FreezingRate}b). 
 Second, thickness variations will drive ice to flow from thick-ice regions to thin-ice regions on million-year time-scales \cite{Tobie-Choblet-Sotin-2003:tidally, Barr-Showman-2009:heat, Ashkenazy-Sayag-Tziperman-2018:dynamics}. To compensate the smoothing effect of the ice flow, ice must form in low latitudes and melt in high latitudes. Assuming an ice rheology, we can calculate ice flow speeds using an upside-down shallow ice model (details are given in the SM section~1D). In this way, we can infer the required freezing/melting rate, as shown by the dashed curve in Fig.\ref{fig:EOS-Hice-FreezingRate}b. This freezing and melting will lead to a meridional salinity gradient over time via brine rejection and fresh water input which, in steady state (assumed), must be balanced by salinity transport in the ocean.
 
\begin{figure*}[htp!]
    \centering
    \includegraphics[page=1,width=0.7\textwidth]{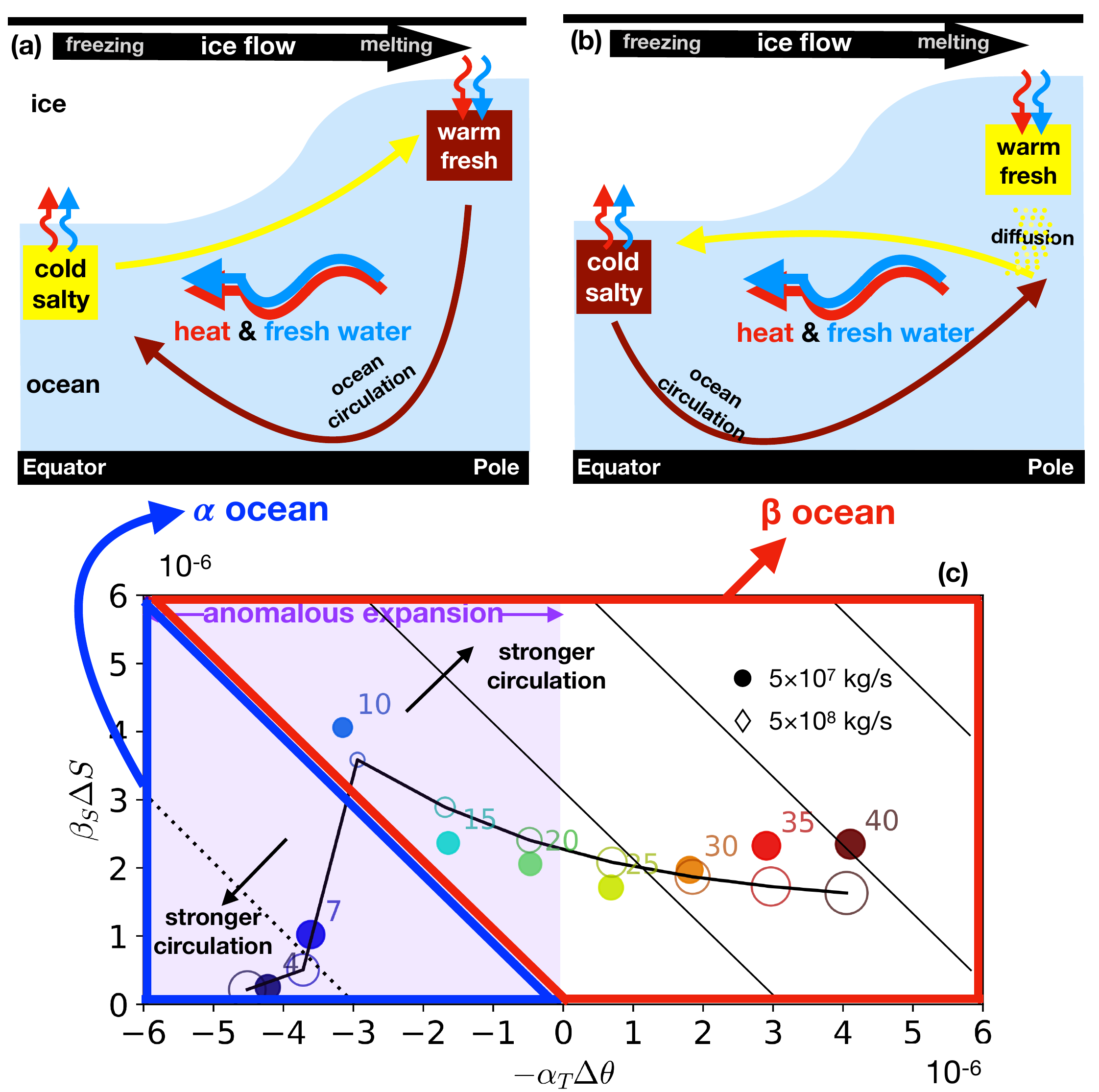}
    
    \caption{\small{At the top we show schematics of ocean circulation and associated transports of heat (red wiggly arrows) and fresh water (blue wiggly arrows) for (a) the fresh alpha ocean and (b) the salty beta ocean. Dark brown arrows denote sinking of dense water, light yellow arrows denote rising of buoyant water. The circulations are forced by the freezing/melting required to counterbalance the down-gradient ice flow (thick black arrows marked at the top) and by variations in the freezing point of water due to pressure, as presented in Fig.\ref{fig:EOS-Hice-FreezingRate}b. In panel (c), we present a regime diagram inspired by explicit solutions such as those presented in Fig.\ref{fig:T-S-U-Psi}, showing the influence of temperature and salinity anomalies on density assuming different salinities (the number on the shoulder of each circle gives the $S_0$ used in that experiment), and how the overturning circulation of the ocean responds. Horizontal and vertical axes are the temperature and salinity induced density anomalies at the equator, relative to the north pole. Note that both $\Delta S$ and $-\Delta T$ are positive (the equator is always saltier and colder than the pole), and the sign of the coordinates reflect the sign of $\alpha_T$ and $\beta_S$: $\beta_S$ is always positive, but $\alpha_T$ increases from negative to positive as $S_0$ increases. In the high/low $S_0$ experiments, the signs of $-\alpha_T\Delta T$ and $\beta_S\Delta S$ are the same/opposite. Red (blue) solid lines delineate the $\beta$ ocean ($\alpha$ ocean) regimes, in which the density is dominated by salinity (temperature) anomalies respectively, as set out in the schematics above. Purple shading highlights the regime where anomalous expansion of seawater is present with negative $\alpha_T$ so that warming leads to sinking. The size of each circle represents the amplitude of the overturning circulation (the peak $\Psi$ occurs in the northern hemisphere, exact values are listed in Tab.~2 in the SM). The 45$^\circ$ tilted black lines are isolines of the equator-to-pole density difference $\Delta \rho$. Solid lines denote dense water near the equator and dotted lines denote dense water over the poles. As illustrated by the black arrows, circulation strengthens with $\Delta \rho$ moving away from the transition line between $\alpha$ ocean and $\beta$ ocean. The empty circles connected by a black solid curve show the fit of the conceptual model developed in Section 4 which broadly captures the behavior of the explicit calculation using our full model.
    }}
    \label{fig:Schematics}
  \end{figure*}

 The combined effect of these temperature and salinity gradients just beneath the ice will be to make equatorial waters saltier and colder than polar waters. In a salty ocean, where water volume contracts when it is cold (known as the beta ocean), we expect the ocean to sink at the cold low latitudes, because the water density is high there (see Fig.~\ref{fig:Schematics}b). In contrast, in a fresh ocean (known as the alpha ocean), the opposite is possible because of seawater's anomalous expansion upon cooling (see Fig.~\ref{fig:Schematics}a). Thus the overturning circulation in alpha and beta oceans can be expected to be of opposite sign. However, no matter in which direction the ocean circulates, heat will be converged toward the equator, because of the tendency of cold equatorial water and warm polar water to mix together. Limited by the efficiency of conductive heat loss through the thick equatorial ice, the equatorward heat convergence cannot be arbitrarily strong. By examining the heat budget of the ice, then, knowledge of ocean heat transport under various salinities and core-shell heat partitions can be used to discriminate between different scenarios.

 In order to study the possible ocean circulation and heat transport on Enceladus in this way, we set up a zonally-averaged ocean circulation model to sweep across a range of mean salinities ($S_0=4$, $7$, $10$, $15$, $20$, $25$, $30$, $35$ and $40$~psu) and core-shell heat partitions (0-100\%, 100-0\% and 20-80\%). Our model has its ocean covered by an ice shell that resembles that of the present-day Enceladus \cite{Hemingway-Mittal-2019:enceladuss} (solid curve in Fig.\ref{fig:EOS-Hice-FreezingRate}b), which is assumed to be sustained against the ice flow by a prescribed freezing/melting $q$ (gray dashed curve in Fig.\ref{fig:EOS-Hice-FreezingRate}b). By prescribing $q$, we cut off the positive feedback loop between the ocean heat transport and the ice freezing/melting rates, ensuring that the circulation will always be consistent with those freshwater fluxes. When heat production by the silicate core is assumed to be non-zero, an upward heat flux at the bottom is prescribed. Guided by models of tidal heating described in SM, this is assumed to be slightly polar-amplified (see purple curve in Fig.\ref{fig:EOS-Hice-FreezingRate}d). At the water-ice interface, a downward salinity flux $S_0q$ is imposed to represent the brine rejection and freshwater production associated with freezing/melting. Meanwhile, the ocean temperature there is relaxed toward the local freezing point. Thus the ocean will deposit heat to the ice when its temperature is slightly higher than the freezing point, and vice versa. In order for the heat budget of the ice to close, this ocean-ice heat exchange $\mathcal{H}_{\mathrm{ocn}}$, together with the tidal heat produced in the ice $\mathcal{H}_{\mathrm{ice}}$ (red curve in Fig.~\ref{fig:EOS-Hice-FreezingRate}d) and the latent heat released $\mathcal{H}_{\mathrm{latent}}$ ($\mathcal{H}_{\mathrm{latent}}=\rho L_fq$, where $\rho$ and $L_f$ are the density and fusion energy of ice, see the gray curve in Fig.~\ref{fig:EOS-Hice-FreezingRate}d) should balance the conductive \footnote{The ice shell on Enceladus is relatively thin and so unlikely to be convective \cite{Barr-McKinnon-2007:convection}.} heat loss through the ice shell $\mathcal{H}_{\mathrm{cond}}$ (green curve in Fig.~\ref{fig:EOS-Hice-FreezingRate}d). The degree to which this heat budget is in balance informs us of the plausibility of the assumed salinity and heat partition.\footnote{The global heat budget is closed by design, i.e., $\overline{\mathcal{H}_{\mathrm{cond}}}=\overline{\mathcal{H}_{\mathrm{ice}}}+\overline{\mathcal{H}_{\mathrm{core}}}$. Latent heat's global average vanishes, because $\overline{q}=\overline{\nabla\cdot \mathcal{Q}}=0$, where $\mathcal{Q}$ is the ice flow.}
 
 Before going on to describe our results, we emphasize that we have adopted a zonally-averaged modeling framework so that we can readily explore parameter space whilst integrating our models out to an equilibrium state, which takes about 10,000 model years. This necessarily implies that our ocean model is highly parameterized -- as are the models of tidal heating and ice flows that are used to provide the forcing at the boundaries that drive it -- and so have many unavoidable uncertainties. In particular, and as described in detail in SM and just as in terrestrial ocean models, processes such as convection, diapycnal mixing and baroclinic instability are parameterized guided by our knowledge of the mechanisms that underlie them.

\section{Patterns of ocean circulation, temperature and salinity}

\begin{figure*}[htp!]
    \centering
    \includegraphics[page=3,width=\textwidth]{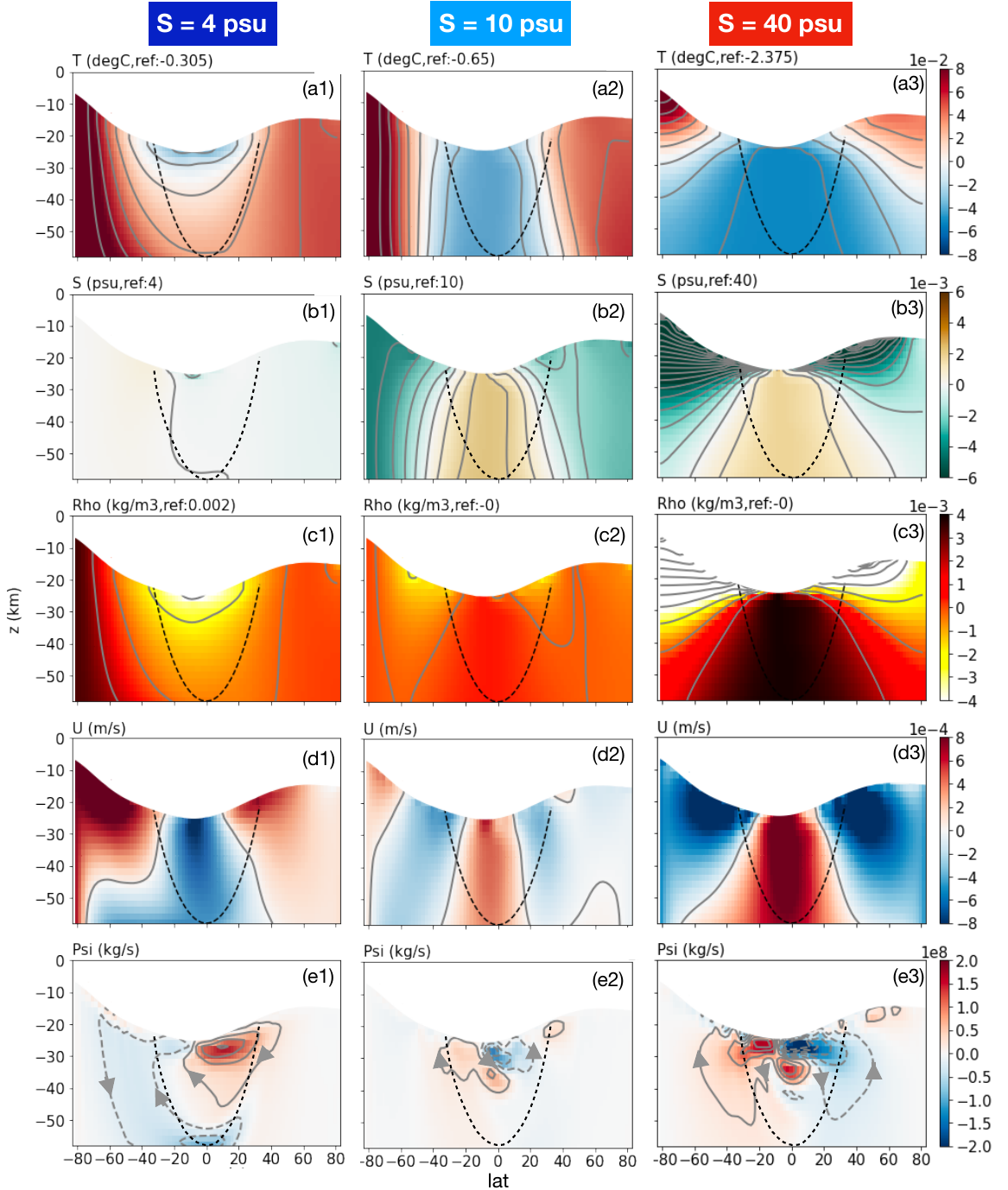}
    
    \caption{\small{Ocean circulation and thermodynamic state for experiments driven by the glacial melt and under-ice temperature distributions shown in Fig.~\ref{fig:EOS-Hice-FreezingRate}b for oceans with various mean salinities. Moving from top to bottom present temperature $T$, salinity $S$, density anomaly $\Delta\rho$, zonal flow speed $U$ and meridional overturning streamfunction $\Psi$ with arrows indicating the sense of flow. The left column presents results for a low salinity ocean ($S_0=4$~psu), the right column for high salinity ($S_0=40$~psu), and the middle column for an ocean with intermediate salinity ($S_0=10$~psu). The reference temperature and salinity (marked at the top of each plot) are subtracted from $T$ and $S$ to better reveal spatial patterns. Positive $U$ indicates flow to the east and positive $\Psi$ indicates a clockwise overturning circulation. Black dashed lines mark the position of the tangent cylinder. The bottom row shows the vertically-integrated meridional heat transport against latitude.}}
    
    \label{fig:T-S-U-Psi}
  \end{figure*}

Due to the relatively low freezing point (Fig.~\ref{fig:EOS-Hice-FreezingRate}c) and elevated freezing rate (Fig.~\ref{fig:EOS-Hice-FreezingRate}b) of low latitudes, water just under the ice is colder and saltier than near the poles, regardless of the mean salinity. This pole-to-equator temperature and salinity contrast leads to variations in density, which in turn drive ocean circulation. In Fig.~\ref{fig:T-S-U-Psi}(c,e), we present the density anomaly, $\rho_0(\alpha_T\theta'+\beta_SS')$, and the meridional overturning streamfunction $\Psi(\phi,z) = \int_{-D}^z  \rho(\phi,z') V(\phi,z')\times (2\pi(a-z')\cos\phi)~ dz'$. Here, $\theta'$ and  $S'$ (plotted in Fig.~\ref{fig:T-S-U-Psi}a,b) are the deviation in potential temperature and salinity from the reference, $\alpha_T$ and $\beta_S$ are the thermal expansion and haline contraction coefficient, $V$ is the meridional current, $\rho$ is the water density, and $D$ is the ocean depth, $\phi$ denote latitude and $z$ points upwards.

  Since the density gradient induced by temperature variations can either enhance or diminish that induced by salinity, depending on the mean salinity $S_0$, the overturning circulation can run in either direction. When $S_0$ is greater than $22$~psu, water expands with increasing temperature ($\alpha_T>0$, see reddish curves in Fig.\ref{fig:EOS-Hice-FreezingRate}c, 2~MPa pressure assumed). As a result, the cold and salty water under the thick equatorial ice shell is denser than polar waters, as shown in Fig.\ref{fig:T-S-U-Psi}-c3 and sketched in Fig.~\ref{fig:Schematics}b using the dark brown color. Equatorial waters therefore sink, as shown in Fig.\ref{fig:T-S-U-Psi}-e3 (indicated in Fig.\ref{fig:Schematics}b using the dark brown arrow). This circulation pattern broadly agrees with that suggested by \textit{Lobo et al. 2021}\cite{Lobo-Thompson-Vance-et-al-2021:pole} using a more idealized ocean model.

  However, when $S_0$ is below $22$~psu, the thermal expansion coefficient changes sign ($\alpha_T<0$, as shown by the bluish curves in Fig.\ref{fig:EOS-Hice-FreezingRate}c). This so-called anomalous expansion of water results in the temperature-induced density difference and the salinity-induced density difference partially cancelling one another, giving rise to two possibilities. If the salinity factor dominates ($\beta$ ocean), the overturning circulation becomes one of sinking at the equator, as show in Fig.\ref{fig:T-S-U-Psi}-e2 and sketched in Fig.\ref{fig:Schematics}b using a dark brown arrow. But if temperature dominates ($\alpha$ ocean), the overturning circulation flips direction with sinking over the poles (Fig.\ref{fig:T-S-U-Psi}-e1 and Fig.\ref{fig:Schematics}a) because water is denser there (Fig.\ref{fig:T-S-U-Psi}-c1). The switch in the sense of the overturning circulation with salinity can also occur in models of Earth's ocean \cite{Cullum-Stevens-Joshi-2016:importance, Cael-Ferrari-2017:oceans}, even though Earth's ocean is forced rather differently.

\begin{figure*}[htp!]
    \centering
    \includegraphics[page=24,width=\textwidth]{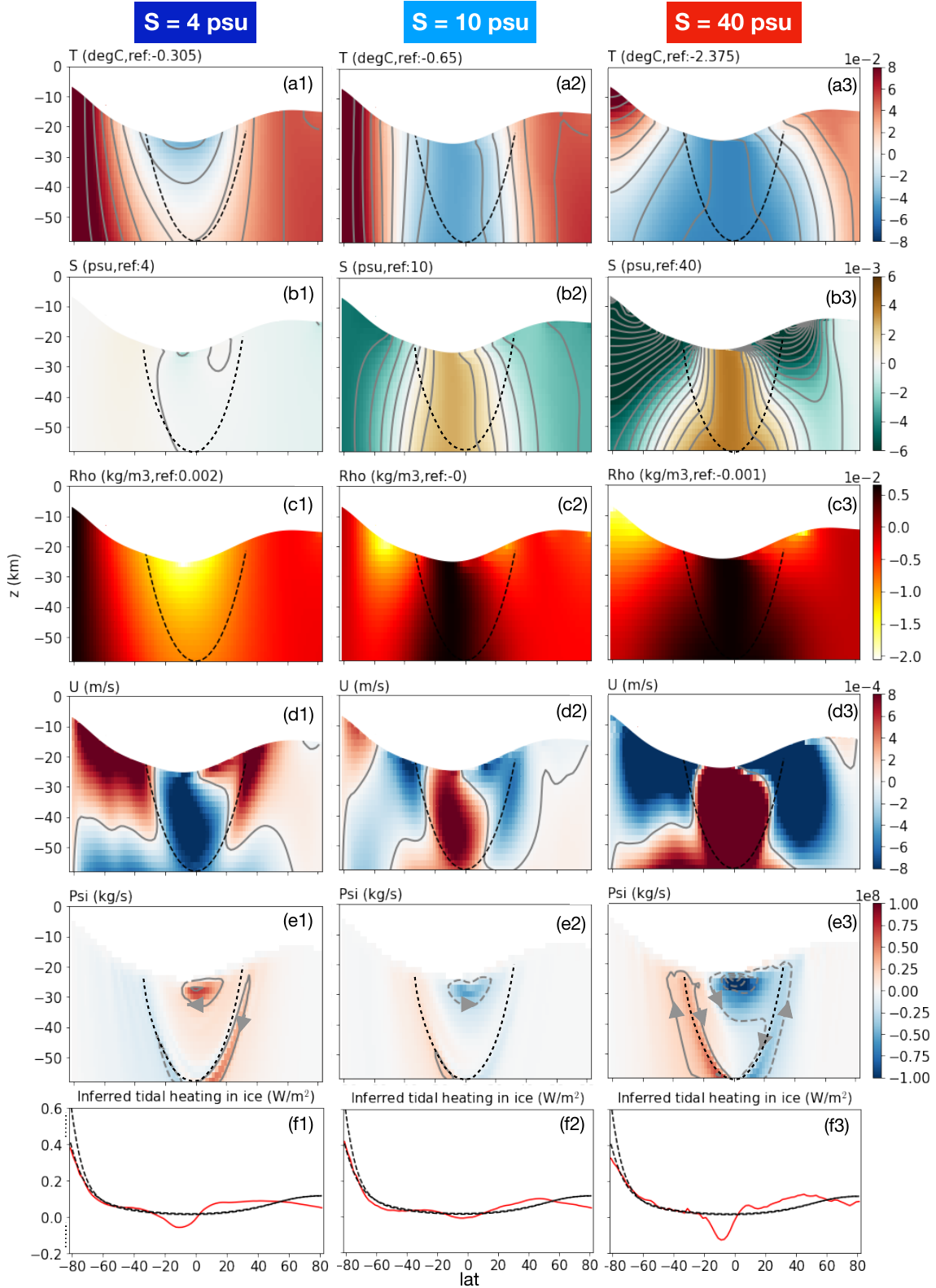}
    
    \caption{\small{Same as Fig.~\ref{fig:T-S-U-Psi}, except for horizontal and vertical viscosity are set to $2$~m$^2$/s instead of $50$~m$^2$/s. 
      }}
    
    \label{fig:T-S-U-Psi-lowvisc}
  \end{figure*}

  The circulations are mostly concentrated near the ocean-ice interface, as suggested by \textit{Lobo et al. 2021}\cite{Lobo-Thompson-Vance-et-al-2021:pole}. However, such circulation can only exist in the Ekman boundary layer. Right below the ice or above the seafloor, friction can balance the westerly/easterly momentum acceleration as water flows toward/away from the rotating axis viscosity. However, in the interior, the balance is maintained by having viscosity transmitting the momentum to the boundaries where there is friction, and viscosity is only effective within the Ekman layer (depth $\sim\sqrt{\nu_v/f}$). In the default setup, viscosity is set to 50~m$^2$/s,  and the Ekman layer is resolved (or permitted) by our 2km vertical resolution. However, when further reducing the viscosity to 1~m$^2$/s, the shallow circulation near the ice disappears, replaced by a deep circulation connecting the two rough boundaries -- the ice shell and the seafloor (see Fig.~\ref{fig:T-S-U-Psi-lowvisc}e). 

The transition from polar to equatorial sinking is governed by the density difference between the poles and the equator. Taking the north pole as a reference, the temperature-related density anomaly at the equator can be written as $-\alpha_T\Delta \theta$, and the salinity-related density anomaly as $\beta_S\Delta S$, where $\Delta\theta$ and $\Delta S$ are the potential temperature and salinity anomaly at the equator relative to the north pole. Fig.\ref{fig:Schematics}c presents the strength of the overturning circulation from all nine experiments in the $(-\alpha_T\Delta T,\ \beta_S\Delta S$) space: the size of the circles are proportional to $\Psi$. The 45 degree tilted line denotes perfect cancellation between the saline and temperature-driven overturning circulations: it passes near $10$~psu, explaining why the $10$~psu experiment has the weakest circulation compared to all others. On moving away from this line in either direction the strength of the overturning circulation increases but is of opposite sign, as represented schematically in the schema above. 

    
%

The sense of the overturning circulation shapes the tracer patterns and associated zonal currents. Downwelling regions (low latitudes for a salty ocean and high latitudes for a fresh ocean) advect density, temperature and salinity anomalies, set at the ocean-ice interface, into the interior ocean. Note the bending of the temperature and salinity contours equatorward (poleward) when downwelling occurs at the poles (equator), as shown in Fig.~\ref{fig:T-S-U-Psi}. This results in meridional density gradients which are in a generalized\footnote{The terms associated with the horizontal component of Coriolis force must be included.} thermal wind balance with zonal currents (Fig.~\ref{fig:T-S-U-Psi}d). 
When the meridional density gradients are weak, the zonal flow is weak, overturning circulation is weak, and the temperature and salinity contours are mostly vertical (see the second column of Fig.~\ref{fig:T-S-U-Psi}).

\section{Implication of ocean heat transport (OHT) for the heat budget of the ice shell}
\begin{figure*}[htp!]
    \centering
    \includegraphics[width=0.8\textwidth]{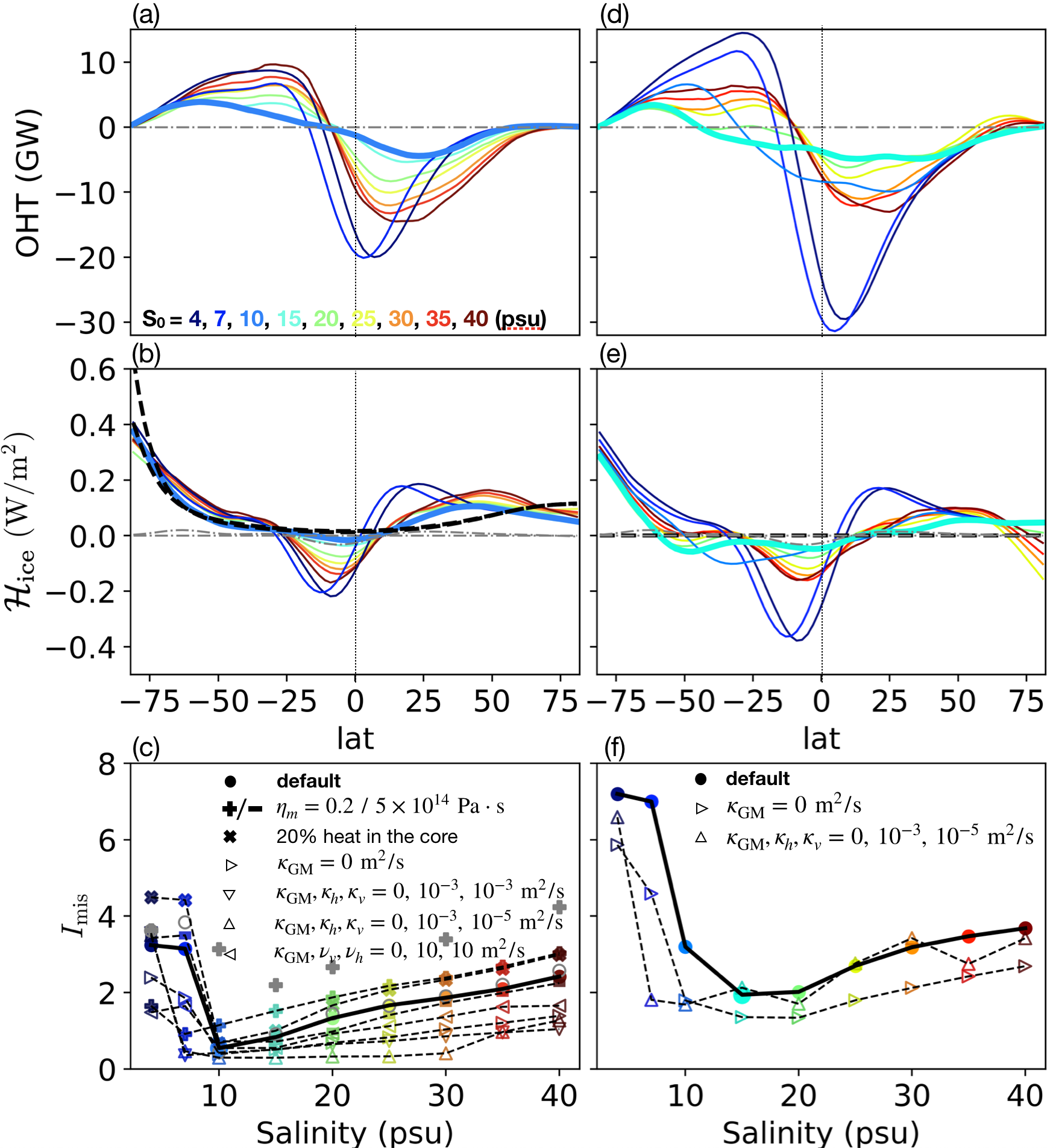}
    
    \caption{\small{Meridional heat transport and heat budget for the shell-heating scenario (left) and the core-heating scenario (right). The top panels show the vertically-integrated meridional OHT for various assumed $S_0$. Positive values denote northward heat transport. The middle panels show the inferred tidal heating $\hat{\mathcal{H}}_{\mathrm{ice}}$. The two black dashed curves in panel (b) are the profiles of $\mathcal{H}_{\mathrm{ice}}$ predicted by a model of tidal heating in the ice shell with $p_\alpha=-2$ and $p_\alpha=-1$, respectively: a more negative $p_\alpha$ indicates a stronger rheology feedback and thus yields a slightly more polar-amplified $\mathcal{H}_{\mathrm{ice}}$ profile. The two gray dot dashed lines denote zero and $-\rho_iL_fq$, respectively. The black dashed curves in panel (e) coincide with the zero line, because $\mathcal{H}_{\mathrm{ice}}=0$ when all the heating is in the core. The bottom panels show the mismatch index $I_{\mathrm{mix}}$, defined in Eq.~(\ref{eq:Imis}). The heat transport and inferred tidal heating profiles corresponding to the best-match experiments are highlighted by thicker curves in top and middle rows. Filled dots connected by a thick solid line corresponds to the default setup (GM diffusivity $\kappa_{\mathrm{GM}}=0.1$~m$^2$/s, horizontal/vertical diffusivity $\kappa_h=\kappa_v=0.005$~m$^2$/s, horizontal/vertical viscosity $\nu_h=\nu_v=50$~m$^2$/s, 100\% heat produced in the ice shell, and melting-point ice viscosity $\eta_m=10^{14}$~Pa$\cdot$s). Other symbols represent sensitivity tests as indicated by perturbations from the default in the legend. Cold to warm colors indicate increasing salinity from $S_0=4$ to $40$~psu. Gray empty circles show the sensitivity test with two times resolution (other parameters are the same as default), and gray plus sign symbols show results from 3D simulations (cube sphere coordinate, other parameters are the same as the $\eta_m=2\times10^{13}$~Pa$\cdot$s experiments). } }
    
    \label{fig:Hy-Htideinfer-mismatchindex}
  \end{figure*}

Ocean and the ice shell form a coupled-system: not only can the ice thickness variation drive ocean circulation, ocean circulation also redistribute heat and shape the ice shell. This connection can potentially allow us to understand the subsurface ocean properties (circulation, salinity etc.) using the well-constrained ice shell geometry \cite{Hemingway-Mittal-2019:enceladuss}. 
Since the temperature variation is dominated by a poleward warming pattern due to the freezing point shift under the ice shell, heat is converged from the poles toward the equator for all salinities, irrespective of the direction and depth of the overturning circulation of the ocean (see Fig.~\ref{fig:Hy-Htideinfer-mismatchindex}a). However, the opposite of what is needed to sustain an ice thickness gradient assuming no ice dissipation, given that heat loss is more efficient through the thin ice shell over the poles \cite{Cadek-Soucek-Behounkova-et-al-2019:long}. Therefore, without the polar-amplified ice dissipation, ice thickness variations will be flattened due to this so-called ice pump mechanism \cite{Lewis-Perkin-1986:ice}. With ice dissipation amplified over the poles, ice thickness variations may be maintained even with equatorward heat convergence, however, there is an upper limit for the amplitude.



To quantify this, we can compute the heat flux transmitted from the ocean to the ice $\mathcal{H}_{\mathrm{ocn}}$, \footnote{$\mathcal{H}_{\mathrm{ocn}}$ was calculated in two ways. First, we directly diagnosed the heat exchange between ice and ocean. Secondly we calculated the meridional heat flux convergence (Fig.\ref{fig:Hy-Htideinfer-mismatchindex}a) and added the bottom heat flux to it (Eq.9 in the SM). They yield identical results, indicating that the ocean was in equilibrium.}, and diagnose how much tidal heating is needed in the ice shell to close the ice's heat budget,
\begin{equation}
  \label{eq:heat-budget}
  \hat{\mathcal{H}}_{\mathrm{ice}}=\mathcal{H}_{\mathrm{cond}}-\mathcal{H}_{\mathrm{ocn}}-\rho_iL_fq.
\end{equation}
The $\hat{\mathcal{H}}_{\mathrm{ice}}$ inferred from our suite of ocean solutions is shown by the solid curves in Fig.\ref{fig:Hy-Htideinfer-mismatchindex}. For comparison, we also present the predicted tidal heating $\mathcal{H}_{\mathrm{ice}}$ given by a tidal dissipation model (details of the model can be found in section~1E of the SM) in the same figure using black dashed curves. However, we see that for many salinities (very fresh and very salty), the implied tidal heating is actually large and negative!, which is obviously not physically possible. The latent heating term on the right-hand-side of Eq.\ref{eq:heat-budget} is subject to significant uncertainties, because the freezing rate $q$ is derived from an idealized ice flow model (see section~1D in the SM) and the ice viscosity is poorly constrained. Assuming an infinitely viscous ice shell, $q$ would approach zero, and the equatorial $\hat{\mathcal{H}}_{\mathrm{ice}}$ would be $\sim$20~mW/m$^2$ higher ($-rho_iL_fq$ is shown by a gray dot dashed line in Fig.~\ref{fig:Hy-Htideinfer-mismatchindex}b). Taking this limit implies $\hat{\mathcal{H}}_{\mathrm{ice}}$ is positive everywhere for the 10~psu case. Instead, other cases exhibit negative heating rates near the equator because the heat convergence there exceeds that which can be lost by heat conduction through the thick equatorial ice shell.

The mismatch between $\hat{\mathcal{H}}_{\mathrm{ice}}$ and $\mathcal{H}_{\mathrm{ice}}$ can be measured by the following index 
\begin{equation}
  \label{eq:Imis}
  I_{\mathrm{mis}}=\overline{\left(\frac{\left|\hat{\mathcal{H}}_{\mathrm{ice}}-\mathcal{H}_{\mathrm{ice}}\right|}{\max\{\mathcal{H}_{\mathrm{ice}}, 20~\mathrm{mW/m^2}\}}\right)},
\end{equation}
where the over-bar represents a global area-weighted average. The max function in the denominator helps avoid the singularity when $\mathcal{H}_{\mathrm{ice}}\rightarrow 0$.

\paragraph{Dependence on ocean salinity.} Although ocean heat transport (OHT) is always equatorward, its amplitude co-varies with the circulation strength. As can be seen in Fig.~\ref{fig:Hy-Htideinfer-mismatchindex}a, the heat convergence in an ocean with an intermediate salinity is a small fraction of that in the end-member cases. Such sensitivity is unsurprising because the heat flux is proportional to the product of the overturning strength multiplied by a temperature contrast \cite{Czaja-Marshall-2006:partitioning}: the latter is broadly the same in all experiments but the former strongly depends on the mean salinity \footnote{The coldest and warmest locations are adjacent to the water-ice interface whose temperature is set by the local freezing point of water, meridional variation of which is dominated by pressure effects which are the same in all experiments: see Eq.7 in the SM where the dependence of freezing point on salinity and pressure is explicitly set out.}.

Those ocean solutions with a strong overturning circulation (e.g., $S_0=4,\ 40$~psu) focus a large amount of heat to low latitudes, introducing a heat budget discrepancy as large as 100~mW/m$^2$ \footnote{If used to melt ice, this heat will induce a melting rate of 9.31~km/Myr near the equator, which is twice as strong as the tendency implied by viscous ice flow (shown in Fig.~\ref{fig:EOS-Hice-FreezingRate}b).}, which is reflected by the large $I_{\mathrm{mis}}$. As noted above, near the equator, the inferred $\hat{\mathcal{H}}_{\mathrm{ice}}$ even becomes significantly negative, conflicting with the positive definite nature of tidal dissipation. The best match is achieved in the $S_0=10$~psu scenario (Fig.~\ref{fig:Hy-Htideinfer-mismatchindex}c). In such an intermediate salinity regime, the temperature and salinity-induced density anomalies almost cancel one-another out (see Fig.\ref{fig:Schematics}c), the overturning circulation is weak (Fig.\ref{fig:T-S-U-Psi}-e2), and the heat convergence is broadly consistent with tidal heating rates.

It is interesting to note that the increase in the mismatch is steeper on the fresh side of $10$~psu than the salty side (Fig.\ref{fig:Hy-Htideinfer-mismatchindex}c). This is related to the different energetics of ocean circulation in an $\alpha$ vs a $\beta$-ocean. As pointed out by \textit{Zeng \& Jansen 2021} \cite{Zeng-Jansen-2021:ocean}, in an $\alpha$ ocean, the buoyancy gain at the equator is deeper in the water column than the buoyancy loss at the poles and ocean circulation can always be energized against friction since dense polar water higher up the water column is transported to depth. However, in a $\beta$ ocean, the opposite is true and equatorial dense water cannot easily be drawn upward to the polar ice shell without extra energy input by diffusion \cite{Jeffreys-1925:fluid}. This difference can be seen in Fig.\ref{fig:T-S-U-Psi}-e. The overturning circulation in an $\alpha$ ocean (Fig.\ref{fig:T-S-U-Psi}-e1) can directly connect the water-ice interface at the pole to equatorial regions; in contrast in a $\beta$ ocean (Fig.\ref{fig:T-S-U-Psi}-e3), the circulation weakens moving poleward and almost completely vanishes in the fresh water lens formed under the polar ice shell. Strong stratification develops in the diffusive layer (Fig.\ref{fig:T-S-U-Psi}-c3) which sustains an upward buoyancy flux without strong circulation, as indicated in the schematic diagram Fig.\ref{fig:Schematics}b.

\paragraph{Dependence on the core-shell heat partition.} Heat generation occurs in the core does not seem to significantly change the overall circulation patterns, temperature/salinity profiles, and meridional heat transport, as can be seen from Fig.~\ref{fig:Hy-Htideinfer-mismatchindex}(d,e), Fig.S3 and Fig.S4. This is because the heating-induced bottom-to-top temperature difference is typically only a few tens of milliKelvin \footnote{The vertical temperature gradient induced by bottom heating is larger in a fresh ocean because of the suppression of convection.}, much smaller than the equator-to-pole temperature difference induced by the freezing point variations (Fig.\ref{fig:T-S-U-Psi}) which is order 0.1 Kelvin --- see Fig.\ref{fig:EOS-Hice-FreezingRate}b,c. However, the predicted tidal heating rates in the ice shell approach zero as the core becomes the dominant heat source. As a result, the equatorward OHT can no longer be effectively compensated by polar-amplified dissipation in the ice shell, and the discrepancy in the heat budget increases. $I_{\mathrm{mis}}$ for the 20\% core-heating scenario is plotted with a dashed line marked by crosses in Fig.~\ref{fig:Hy-Htideinfer-mismatchindex}c, as a sensitivity test. $I_{\mathrm{mis}}$ for the 100\% core-heating scenario plotted in Fig.~\ref{fig:Hy-Htideinfer-mismatchindex}f. Over the south pole where the ice shell is thin, heat is transported equatorward by ocean circulation, implying that polar ice will accumulate over time in the absence of local heating within the ice (black dashed curved in panel e). We also observe that for the pure core-heating scenario, the best-matching salinity is again 10-20~psu. More detailed discussions of the bottom heating solutions can be found in section~2A of the SM.

\paragraph{Sensitivity tests.} To explore sensitivity to parameter choices, we carried out many sets of experiments changing the assumed ice rheology and mixing rates in the ocean. By default, the melting point ice viscosity $\eta_m$ is set to $10^{14}$~Pa$\cdot$s, an intermediate choice between an estimated lower bound of $10^{13}$~Pa$\cdot$s and an upper bound of $10^{15}$~Pa$\cdot$s \cite{Tobie-Choblet-Sotin-2003:tidally}. In the ice rheology sensitivity test, we examined $\eta_m=2\times 10^{13}$~Pa$\cdot$s and $\eta_m=5\times 10^{14}$~Pa$\cdot$s. A lower (higher) ice viscosity induces stronger (weaker) ice flows, which require a greater (smaller) balancing freezing/melting rate; this in turn enhances (suppresses) the salinity flux imposed upon the ocean, giving rise to larger (weaker) salinity gradients. Compensating the density anomaly implied by this salinity gradient thus requires a more (less) negative $\alpha_T$ and lower $S_0$. As shown by the plus signs in Fig.~\ref{fig:Hy-Htideinfer-mismatchindex}c, the best matching $S_0$ is indeed reduced from 10~psu to 7~psu (the full solution is summarized in Fig.S11 of the SM) with $\eta_m=2\times 10^{13}$~Pa$\cdot$s and increased from 10~psu to 15 psu (the full solution is summarized in Fig.S12 of the SM) with $\eta_m=5\times 10^{14}$~Pa$\cdot$s.

The dissipation rate within the ocean driven by libration/tidal motions is also under debate \cite{Chen-Nimmo-Glatzmaier-2014:tidal, Hay-Matsuyama-2019:nonlinear, Rekier-Trinh-Triana-et-al-2019:internal} leading to a wide range of possible diapycnal diffusivities. Assuming a dissipation rate given by \textit{Rekier et al. 2019} \cite{Rekier-Trinh-Triana-et-al-2019:internal}, we find the explicit vertical diffusivity in Enceladus to be in order of $10^{-3}$~m$^2$/s (derivations can be found in the SM section~1B), which is orders of magnitude greater than the molecular diffusivity. As a comparison, \textit{Zeng \& Jansen 2021} \cite{Zeng-Jansen-2021:ocean} estimated the vertical diffusivity to be between $3\times10^{-10}$ and $3\times10^{-3}$~m$^2$/s. We apply this diffusivity to both vertical and horizontal directions in our default setup. In addition to the mixing induced by libration/tidal motions, baroclinic eddies will tend to flatten the isopycnals. Instead of resolving baroclinic eddies, we parameterize them using the Gent-McWilliams scheme commonly employed in terrestrial ocean models \cite{Gent-Mcwilliams-1990:isopycnal}. In SM section~1B, we estimate the GM diffusivity using scaling laws and find it to be around $0.1$~m$^2$/s. We note in passing that this diffusivity is orders of magnitude smaller than assumed in \textit{Lobo et al. 2021} \cite{Lobo-Thompson-Vance-et-al-2021:pole}, who used values more typical of Earth's ocean. To test our solutions' sensitivity to mixing parameters, we carried out several sets of experiments for all salinity scenarios using different Gent-McWilliams diffusivity $\kappa_{\mathrm{GM}}$, horizontal/vertical explicit diffusivity $\kappa_h,\ \kappa_v$ and horizontal/vertical viscosity $\nu_h, \nu_v$. As shown by the triangular markers in Fig.~\ref{fig:Hy-Htideinfer-mismatchindex}c, all sensitivity tests have $I_{\mathrm{mis}}$ first decrease then increase with the assumed ocean salinity, with a minimum achieved near $10$~psu, just as in the control experiments (solid line with filled dots). On top of the general trend, the equatorward heat transport decreases when smaller $\nu$ and $\kappa$ are used, making the heat budget more in balance. This happens because diffusive processes are responsible for pumping dense water back to the surface to finish the circulation loop and viscous processes help balance the Coriolis term in the zonal momentum equation so that the shallow circulation can form near the ice. When both viscosity and diffusivity are weak, the circulation and heat transport slow down as can be seen by comparing Fig.~\ref{fig:T-S-U-Psi-lowvisc} with Fig.~\ref{fig:T-S-U-Psi}. For more detailed discussions about the sensitivity tests, readers are referred to SM section~2B. Given this sensitivity, more understanding of the ocean stratification and dissipation rate is therefore needed in order to make better constraints on ocean salinity.

Thus far, all experiments are run under the same resolution with only two dimensions. In order to examine the sensitivity to resolution and 3D dynamics, we carried out two sets of experiments, one with twice the resolution (other parameters are default values) and the other with 3D configuration (other parameters are the same as the low-$\eta_m$ experiments). The results are shown by gray circles and gray plus sign symbols, respectively, in Fig.~\ref{fig:Hy-Htideinfer-mismatchindex}c. Doubling the resolution leads to almost identical results and 3D configuration doesn't lead to qualitative changes either. The solutions for these sensitivity tests can be found in Fig.~S13 and Fig.S14 in SM section~2D. Although the 3D dynamics is on in the sensitivity test, the simulation is run under a fairly moderate resolution (64x128 grid). This is necessary for us to integrate tens of thousands of years until equilibrium is reached, but compromise the dynamics. More work is therefore required to quantify the contribution of sub-gridscale eddies.


 \section{Exploring a wider parameter regime using a conceptual model} 
  The numerical solutions presented above suggest that if Enceladus' ocean is of intermediate salinity, near the transition between an $\alpha$ ocean and a $\beta$ ocean, then equatorial convergence of heat is minimized, allowing a thick equatorial ice shell to be maintained. This is much less likely in very fresh or very salty oceans. Here, we use a conceptual model that is similar to that of Stommel \cite{Marotzke-2000:abrupt} to highlight the physical processes that controls the circulation strength and explore a wider range of parameter space that can be applied to other icy moons. 

 We represent the overall density contrast using the equator minus north pole density difference $\Delta \rho$. The temperature-related density anomaly is $-\alpha_T\Delta T$, and salinity-related one is $\beta_S\Delta S$, where $\Delta T$ and $\Delta S$ are the potential temperature and salinity anomaly at the equator relative to the north pole. We expect the mass transport by overturning circulation $\Psi$ to vary proportionally with $\Delta \rho$. For simplicity, we assume a linear form
  \begin{equation}
    \label{eq:Psi-drho}
    \Psi=A(-\alpha_T\Delta T+\beta_S\Delta S),
  \end{equation}
  where the constant $A$ maps the density contrast on to the vigor of the overturning circulation, $\beta_S\approx 8\times10^{-4}$/psu for all $S_0$, but $\alpha_T$ depends sensitively on $S_0$, as given by the Gibbs Seawater Toolbox \cite{McDougall-Barker-2011:getting}.

 The temperature contrast $\Delta T$ is determined by the pressure-induced freezing point shift from the north pole to the equator\footnote{Ocean temperature at the water-ice interface is relaxed tightly toward the local freezing point, which varies by around 0.1K across latitude. Reducing the meridional temperature contrast by only 10\% (ocean temperature deviates $\delta T=$0.01K from the freezing point) would induce a heat flux of $\gamma_TC_p\rho_0\delta T = $400~mW/m$^2$ ($\gamma_T=10^{-5}$~m/s is the water-ice exchange coefficient for temperature, $C_p=4000$~J/kg/K is the heat capacity of water and $\rho_0$ is the reference water density), which is unacceptably large.},
  \begin{equation}
    \label{eq:deltaT}
    \Delta T=b_0\Delta P=b_0\rho_i g \Delta H,
  \end{equation}
  where $b_0=-7.61\times10^{-4}$~K/dbar, $\rho_i=917$~kg/m$^3$ is the ice density, $g=0.113$~m/s$^2$ is the surface gravity of Enceladus and $\Delta H=11$~km is the difference in ice thickness between the equator and the north pole.

 The lateral salinity flux is given by the product of $\Psi$ and a salinity contrast $\Delta S$ and balances the salinity flux due to freezing and melting yielding (see a detailed derivation in \textit{Marshall \& Radko 2003} \cite{Marshall-Radko-2003:residual}):
  \begin{equation}
    \label{eq:deltaS}
    \left|\Psi\right|\Delta S=\rho_0S_0\Delta q \times (\pi a^2)
  \end{equation}
  Here, $\Delta q$, the difference in the freezing rate between equator and pole, is chosen to be $4$~km/Myr based on Fig.\ref{fig:EOS-Hice-FreezingRate}b. $a=250$~km is the radius of Enceladus, and $S_0$ is the mean salinity. The fact that $\Psi$ and $\Delta S$ appear as a product indicates that the salinity gradient will weaken as the overturning circulation strengthens for fixed salinity forcing. 

  Combining Eq~(\ref{eq:Psi-drho}), Eq~(\ref{eq:deltaT}) and Eq~(\ref{eq:deltaS}), we can solve for $\Delta S$ and $\Psi$. The only tunable parameter here is $A$, which controls the strength of the overturning circulation and can be adjusted to fit that obtained in our ocean model. With $A$ set to $4.5\times10^{13}$~kg/s, we obtain the solutions shown by the open circles in Fig.\ref{fig:Schematics}c (the size of the circle reflect the amplitude of $\Psi$). The conceptual model solution broadly captures the behavior of the numerical simulations (filled circles), including the strengthening of the overturning circulation and the weakening of salinity gradient away from the transition zone separating the alpha ocean and beta ocean. 

 When $S_0<22$~psu, $\alpha_T\Delta T$ and $\beta_S\Delta S$ take opposite signs, and depending on which one has a greater absolute value, the circulation $\Psi$ can be in either direction. Each possibility corresponds to one solution. The solution in the alpha ocean regime matches the numerical model results. The solution in the beta ocean regime (overlapped by a cross mark in Fig.\ref{fig:Schematics}) requires an extraordinarily strong salinity gradient to dominate the negative $\alpha_T\Delta T$. This is only possible when the salinity- and temperature-induced density variations exactly cancel one-another and when there is no other form of mixing, and so implausible.
  
  What is the all-important heat flux implied by our conceptual model? Analogously to Eq~(\ref{eq:deltaS}), the meridional heat transport can be written
  \begin{equation}
    \label{eq:water-ice-heat-exchange}
    \mathcal{H}_{\mathrm{ocn}}=\frac{C_p|\Psi|\Delta T}{4\pi a^2}.
  \end{equation}
  This is shown as a function of salinity and equator-to-pole thickness variations in Fig.\ref{fig:other-moons}a. Recall that the water-ice heat exchange must be smaller than the heat conduction rate of 50~mW/m$^2$ to maintain observed thickness variations of the Enceladus ice shell. The likely/unlikely parameter regime is shaded white/red \cite{Iess-Stevenson-Parisi-et-al-2014:gravity, Beuthe-Rivoldini-Trinh-2016:enceladuss, Tajeddine-Soderlund-Thomas-et-al-2017:true, Cadek-Soucek-Behounkova-et-al-2019:long, Hemingway-Mittal-2019:enceladuss}. We see that a salinity between roughly 10-22~psu (marked by two vertical blue dashed lines) is required to maintain ice thickness variations as large as are seen on Enceladus.


 \section{Concluding Remarks}
 In conclusion, from knowledge of the geometry of the ice shell on Enceladus we have deduced likely patterns of (i) salinity gradients associated with freezing and melting and (ii) under-ice temperature gradients due to the depression of the freezing point of water due to pressure. We have considered the resulting ocean circulation driven by these boundary conditions, along with the effect of putative heat fluxes emanating from the bottom if tidal dissipation in the core is significant. We find that the ocean circulation strongly depends on its assumed salinity. If the ocean is fresh, sinking occurs at the poles driven by the meridional temperature gradient (Fig.\ref{fig:T-S-U-Psi} first column); if the ocean is salty, sinking occurs at the equator driven by the salinity gradient (Fig.\ref{fig:T-S-U-Psi} third column). Either way, heat is converged toward the equator as the warm polar water is mixed with the cold equatorial water.

 Without a polar-amplified ice dissipation to counterbalance the equatorward heat transport, the polar (equatorial) ice shell will inevitably freeze (melt), because the conductive heat loss through the ice shell also tends to cool the polar regions more. This, together with the ice flow, will flatten ice geometry in the core-heating scenarios. Ocean salinity affects the ocean circulation and thereby the heat budget, along with the heat partition between the core and the ice shell -- this opens up a chance for us to infer these properties using the relatively well-constrained ice shell geometry. We carry out the exercise here to demonstrate the procedure. Based on the results from our idealized 2D circulation model, scenarios without plenty of heat production in the shell cannot save the equatorial ice shell from being thinned by the equatorward heat convergence and the ice flow. Even with all heat produced in the ice shell, the equatorward heat convergence in a very salty or very fresh ocean seems too strong to maintain the heat budget in balance (more heat is converged to the equator than what can be lost through heat conduction).
 Only when heat production occurs primarily in the ice shell and salinity is in an intermediate range (our calculations suggest the most plausible range is 10-30~psu), the temperature and salinity-driven overturning circulation largely cancel one-another and the equatorward heat transport is small enough that the polar-amplified ice dissipation can sustain a broadly balanced heat budget. 
 As discussed in the introduction, such lower salinities are consistent with the predictions from chemical equilibrium models of the interaction between the rocky core and the ocean \cite{Zolotov-2007:oceanic, zolotov2014can, Glein-Postberg-Vance-2018:geochemistry}. 
  
\begin{figure*}[htp!]
    \centering
    \includegraphics[page=6,width=0.9\textwidth]{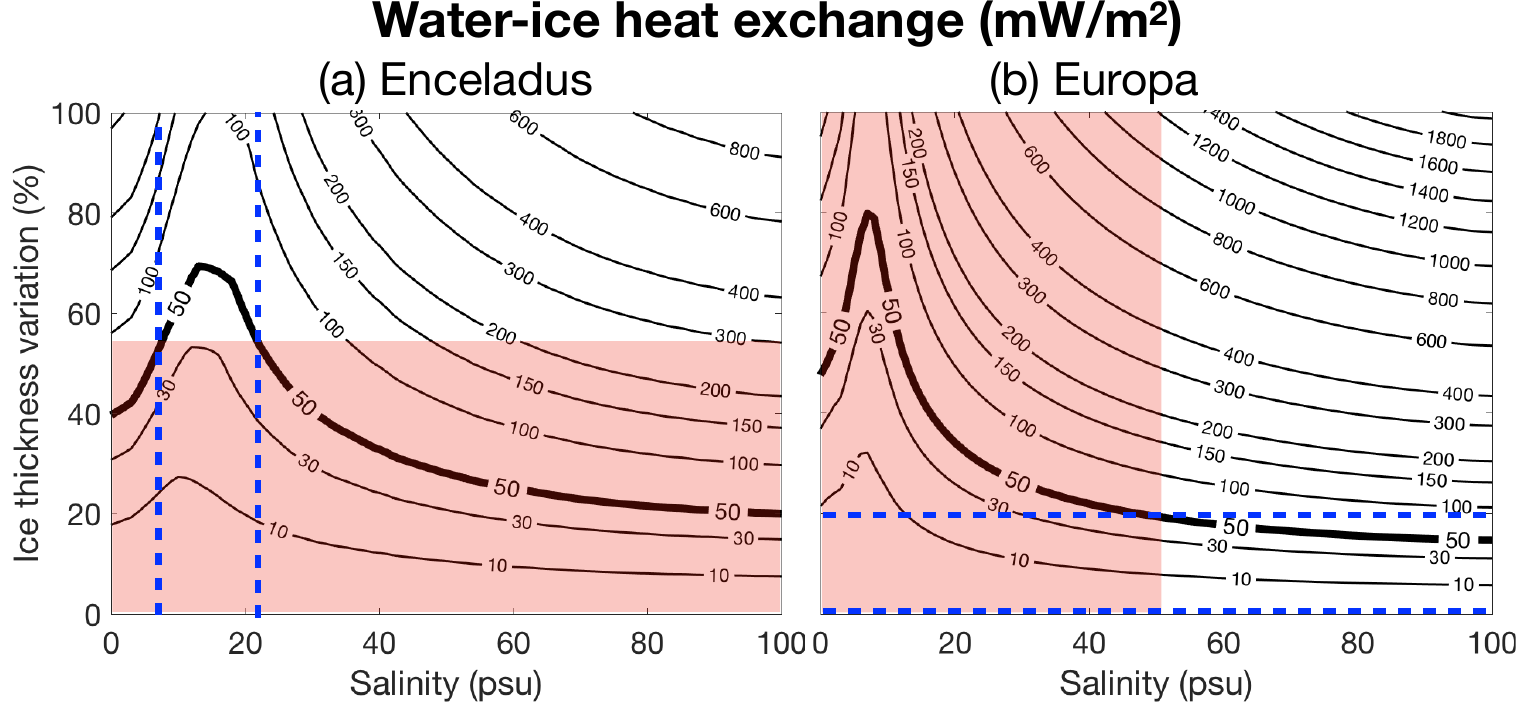}
    
    \caption{\small{The water-ice heat exchange in equatorial regions for Enceladus (left) and Europa (right) predicted by our conceptual model (Eq.\ref{eq:water-ice-heat-exchange}) as a function of salinity and equator-to-pole percentage ice thickness variation (equatorial minus polar ice thickness divided by the mean). A degree-2 poleward-thinning structure is assumed and physical parameters are defined in the SM Table~S1. Parameter regimes that are not consistent with observations are shaded in red: the ice shell of Enceladus is thought to have large thickness variations \cite{Iess-Stevenson-Parisi-et-al-2014:gravity, Beuthe-Rivoldini-Trinh-2016:enceladuss, Tajeddine-Soderlund-Thomas-et-al-2017:true, Cadek-Soucek-Behounkova-et-al-2019:long, Hemingway-Mittal-2019:enceladuss} and the salinity of Europa is thought to have a salinity likely beyond 50~psu \cite{Hand-Chyba-2007:empirical}. The 50~mW/m$^2$ contour is highlighted by a thicker curve; heat exchange rates that exceed this are considered unphysical as the equatorial ice sheet of both Enceladus and Europa only allow $\sim$40~mW/m$^2$ or so of heat flux to conduct through. The equatorial heat loss rates are similar on the two icy moons because the likely thin ice shell on Europa (15km) compensates for its relatively high surface temperature (110K). Our simplified model suggests that salinities on Enceladus and ice thickness variations for Europa lie in the region enclosed by the blue dashed lines.  The most plausible ice-thickness variations and salinity on Enceladus and Europa thus lie in the white areas between the blue dashed lines.} }
    
    \label{fig:other-moons}
  \end{figure*}

  Insights from this study may also have implications for other icy moons. For example, Europa may have a salinity above $50$ psu, as implied by the strong magnetic induction field measured by the Galileo mission \cite{Hand-Chyba-2007:empirical} -- see \textit{Zolotov \& Shock 2001} \cite{Zolotov-Shock-2001:composition}, \textit{Khurana et al. 2009} \cite{khurana2009electromagnetic}, \textit{Vance et al. 2020} \cite{vance2020magnetic} for discussion of possible ocean compositions together with uncertainties. At salinities as high as $50$~psu, the ice pump mechanism and associated dynamics (as discussed above) would remove any fluctuations in the ice shell thickness, leaving a relatively flat ice sheet. Applying our simplified model to Europa in Fig.\ref{fig:other-moons}b, the predicted water-ice heat exchange becomes unphysically large (marked by the thick contour) when the equator-to-pole thickness variation is beyond 20\% (indicated by the two horizontal blue dashed lines). Due to the large gravity on Europa, the circulation coefficient $A$ is likely to be even greater \footnote{Diffusivity is another factor that can affect $A$.}, which indicates even lower tolerance for the thickness variation. This is consistent with the observation that the mean ice thickness is less than $15$~km (best match at $4$~km, see also \cite{Howell-2021:likely} for other estimates) \cite{Hand-Chyba-2007:empirical, Park-Bills-Buffington-et-al-2015:improved} and that no fissures that mimic the ``tiger stripes'' of Enceladus have been found on Europa. For icy moons with thicker ice shells, such as Dione, Titan, Ganymede and Callisto, the high pressure under the ice shell would remove the anomalous expansion for all salinities, making it impossible for the alpha and beta effects to cancel one another. Furthermore, the ice flow becomes very efficient, because it is proportional to the ice thickness cubed (see Eq.17 in SM). Our conceptual model indeed indicates that icy ocean worlds with thick ice shells are likely to have small spatial shell thickness variations. This is consistent with shell thickness reconstructions based on gravity and shape measurements \cite{Zannoni-Hemingway-Casajus-et-al-2020:gravity, Durante-Hemingway-Racioppa-et-al-2019:titans, Nimmo-Bills-2010:shell, Vance-Bills-Cochrane-et-al-2020:magnetic}. With improved measurements of gravity, topography, and induced magnetic fields for icy moons by future space missions (e.g., Europa Clipper), our framework can be used to examine the consistency between observations. 

\begin{thebibliography}{10}

\bibitem{MITgcm-group-2010:mitgcm}
{MITgcm-group}, {MITgcm} {U}ser {M}anual, {\it Online documentation\/},
  {MIT}/{EAPS}, Cambridge, MA 02139, USA (2010).
  \rm{http://mitgcm.org/public/r2\_manual/latest/online\_documents/manual.html}.

\bibitem{Marshall-Adcroft-Hill-et-al-1997:finite}
J.~Marshall, A.~Adcroft, C.~Hill, L.~Perelman, C.~Heisey, {\it J. Geophys.
  Res.\/} {\bf 102}, 5,753 (1997).

\bibitem{Losch-2008:modeling}
M.~Losch, {\it J. Geophys. Res.\/} {\bf 113}, 10.1029/2007JC004368 (2008).

\bibitem{Hemingway-Mittal-2019:enceladuss}
D.~J. Hemingway, T.~Mittal, {\it Icarus\/} {\bf 332}, 111 (2019).

\bibitem{Rekier-Trinh-Triana-et-al-2019:internal}
J.~Rekier, A.~Trinh, S.~Triana, V.~Dehant, {\it Journal of Geophysical
  Research: Planets\/} {\bf 124}, 2198 (2019).

\bibitem{Wunsch-Ferrari-2004:vertical}
C.~Wunsch, R.~Ferrari, {\it Ann. Rev. Fluid Mech.\/} {\bf 36}, 281 (2004).

\bibitem{Klinger-Marshall-Send-1996:representation}
B.~A. Klinger, J.~Marshall, U.~Send, {\it Journal of Geophysical Research:
  Oceans\/} {\bf 101}, 18175 (1996).

\bibitem{Kang-Bire-Campin-et-al-2020:differing}
W.~Kang, {\it et~al.\/}, {\it arXiv preprint arXiv:2008.03764\/}  (2020).

\bibitem{Jones-Marshall-1993:convection}
H.~Jones, J.~Marshall, {\it J. Phys. Oceanogr.\/} {\bf 23}, 1009 (1993).

\bibitem{Redi-1982:oceanic}
M.~H. Redi, {\it J. Phys. Oceanogr.\/} {\bf 12}, 1154 (1982).

\bibitem{Gent-Mcwilliams-1990:isopycnal}
P.~R. Gent, J.~C. Mcwilliams, {\it Journal of Physical Oceanography\/} {\bf
  20}, 150 (1990).

\bibitem{Visbeck-Marshall-Haine-et-al-1997:specification}
M.~Visbeck, J.~Marshall, T.~Haine, M.~Spall, {\it J. Phys. Oceanogr.\/} {\bf
  27}, 381 (1997).

\bibitem{Lobo-Thompson-Vance-et-al-2021:pole}
A.~H. Lobo, A.~F. Thompson, S.~D. Vance, S.~Tharimena, {\it Nature
  Geoscience\/} pp. 1--5 (2021).

\bibitem{McDougall-Barker-2011:getting}
T.~J. McDougall, P.~M. Barker, {\it SCOR/IAPSO WG\/} {\bf 127}, 1 (2011).

\bibitem{McDougall-Jackett-Wright-et-al-2003:accurate}
T.~J. McDougall, D.~R. Jackett, D.~G. Wright, R.~Feistel, {\it Journal of
  Atmospheric and Oceanic Technology\/} {\bf 20}, 730 (2003).

\bibitem{Tajeddine-Soderlund-Thomas-et-al-2017:true}
R.~Tajeddine, {\it et~al.\/}, {\it Icarus\/} {\bf 295}, 46 (2017).

\bibitem{Chen-Nimmo-2011:obliquity}
{Chen, E M A}, {Nimmo, F}, {\it Icarus\/} {\bf 214}, 779 (2011).

\bibitem{Beuthe-2016:crustal}
M.~Beuthe, {\it Icarus\/} {\bf 280}, 278 (2016).

\bibitem{Hay-Matsuyama-2019:nonlinear}
H.~C. F.~C. Hay, I.~Matsuyama, {\it Icarus\/} {\bf 319}, 68 (2019).

\bibitem{Beuthe-2019:enceladuss}
M.~Beuthe, {\it Icarus\/} {\bf 332}, 66  (2019).

\bibitem{Choblet-Tobie-Sotin-et-al-2017:powering}
G.~Choblet, {\it et~al.\/}, {\it Nature Astronomy\/} {\bf 1}, 841 (2017).

\bibitem{Holland-Jenkins-1999:modeling}
D.~M. Holland, A.~Jenkins, {\it J. Phys. Oceanogr.\/} {\bf 29}, 1787 (1999).

\bibitem{Beuthe-2018:enceladuss}
M.~Beuthe, {\it Icarus\/} {\bf 302}, 145 (2018).

\bibitem{Ashkenazy-Sayag-Tziperman-2018:dynamics}
Y.~Ashkenazy, R.~Sayag, E.~Tziperman, {\it Nature Astronomy\/} {\bf 2}, 43
  (2018).

\bibitem{Tobie-Choblet-Sotin-2003:tidally}
G.~Tobie, G.~Choblet, C.~Sotin, {\it J. Geophys. Res - Atmospheres\/} {\bf
  108}, 219 (2003).

\bibitem{Kang-Flierl-2020:spontaneous}
W.~Kang, G.~Flierl, {\it PNAS\/} {\bf 117}, 14764 (2020).

\bibitem{McCarthy-Cooper-2016:tidal}
C.~McCarthy, R.~F. Cooper, {\it Earth and Planetary Science Letters\/} {\bf
  443}, 185 (2016).

\bibitem{Renaud-Henning-2018:increased}
J.~P. Renaud, W.~G. Henning, {\it Astrophysical Journal\/} {\bf 857}, 98
  (2018).

\bibitem{Hand-Chyba-2007:empirical}
K.~Hand, C.~Chyba, {\it Icarus\/} {\bf 189}, 424 (2007).

\bibitem{Petrenko-Whitworth-1999:physics}
V.~Petrenko, R.~Whitworth, {\it Physics of Ice\/} (OUP Oxford, 1999).

\bibitem{Zeng-Jansen-2021:ocean}
Y.~Zeng, M.~F. Jansen, {\it arXiv preprint arXiv:2101.10530\/}  (2021).

\bibitem{Lemasquerier-Grannan-Vidal-et-al-2017:libration}
D.~Lemasquerier, {\it et~al.\/}, {\it Journal of Geophysical Research:
  Planets\/} {\bf 122}, 1926 (2017).

\end{thebibliography}


\begin{thebibliography}{10}

\bibitem{Postberg-Kempf-Schmidt-et-al-2009:sodium}
F.~Postberg, {\it et~al.\/}, {\it Nature\/} {\bf 459}, 1098 (2009).

\bibitem{Thomas-Tajeddine-Tiscareno-et-al-2016:enceladus}
P.~Thomas, {\it et~al.\/}, {\it Icarus\/} {\bf 264}, 37 (2016).

\bibitem{Choblet-Tobie-Sotin-et-al-2017:powering}
G.~Choblet, {\it et~al.\/}, {\it Nature Astronomy\/} {\bf 1}, 841 (2017).

\bibitem{Beuthe-2019:enceladuss}
M.~Beuthe, {\it Icarus\/} {\bf 332}, 66  (2019).

\bibitem{Waite-Combi-Ip-et-al-2006:cassini}
J.~H. Waite, {\it et~al.\/}, {\it Science\/} {\bf 311}, 1419 (2006).

\bibitem{Hsu-Postberg-Sekine-et-al-2015:ongoing}
H.-W. Hsu, {\it et~al.\/}, {\it Nature\/} {\bf 519}, 207 (2015).

\bibitem{Waite-Glein-Perryman-et-al-2017:cassini}
J.~H. Waite, {\it et~al.\/}, {\it Science\/} {\bf 356}, 155 (2017).

\bibitem{Postberg-Khawaja-Abel-et-al-2018:macromolecular}
F.~Postberg, {\it et~al.\/}, {\it Nature\/} {\bf 558}, 564 (2018).

\bibitem{Taubner-Pappenreiter-Zwicker-et-al-2018:biological}
R.-S. Taubner, {\it et~al.\/}, {\it Nature communications\/} {\bf 9}, 1 (2018).

\bibitem{Glein-Waite-2020:carbonate}
C.~R. Glein, J.~H. Waite, {\it Geophysical Research Letters\/} {\bf 47}, 591
  (2020).

\bibitem{Hansen-Esposito-Stewart-et-al-2006:enceladus}
C.~J. Hansen, {\it et~al.\/}, {\it Science\/} {\bf 311}, 1422 (2006).

\bibitem{Howett-Spencer-Pearl-et-al-2011:high}
C.~J.~A. Howett, J.~R. Spencer, J.~Pearl, M.~Segura, {\it Journal of
  Geophysical Research-Atmospheres\/} {\bf 116}, 189 (2011).

\bibitem{Spencer-Howett-Verbiscer-et-al-2013:enceladus}
J.~R. {Spencer}, {\it et~al.\/}, {\it European Planetary Science Congress\/}
  {\bf 8}, EPSC2013 (2013).

\bibitem{McCarthy-Cooper-2016:tidal}
C.~McCarthy, R.~F. Cooper, {\it Earth and Planetary Science Letters\/} {\bf
  443}, 185 (2016).

\bibitem{Tobie-Mocquet-Sotin-2005:tidal}
G.~Tobie, A.~Mocquet, C.~Sotin, {\it Icarus\/} {\bf 177}, 534 (2005).

\bibitem{Rekier-Trinh-Triana-et-al-2019:internal}
J.~Rekier, A.~Trinh, S.~Triana, V.~Dehant, {\it Journal of Geophysical
  Research: Planets\/} {\bf 124}, 2198 (2019).

\bibitem{Soderlund-Kalousova-Buffo-et-al-2020:ice}
K.~M. Soderlund, {\it et~al.\/}, {\it Space Science Reviews\/} {\bf 216}, 1
  (2020).

\bibitem{Hay-Matsuyama-2019:nonlinear}
H.~C. F.~C. Hay, I.~Matsuyama, {\it Icarus\/} {\bf 319}, 68 (2019).

\bibitem{McDougall-Barker-2011:getting}
T.~J. McDougall, P.~M. Barker, {\it SCOR/IAPSO WG\/} {\bf 127}, 1 (2011).

\bibitem{Zeng-Jansen-2021:ocean}
Y.~Zeng, M.~F. Jansen, {\it arXiv preprint arXiv:2101.10530\/}  (2021).

\bibitem{Cullum-Stevens-Joshi-2016:importance}
J.~Cullum, D.~P. Stevens, M.~M. Joshi, {\it Proceedings of the National Academy
  of Sciences\/} {\bf 113}, 4278 (2016).

\bibitem{Cael-Ferrari-2017:oceans}
B.~Cael, R.~Ferrari, {\it Geophysical Research Letters\/} {\bf 44}, 1886
  (2017).

\bibitem{Travis-Schubert-2015:keeping}
B.~J. Travis, G.~Schubert, {\it Icarus\/} {\bf 250}, 32 (2015).

\bibitem{Hemingway-Mittal-2019:enceladuss}
D.~J. Hemingway, T.~Mittal, {\it Icarus\/} {\bf 332}, 111 (2019).

\bibitem{Soucek-Behounkova-Cadek-et-al-2019:tidal}
O.~Soucek, {\it et~al.\/}, {\it Icarus\/} {\bf 328}, 218 (2019).

\bibitem{Robuchon-Choblet-Tobie-et-al-2010:coupling}
G.~Robuchon, {\it et~al.\/}, {\it Icarus\/} {\bf 207}, 959 (2010).

\bibitem{Shoji-Hussmann-Kurita-et-al-2013:ice}
D.~Shoji, H.~Hussmann, K.~Kurita, F.~Sohl, {\it Icarus\/} {\bf 226}, 10 (2013).

\bibitem{Behounkova-Tobie-Choblet-et-al-2013:impact}
M.~B{\v{e}}hounkov{\'a}, G.~Tobie, G.~Choblet, O.~{\v{C}}adek, {\it Icarus\/}
  {\bf 226}, 898 (2013).

\bibitem{Gevorgyan-Boue-Ragazzo-et-al-2020:andrade}
Y.~Gevorgyan, G.~Bou{\'e}, C.~Ragazzo, L.~S. Ruiz, A.~C. Correia, {\it
  Icarus\/} {\bf 343}, 113610 (2020).

\bibitem{Zolotov-2007:oceanic}
M.~Y. Zolotov, {\it Geophysical Research Letters\/} {\bf 34} (2007).

\bibitem{zolotov2014can}
M.~Y. Zolotov, F.~Postberg, {\it LPI\/} p. 2496 (2014).

\bibitem{Glein-Postberg-Vance-2018:geochemistry}
C.~Glein, F.~Postberg, S.~Vance, {\it Enceladus and the icy moons of Saturn\/}
  {\bf 39} (2018).

\bibitem{Ingersoll-Nakajima-2016:controlled}
A.~P. Ingersoll, M.~Nakajima, {\it Icarus\/} {\bf 272}, 319 (2016).

\bibitem{Fox-Powell-Cousins-2021:partitioning}
M.~G. Fox-Powell, C.~R. Cousins, {\it Journal of Geophysical Research:
  Planets\/} {\bf 126}, e2020JE006628 (2021).

\bibitem{Iess-Stevenson-Parisi-et-al-2014:gravity}
L.~Iess, {\it et~al.\/}, {\it Science\/} {\bf 344}, 78 (2014).

\bibitem{Beuthe-Rivoldini-Trinh-2016:enceladuss}
M.~Beuthe, A.~Rivoldini, A.~Trinh, {\it Geophysical Research Letters\/} {\bf
  43}, 10,088 (2016).

\bibitem{Tajeddine-Soderlund-Thomas-et-al-2017:true}
R.~Tajeddine, {\it et~al.\/}, {\it Icarus\/} {\bf 295}, 46 (2017).

\bibitem{Cadek-Soucek-Behounkova-et-al-2019:long}
O.~{\v{C}}adek, {\it et~al.\/}, {\it Icarus\/} {\bf 319}, 476 (2019).

\bibitem{Tobie-Choblet-Sotin-2003:tidally}
G.~Tobie, G.~Choblet, C.~Sotin, {\it J. Geophys. Res - Atmospheres\/} {\bf
  108}, 219 (2003).

\bibitem{Barr-Showman-2009:heat}
A.~C. Barr, A.~P. Showman, {\it Europa\/} (Univ. Arizona Press, 2009), pp.
  405--430.

\bibitem{Ashkenazy-Sayag-Tziperman-2018:dynamics}
Y.~Ashkenazy, R.~Sayag, E.~Tziperman, {\it Nature Astronomy\/} {\bf 2}, 43
  (2018).

\bibitem{Barr-McKinnon-2007:convection}
A.~C. Barr, W.~B. McKinnon, {\it Geophysical Research Letters\/} {\bf 34}
  (2007).

\bibitem{Lobo-Thompson-Vance-et-al-2021:pole}
A.~H. Lobo, A.~F. Thompson, S.~D. Vance, S.~Tharimena, {\it Nature
  Geoscience\/} pp. 1--5 (2021).

\bibitem{Lewis-Perkin-1986:ice}
E.~Lewis, R.~Perkin, {\it J. Geophys. Res\/} {\bf 91}, 756 (1986).

\bibitem{Kang-Flierl-2020:spontaneous}
W.~Kang, G.~Flierl, {\it PNAS\/} {\bf 117}, 14764 (2020).

\bibitem{Czaja-Marshall-2006:partitioning}
A.~Czaja, J.~Marshall, {\it Journal of the atmospheric sciences\/} {\bf 63},
  1498 (2006).

\bibitem{Jeffreys-1925:fluid}
H.~Jeffreys, {\it Quarterly Journal of the Royal Meteorological Society\/} {\bf
  51}, 347 (1925).

\bibitem{Chen-Nimmo-Glatzmaier-2014:tidal}
E.~M.~A. Chen, F.~Nimmo, G.~A. Glatzmaier, {\it Icarus\/} {\bf 229}, 11 (2014).

\bibitem{Gent-Mcwilliams-1990:isopycnal}
P.~R. Gent, J.~C. Mcwilliams, {\it Journal of Physical Oceanography\/} {\bf
  20}, 150 (1990).

\bibitem{Marotzke-2000:abrupt}
J.~Marotzke, {\it Proc. Natl. Acad. Sci. U.S.A.\/} {\bf 97}, 1347 (2000).

\bibitem{Marshall-Radko-2003:residual}
J.~Marshall, T.~Radko, {\it J. Phys. Oceanogr.\/} {\bf 33}, 2341 (2003).

\bibitem{Hand-Chyba-2007:empirical}
K.~Hand, C.~Chyba, {\it Icarus\/} {\bf 189}, 424 (2007).

\bibitem{Zolotov-Shock-2001:composition}
M.~Y. Zolotov, E.~L. Shock, {\it Journal of Geophysical Research: Planets\/}
  {\bf 106}, 32815 (2001).

\bibitem{khurana2009electromagnetic}
K.~K. Khurana, M.~G. Kivelson, K.~P. Hand, C.~T. Russell, {\it Robert. T.
  Pappalardo, William. B. McKinnon, and K. Khurana, Editors. Europa, University
  of Arizona Press, Tucson\/} pp. 572--586 (2009).

\bibitem{vance2020magnetic}
S.~D. Vance, {\it et~al.\/}, {\it arXiv preprint arXiv:2002.01636\/}  (2020).

\bibitem{Howell-2021:likely}
S.~M. Howell, {\it The Planetary Science Journal\/} {\bf 2}, 129 (2021).

\bibitem{Park-Bills-Buffington-et-al-2015:improved}
R.~S. Park, {\it et~al.\/}, {\it Planetary and Space Science\/} {\bf 112}, 10
  (2015).

\bibitem{Zannoni-Hemingway-Casajus-et-al-2020:gravity}
M.~Zannoni, D.~Hemingway, L.~G. Casajus, P.~Tortora, {\it Icarus\/} p. 113713
  (2020).

\bibitem{Durante-Hemingway-Racioppa-et-al-2019:titans}
D.~Durante, D.~Hemingway, P.~Racioppa, L.~Iess, D.~Stevenson, {\it Icarus\/}
  {\bf 326}, 123 (2019).

\bibitem{Nimmo-Bills-2010:shell}
F.~Nimmo, B.~Bills, {\it Icarus\/} {\bf 208}, 896 (2010).

\bibitem{Vance-Bills-Cochrane-et-al-2020:magnetic}
S.~D. Vance, {\it et~al.\/}, {\it arXiv preprint arXiv:2002.01636\/}  (2020).


\bibitem{MITgcm-group-2010:mitgcm}
{MITgcm-group}, {MITgcm} {U}ser {M}anual, {\it Online documentation\/},
  {MIT}/{EAPS}, Cambridge, MA 02139, USA (2010).
  \rm{http://mitgcm.org/public/r2\_manual/latest/online\_documents/manual.html}.

  \bibitem{Marshall-Adcroft-Hill-et-al-1997:finite}
J.~Marshall, A.~Adcroft, C.~Hill, L.~Perelman, C.~Heisey, {\it J. Geophys.
  Res.\/} {\bf 102}, 5,753 (1997).

\bibitem{Losch-2008:modeling}
M.~Losch, {\it J. Geophys. Res.\/} {\bf 113}, 10.1029/2007JC004368 (2008).



\bibitem{Wunsch-Ferrari-2004:vertical}
C.~Wunsch, R.~Ferrari, {\it Ann. Rev. Fluid Mech.\/} {\bf 36}, 281 (2004).

\bibitem{Klinger-Marshall-Send-1996:representation}
B.~A. Klinger, J.~Marshall, U.~Send, {\it Journal of Geophysical Research:
  Oceans\/} {\bf 101}, 18175 (1996).

\bibitem{Kang-Bire-Campin-et-al-2020:differing}
W.~Kang, {\it et~al.\/}, {\it arXiv preprint arXiv:2008.03764\/}  (2020).

\bibitem{Jones-Marshall-1993:convection}
H.~Jones, J.~Marshall, {\it J. Phys. Oceanogr.\/} {\bf 23}, 1009 (1993).

\bibitem{Redi-1982:oceanic}
M.~H. Redi, {\it J. Phys. Oceanogr.\/} {\bf 12}, 1154 (1982).


\bibitem{Visbeck-Marshall-Haine-et-al-1997:specification}
M.~Visbeck, J.~Marshall, T.~Haine, M.~Spall, {\it J. Phys. Oceanogr.\/} {\bf
  27}, 381 (1997).



\bibitem{McDougall-Jackett-Wright-et-al-2003:accurate}
T.~J. McDougall, D.~R. Jackett, D.~G. Wright, R.~Feistel, {\it Journal of
  Atmospheric and Oceanic Technology\/} {\bf 20}, 730 (2003).


\bibitem{Chen-Nimmo-2011:obliquity}
{Chen, E M A}, {Nimmo, F}, {\it Icarus\/} {\bf 214}, 779 (2011).

\bibitem{Beuthe-2016:crustal}
M.~Beuthe, {\it Icarus\/} {\bf 280}, 278 (2016).




\bibitem{Holland-Jenkins-1999:modeling}
D.~M. Holland, A.~Jenkins, {\it J. Phys. Oceanogr.\/} {\bf 29}, 1787 (1999).

\bibitem{Beuthe-2018:enceladuss}
M.~Beuthe, {\it Icarus\/} {\bf 302}, 145 (2018).





\bibitem{Renaud-Henning-2018:increased}
J.~P. Renaud, W.~G. Henning, {\it Astrophysical Journal\/} {\bf 857}, 98
  (2018).


\bibitem{Petrenko-Whitworth-1999:physics}
V.~Petrenko, R.~Whitworth, {\it Physics of Ice\/} (OUP Oxford, 1999).




\end{thebibliography}
\bibliographystyle{Science}

\section*{Acknowledgments}
This work was carried out in the Department of Earth, Atmospheric and Planetary Science (EAPS) in MIT. WK and TM acknowledges support as a Lorenz/Crosby Fellow supported by endowed funds in EAPS. SB, JC and JM acknowledge part-support from NASA Astrobiology Grant 80NSSC19K1427 “Exploring Ocean Worlds”. We thank Mikael Beuthe and Malte Jansen for advice and discussions.

\section*{Supplementary materials}
Model description\\
Exploring the sensitivity of ocean model solutions to parameters\\
Figs. S1 to S15\\
Tables S1 to S2\\
References \textit{(62-77)}


\clearpage


\end{document}


\baselineskip24pt
\maketitle

\section{Model description}
\label{sec:model-description}
\subsection{An overview of the General Circulation Model}
\label{sec:ocean-model}
  
  Our simulations are carried out using the Massachusetts Institute of Technology OGCM (MITgcm \cite{MITgcm-group-2010:mitgcm, Marshall-Adcroft-Hill-et-al-1997:finite}) configured for application to icy moons. Our purpose is to 1) simulate the large-scale circulation and tracer transport driven by under-ice salinity gradients induced by patterns of freezing and melting, under-ice temperature gradients due to the pressure-dependence of the freezing point of water and bottom heat fluxes associated with tidal dissipation in the core, 2) diagnose the water-ice heat exchange rate and, 3) examine whether this heat exchange is consistent with the heat budget of the ice sheet, comprising heat loss due to conduction, tidal heating in the ice sheet, and heating due to latent heat release on freezing, as presented graphically in Fig.~1 of the main text.
  
  In our calculations the ice shell freezing/melting rate is derived from a model of ice flow (described below), based on observational inferences of ice shell thickness, prescribed and held constant: it is not allowed to respond to the heat/salinity exchange with the ocean underneath. To enable us to integrate our ocean model out to equilibrium on a 10,000 year timescales and to explore a wide range of parameters, we employ a zonally-symmetric configuration at relatively coarse resolution, and parameterize the diapycnal mixing, convection and baroclinic instability of small-scale turbulent processes that cannot be resolved. Each experiment is initialized from rest and a constant salinity distribution. The initial potential temperature at each latitude is set to be equal to the freezing point at the water-ice interface. The simulations are then launched for 10,000 years. By the end of 10,000 years of integration thermal equilibrium has been reached.
  
  The model integrates the non-hydrostatic primitive equations for an incompressible fluid in height coordinates, including a full treatment of the Coriolis force in a deep fluid, as described in \cite{MITgcm-group-2010:mitgcm, Marshall-Adcroft-Hill-et-al-1997:finite}. Such terms are typically neglected when simulating Earth's ocean because the ratio between the fluid depth and horizontal scale is small. Instead Enceladus' aspect ratio is order $40$km$/252$km$\sim0.16$ and so not negligibly small. The size of each grid cell shrinks with depth due to spherical geometry and is accounted for by switching on the ``deepAtmosphere'' option of MITgcm. Since the depth of Enceladus' ocean is comparable to its radius, the variation of gravity with depth is significant. The vertical profile of gravity in the ocean and ice shell is given by, assuming a bulk density of $\rho_{\mathrm{out}}=1000$~kg/m$^3$:
  \begin{equation}
    \label{eq:g-z}
    g(z)=\frac{G\left[M-(4\pi/3)\rho_{\mathrm{out}}(a^3-(a-z)^3)\right]}{(a-z)^2}.
  \end{equation}
  In the above equation, $G=6.67\times10^{-11}$~N/m$^2$/kg$^2$ is the gravitational constant and $M=1.08\times 10^{20}$~kg and $a=252$~km are the mass and radius of Enceladus.
  
  Since it takes several tens of thousands of years for our solutions to reach equilibrium, we employ a moderate resolution of $2$~degree ($8.7$~km) and run the model in a 2D, zonal-average configuration whilst retaining full treatment of Coriolis terms. By doing so, the zonal variations are omitted (the effects of 3D dynamics are to be explored in future studies). In the vertical direction, the $60$~km ocean-ice layer is separated into $30$ layers, each of which is $2$~km deep. The ocean is encased by an ice shell with meridionally-varying thickness using MITgcm's ``shelfice'' and ice ``boundary layer'' module \cite{Losch-2008:modeling}. We set the ice thickness $H$ using the zonal average of the thickness map given by \textit{Hemingway \& Mittal 2019} \cite{Hemingway-Mittal-2019:enceladuss}, as shown by a solid curve in Fig.1b in the main text, and assume hydrostacy (i.e., ice is floating freely on the water). We employ partial cells to better represent the ice topography: water is allowed to occupy a fraction of the height of a whole cell with an increment of 10\%.

  \subsection{Parameterization of subgridscale processes}
  Key processes that are not explicitly resolved in our model are diapycnal mixing, convection and baroclinic instability. Here we review the parameterizations and mixing schemes used in our model to parameterize them. Sensitivity tests of our solutions when mixing parameters are varied about reference values are presented in Section~\ref{sec:sensitivity}\ref{sec:sensitivity-diffusivity}.
  
  \underline{Vertical mixing of tracers and momentum}
  
  To account for the mixing of momentum, heat and salinity by unresolved turbulence, in our reference calculation we set the explicit horizontal/vertical diffusivity to $0.005$~m$^2$/s. This is roughly 3 orders of magnitude greater than molecular diffusivity, but broadly consistent with dissipation rates suggested by \textit{Rekier et al. 2019} for Enceladus \cite{Rekier-Trinh-Triana-et-al-2019:internal}, where both libration and tidal forcing are taken into account. According to \cite{Rekier-Trinh-Triana-et-al-2019:internal}, the tidal dissipation in the ocean is mostly induced by libration implying a global dissipation rate $E$ of order 1~MW, but with considerable uncertainty. As reviewed by \textit{Wunsch \& Ferrari 2004} \cite{Wunsch-Ferrari-2004:vertical}, this suggests a vertical diffusivity given by
  \begin{equation}
    \label{eq:kappav}
    \kappa_{v}=\frac{\Gamma \varepsilon}{\rho_0 N^{2}},
  \end{equation}
  where $\Gamma\sim 0.2$ is the efficiency at which dissipation of kinetic energy is available for production of potential energy. Here, $\varepsilon=E/V$ is the dissipation rate per volume, $V\approx 4\pi (a-H_0-D/2)^2D$ is the total volume of the ocean ($H_0$ and $D$ are the mean thickness of the ice layer and ocean layer, and $a$ is the moon's radius) and $\rho_0\sim 1000$~kg/m$^3$ is the density of water. $N^2=g(\partial \ln \rho/\partial z)\sim g (\Delta \rho/\rho_0)/D$ is the Brunt-Vaisala frequency, where $g$ is the gravity constant. $\Delta \rho/\rho_0$ can be estimated from $\alpha_T \Delta T_f$, where $\alpha_T$ is the thermal expansion coefficient near the freezing point and $\Delta T_f$ is the freezing point difference between the underside of the equatorial and the north polar ice shell. Here we take $|\alpha_T|\sim 1\times10^{-5}$/K (corresponding to $S_0=27$ and $S_0=17$~psu), and $|\Delta T_f\sim 0.07|$K (a measure of the overall vertical temperature gradients in our default set of experiments). Substituting into Eq.\ref{eq:kappav}, yields $\kappa_v\sim 0.005$~m$^2$/s, which we choose to be the default horizontal and vertical diffusivity used in our experiments. The diffusivity for temperature and salinity are set to be the same, so that double diffusive effects are excluded. Uncertainties stem from both $E$ and $N^2$ and show considerable spatial variability in our experients -- see the discussion in \cite{Rekier-Trinh-Triana-et-al-2019:internal}. One might expect $N^2$ to be smaller ($\kappa$ larger) in cases where temperature- and salinity-induced density gradients cancel one-another, and vice versa; the former scenario seems to be more plausible, a main conclusion of our study. It is for the reason that we set our default diffusivities to the above high values in all our reference experiments and explore the impact of lower diffusivities as sensitivity tests. 
  
  The horizontal and vertical viscosity $\nu_h,\nu_v$ are set to $50$~m$^2$/s. This value is minimum needed to control grid-scale noise and to make the Ekman boundary layer (thickness $\sim \sqrt{\nu_v/f}$, where $f$ is the Coriolis coefficient) thick enough so that the 2-km grid height can resolve it. In addition, to damp numerical noise induced by our use of stair-like ice topography, we employ a bi-harmonic hyperviscosity of $3\times 10^9$~m$^4$/s and a bi-harmonic hyperdiffusivity of $5\times 10^7$~m$^4$/s.

\begin{figure*}[htbp!]
    \centering
    \includegraphics[page=7,width=0.9\textwidth]{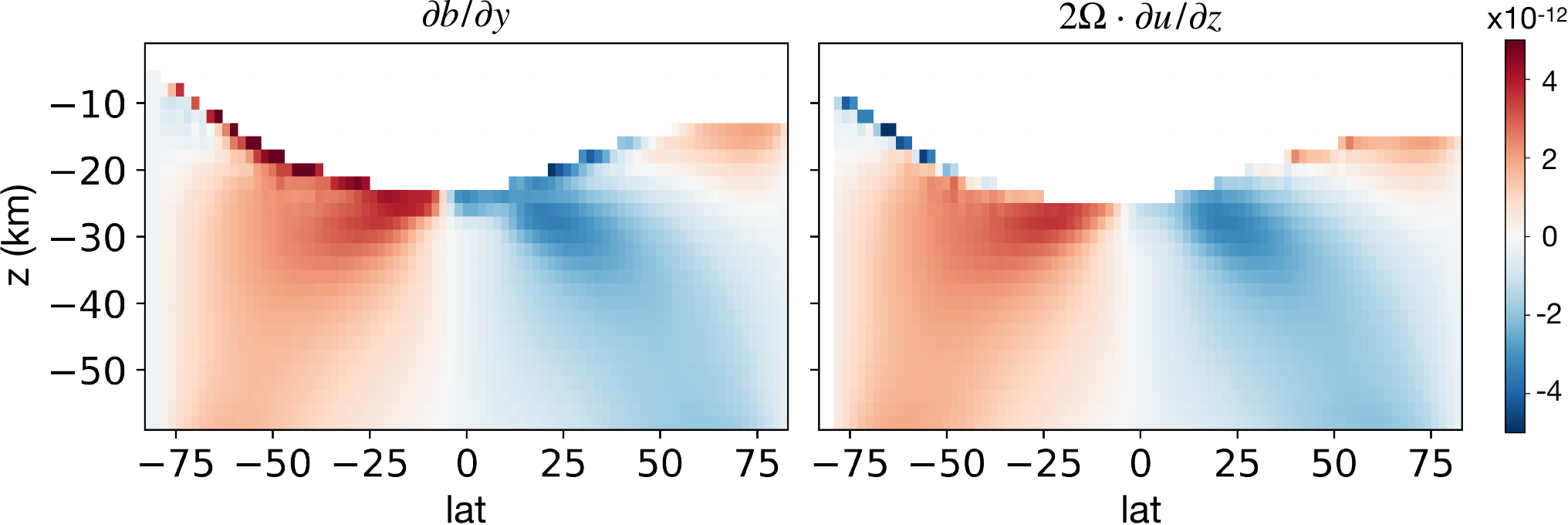}
    \caption{Thermal wind balance in the control simulation. Panels shows the two terms in the thermal wind balance, $2\mathbf{\Omega}\cdot \nabla U$ and $\partial b/a\partial \phi$, respectively. Here $\Omega$ is the rotation rate of the moon, $U$ is the zonal flow speed, $b=-g(\rho-\rho_0)/\rho_0$ is buoyancy, $a$ is the moon's radius and $\phi$ is latitude. }
    \label{fig:thermal-wind-balance}
  \end{figure*}

  Despite use of these viscous and smoothing terms, the dominant balance in the momentum equation is between the Coriolis force and the pressure gradient force and so zonal currents on the large-scale remain in thermal wind balance, especially in the interior of the ocean. As shown by Fig.~\ref{fig:thermal-wind-balance}, the two-term balance in the thermal wind equation, $2\mathbf{\Omega}\cdot \nabla U=\partial b/a\partial \theta$, are almost identical. Since thermal wind balance is a consequence of geostrophic and hydrostatic balance and the latter is always a good approximation on the large scale, geostrophic balance is indeed well satisfied.
  
  \underline{Convection}
  
  Due to the coarse resolution of our model, convection cannot be resolved and must be parameterised. In regions that are convectively unstable, we set the diffusivity to a much larger value, $1$~m$^2$/s, to represent the vertical mixing associated with convective overturns.  Similar approaches are widely used to parameterize convection in coarse resolution ocean models (see, e.g. \textit{Klinger and Marshall 1996} \cite{Klinger-Marshall-Send-1996:representation}) and belong to a family of convective adjustment schemes. This value is obtained based on the equilibrium top-to-bottom temperature gradient in a high-resolution Enceladus simulation \cite{Kang-Bire-Campin-et-al-2020:differing}, where we assume a salty ocean (40~psu) and enforce $\sim 50$~mW/m$^2$ of heat from the bottom. Scaling argument would lead to similar results. According to \textit{Jones and Marshall 1993} \cite{Jones-Marshall-1993:convection}, the velocity in a rotation-dominated regime scales with $\sqrt{B/f}$, where $B$ is the buoyancy flux and $f$ is the Coriolis coefficient. Utilizing the fact that convective plumes/rolls should occupy the whole ocean depth $D$, a diffusivity can be estimated by multiplying the length scale and velocity scale together
  \begin{equation}
    \label{eq:kappa-conv}
    \kappa_{\mathrm{conv}}\sim \sqrt{B/f}D\sim 1~m^2/s.
  \end{equation}
  Here we have chosen $B$ to be $10^{-13}$~m$^3$/s$^2$, which is the buoyancy flux produced by a 50~mW/m$^2$ bottom heat flux, or equivalently, the buoyancy flux induced by a 1~km/My freezing rate, in an ocean with 40~psu salinity.
  
  This equivalent diffusivity is expected to vary with the buoyancy flux, which include the contributions from bottom heating as well as top freezing. However, our results are not found to be sensitive to this choice provided the associated diffusive time scale $D^2/\nu_{\mathrm{conv}}\approx 0.5$~yr is much shorter than the advective time scale $M_{\mathrm{half}}/\Psi\approx 2000$~yrs ($M_{\mathrm{half}}$ is half of the total mass of the ocean and $\Psi$ is the maximum meridional streamfunction in $kg/s$). It should be emphasized that, as noted above, away from boundary layers our solutions are close to geostrophic, hydrostatic and thermal wind balance and are not convectively unstable. However, convective heating from the bottom and/or salinization of water at the top can and do lead to convective instability which are mixed away diffusively.
  
    \underline{Baroclinic instability}

The large-scale currents set up in our model are in thermal wind balance with horizontal density gradients induced by under-ice temperature and salinity gradients. There is thus a store of available potential energy which will be tapped by baroclinic instability, a process which is not resolved in our model because of its zonally-symmetric configuration. Following an approach widely used in modeling Earth's ocean, we use the Gent-McWilliams (GM) scheme \cite{Redi-1982:oceanic, Gent-Mcwilliams-1990:isopycnal} to parameterize the associated eddy-induced circulation and mixing of tracers along isopycnal surfaces. The key parameter that characterize the efficiency of the along-isopycnal mixing is the GM diffusivity $\kappa_{\mathrm{GM}}$. According to \textit{Visbeck et al. 1997} \cite{Visbeck-Marshall-Haine-et-al-1997:specification}, $\kappa_{\mathrm{GM}}$ can be estimated by
     \begin{equation}
       \label{eq:kappa-GM}
       \kappa_{\mathrm{GM}}=\alpha l^{2} \frac{f}{\sqrt{\mathrm{Ri}}},
     \end{equation}
where $\frac{f}{\sqrt{\mathrm{Ri}}}$ is proportional to the Eady growth rate, $l$ is the width of the baroclinic zone, $\alpha$=0.015 is a universal constant, $f$ is the Coriolis parameter and $\mathrm{Ri}=N^2/U_z$ is the Richardson number. We estimate $l$ using the Rhine's scale $\sqrt{U/\beta}$, where $U$ is the zonal flow speed and $\beta$ is the meridional gradient of the Coriolis parameter. Substituting $N^2\sim 10^{-11}$~s$^{-2}$, $f\sim 10^{-4}$~s$^{-1}$, $U\sim 10^{-3}$~m, and $\beta\sim 4\times10^{-10}$~s$^{-1}$m$^{-1}$, we get $\kappa_{\mathrm{GM}}$ $\sim 0.1$~m$^2$/s. It is notable that this is 2-3 orders of magnitude smaller than the value used for Earth's ocean and those adopted by \textit{Lobo et al. 2021} \cite{Lobo-Thompson-Vance-et-al-2021:pole}.

  \subsection{Equation of state and the freezing point of water}
  To make the dynamics as transparent as possible, we adopt a linear equation of state (EOS) to determine how density depends on temperature, salinity and pressure. The dependence of potential density $\rho$ on potential temperature $\theta$ and salinity $S$ is determined as follows:
  \begin{eqnarray}
    \label{eq:EOS-linear}
  \rho(\theta,S)&=&\rho_0\left(1-\alpha_T(\theta-\theta_0)+\beta_S
                    (S-S_0)\right)\\
    \rho_0&=&\rho(\theta_0,S_0).\label{eq:rho0}
  \end{eqnarray}
  Here, $\rho_0,\ \theta_0$ and $S_0$ are the reference potential density, potential temperature and salinity. $\alpha_T$ and $\beta_S$, the thermal expansion coefficient and the haline contraction coefficient, are set to the first derivative of density with respect to potential temperature and salinity at the reference point using the Gibbs Seawater Toolbox \cite{McDougall-Barker-2011:getting}. We carried out two test experiments (one with $S_0=10$~psu and the other with $S_0=20$~psu) using the full ``MDJWF'' equation of state \cite{McDougall-Jackett-Wright-et-al-2003:accurate} and obtained almost identical results. To explore a wide range of background salinity, $S_0$ is prescribed to values between $4$~psu and $40$~psu. $\theta_0$ is set to be the freezing temperature at $S_0$ and $P_0=2.2\times10^6$~Pa (this is the pressure under a 20.8~km thick ice sheet on Enceladus). 
  
  The freezing point of water $T_f$ is assumed to depend on local pressure $P$ and salinity $S$ as follows,
  \begin{equation}
    \label{eq:freezing-point}
    T_f(S,P)=c_0+b_0P+a_0S,
  \end{equation}
where $a_0=-0.0575$~K/psu, $b_0=-7.61\times10^{-4}$~K/dbar and $c_0=0.0901$~degC. The pressure $P$ can be calculated using hydrostatic balance $P=\rho_igH$ ($\rho_i=917$~kg/m$^3$ is the density of the ice and $H$ is the ice thickness).

  \subsection{Boundary conditions}
  \label{sec:boundary-conditions}
  
  Our ocean model is forced by heat and salinity fluxes from the ice shell at the top as well as heat fluxes coming from below.
  
  \underline{Diffusion of heat through the ice}
  
\begin{figure*}[htbp!]
    \centering \includegraphics[width=0.62\textwidth]{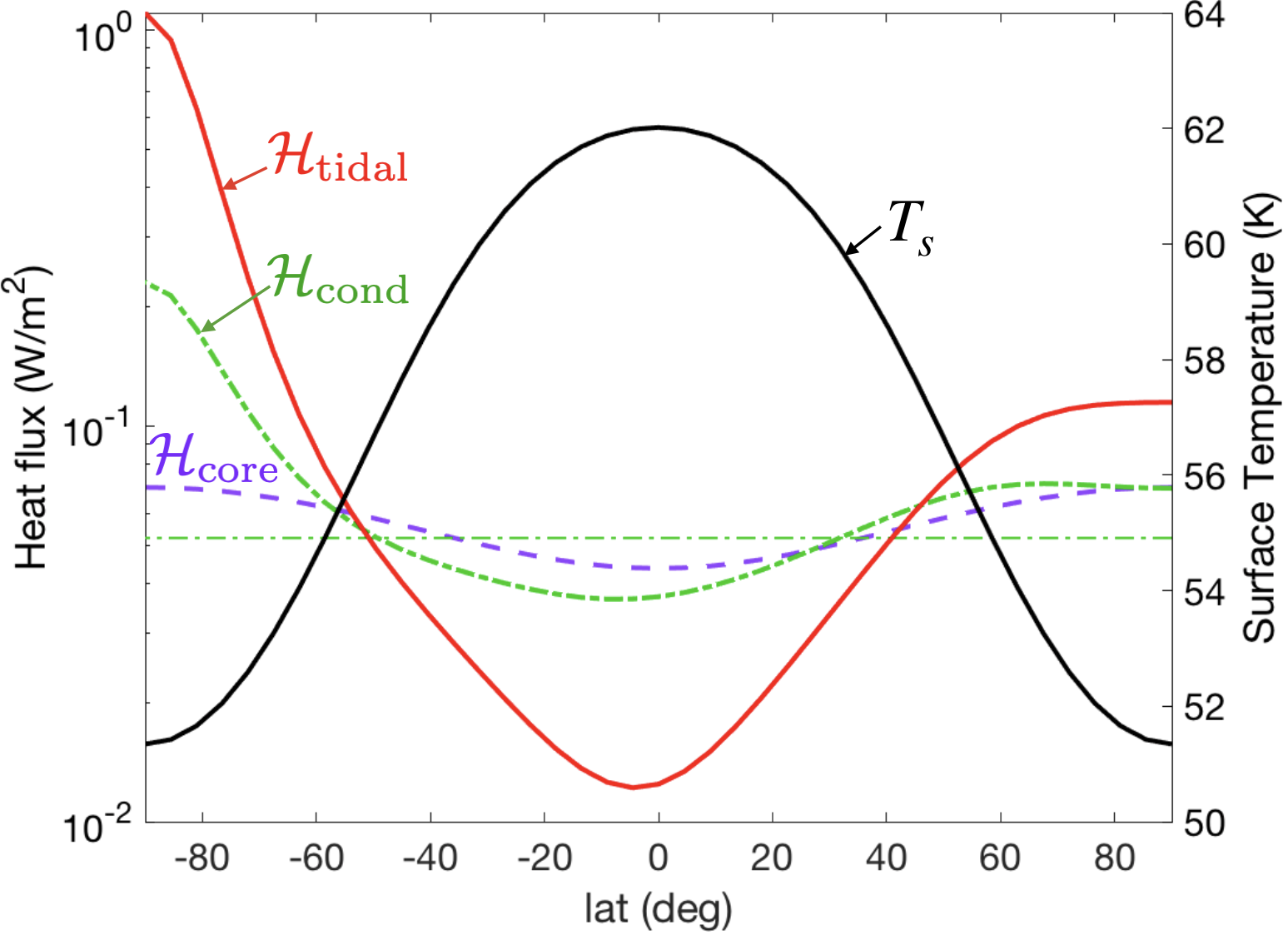}
    
    \caption{\small{Meridional profiles of heat fluxes and surface temperature. Heat fluxes are plotted using colored curves, with a scale on the left. Conductive heat loss $\mathcal{H}_{\mathrm{cond}}$ (Eq.~\ref{eq:H-cond}) is shown by a thick green dash-dotted line which, in the global average, is balanced by heat generation in the silicate core $\mathcal{H}_{\mathrm{core}}$ (purple dashed line, Eq.~\ref{eq:H-core}) and $\mathcal{H}_{\mathrm{ice}}$ (red solid line, Eq.~\ref{eq:H-tide}). All heat fluxes are normalized to have the same global mean value of $\mathcal{H}_{\mathrm{cond}}$. The surface temperature $T_s$ (black solid line, axis on the right) is set to be in radiative equilibrium with the solar radiation and is warmer at the equator.}}
    
    \label{fig:heating-profiles-Ts}
  \end{figure*}

  Heat loss to space by heat conduction through the ice $\mathcal{H}_{\mathrm{cond}}$ is represented using a 1D vertical heat conduction model,
\begin{equation}
  \mathcal{H}_{\mathrm{cond}}=\frac{\kappa_{0}}{H} \ln \left(\frac{T_{f}}{T_{s}}\right),
  \label{eq:H-cond}
  \end{equation}
  where $H$ is the thickness of ice  (solid curve in Fig.1b of the main text), the surface temperature is $T_s$ and the ice temperature at the water-ice interface is the local freezing point $T_f$ (Eq.~\ref{eq:freezing-point}).
  We approximate the surface temperature $T_s$ using radiative equilibrium based on the incoming solar radiation and obliquity ($\delta=27^\circ$) assuming an albedo of $0.81$. The $T_s$ profile is shown by the black solid curve in Fig.\ref{fig:heating-profiles-Ts}.  Typical heat losses averaged over the globe are $\mathcal{H}_{\mathrm{cond}}$= $50$~mW/m$^2$, broadly consistent with observations \cite{Tajeddine-Soderlund-Thomas-et-al-2017:true}.

 \underline{Tidal heating in the core}
 
 Conductive heat loss is primarily balanced by tidal dissipation in the ice shell $\mathcal{H}_{\mathrm{ice}}$ and the core $\mathcal{H}_{\mathrm{core}}$ (dissipation in the ocean plays a negligible role) \cite{Chen-Nimmo-2011:obliquity, Beuthe-2016:crustal, Hay-Matsuyama-2019:nonlinear, Rekier-Trinh-Triana-et-al-2019:internal}. For each assumed heat partition between the shell and the core, we use the same meridional heating profiles for $\mathcal{H}_{\mathrm{core}}$ and $\mathcal{H}_{\mathrm{ice}}$ (see below). According to \textit{Beuthe 2019}\cite{Beuthe-2019:enceladuss} and \textit{Choblet et al. 2017}\cite{Choblet-Tobie-Sotin-et-al-2017:powering}, the core dissipation $\mathcal{H}_{\mathrm{core}}$ peaks at the two poles. We obtain the meridional heat profile using Eq.60 in \textit{Beuthe 2019}\cite{Beuthe-2019:enceladuss} (Beuthe, personal communication),
  \begin{equation}
    \label{eq:H-core} \mathcal{H}_{\mathrm{core}}(\phi)=\bar{\mathcal{H}}_{\mathrm{core}}\cdot(1.08449 + 0.252257 \cos(2\phi) + 0.00599489 \cos(4\phi)),
  \end{equation}
  where $\phi$ denotes latitude and $\bar{\mathcal{H}}_{\mathrm{core}}$ is the global mean heat flux from the bottom. Since the global surface area shrinks going downward due to the spherical geometry, a factor of $\left.(a-H)^2\right/(a-H-D)^2$ ($H$ is ice thickness, $D$ is ocean depth) needs to be considered when computing $\bar{\mathcal{H}}_{\mathrm{core}}$. The expression within the bracket is normalized for the globe, adjusted to take account of the fact that our model only covers 84S-84N. Using the above formula, the bottom heat flux is twice as strong over the poles than equator, as can be seen in Fig.1d of the main text. We note that the heating profile here is highly idealized and does not have the localized heating stripes seen in \textit{Choblet et al. 2017} \cite{Choblet-Tobie-Sotin-et-al-2017:powering} which arise from the interaction between the porous core and the fluid in the gaps.

  \underline{Ice-ocean fluxes}

    The interaction between ocean and ice is simulated using MITgcm's ``shelf-ice'' package \cite{Losch-2008:modeling, Holland-Jenkins-1999:modeling}. We turn on the ``boundary layer'' option to avoid possible numerical instabilities induced by an ocean layer which is too thin. The code is modified to account for a gravitational acceleration that is very different from that on earth, the temperature dependence of heat conductivity, and the meridional variation of tidal heating generated inside the ice shell and the ice surface temperature. In the description that follows, we begin by introducing the shelf-ice parameterization in a fully coupled ocean-ice system and then make simplifications that fit our goal here. 
    
    Following \textit{Kang et al. 2020}\cite{Kang-Bire-Campin-et-al-2020:differing}, the freezing/melting rate of the ice shell is determined by a heat budget for a thin layer of ice at the base\footnote{This choice is supported by the fact that most tidal heating is generated close to the ocean-ice interface \cite{Beuthe-2018:enceladuss}.}. The budget involves three terms: the heat transmitted upward by ocean $\mathcal{H}_{\mathrm{ocn}}$, the heat loss through the ice shell due to heat conduction $\mathcal{H}_{\mathrm{cond}}$ (Eq.\ref{eq:H-cond}), and the tidal heating generated inside the ice shell $\mathcal{H}_{\mathrm{ice}}$ (Eq.\ref{eq:H-tide}). As elucidated in \textit{Holland and Jenkins 1999} \cite{Holland-Jenkins-1999:modeling} and \textit {Losch 2008} \cite{Losch-2008:modeling}, the continuity of heat flux and salt flux through the ``boundary layer'' gives,
  \begin{eqnarray}
    &~&\mathcal{H}_{\mathrm{ocn}}-\mathcal{H}_{\mathrm{cond}}+\mathcal{H}_{\mathrm{ice}}=-L_fq-C_p(T_{\mathrm{ocn-top}}-T_b)q\label{eq:boundary-heat}\\
&~&\mathcal{F}_{\mathrm{ocn}}=-S_bq-(S_{\mathrm{ocn-top}}-S_b)q, \label{eq:boundary-salinity}   
  \end{eqnarray}
  where $T_{\mathrm{ocn-top}}$ and $S_{\mathrm{ocn-top}}$ denote the temperature and salinity in the top grid of the ocean\footnote{When model resolution is smaller than the boundary layer thickness, the salinity below the upper-most grid cell also contributes to $T_{\mathrm{ocn-top}}$ and $S_{\mathrm{ocn-top}}$.}, $S_b$ denotes the salinity in the ``boundary layer'', and $q$ denotes the freezing rate in $kg/m^2/s$. $C_p=4000$~J/kg/K is the heat capacity of the ocean, $L_f=334000$~J/kg is the latent heat of fusion of ice.
  
$\mathcal{H}_{\mathrm{ocn}}$ and $\mathcal{F}_{\mathrm{ocn}}$ in Eq.\ref{eq:boundary-heat} can be written as
  \begin{eqnarray}
    \mathcal{H}_{\mathrm{ocn}}&=&C_p(\rho_{0}\gamma_T-q)(T_{\mathrm{ocn-top}}-T_b),\label{eq:H-ocn}\\
    \mathcal{F}_{\mathrm{ocn}}&=&(\rho_{0}\gamma_S-q)(S_{\mathrm{ocn-top}}-S_b) \label{eq:S-ocn}
  \end{eqnarray}
  where $\gamma_T=\gamma_S=10^{-5}$~m/s are the exchange coefficients for temperature and salinity, and $T_b$ denotes the and temperature in the ``boundary layer''. The terms associated with $q$ are the heat/salinity change induced by the deviation of $T_{\mathrm{ocn-top}},\ S_{\mathrm{ocn-top}}$ from that in the ``boundary layer'', where melting and freezing occur. $T_b=T_f(S_b,P)$, the freezing temperature at pressure $P$ and salinity $S_b$ (see Eq.\ref{eq:freezing-point}).

  In a fully-coupled system, we would solve $S_b$ and $q$ from Eq.~(\ref{eq:boundary-heat})-(\ref{eq:S-ocn}). When freezing occurs ($q>0$), the salinity flux $\rho_{w0}\gamma_S(S_{\mathrm{ocn-top}}-S_b)$ is negative (downward). This leads to a positive tendency of salinity at the top of the model ocean, together with changes of temperature, thus:
  \begin{eqnarray}
    \frac{dS_{\mathrm{ocn-top}}}{dt}&=&\frac{-\mathcal{F}_{\mathrm{ocn}}}{\rho_{w0}\delta z}=\frac{1}{\rho_{w0}\delta z}(\rho_{w0}\gamma_S-q)(S_b-S_{\mathrm{ocn-top}})=\frac{qS_{\mathrm{ocn-top}}}{\rho_{w0}\delta z},\label{eq:S-tendency}\\
    \frac{dT_{\mathrm{ocn-top}}}{dt}&=&\frac{-\mathcal{H}_{\mathrm{ocn}}}{C_p\rho_{w0}\delta z}=\frac{1}{\rho_{w0}\delta z}(\rho_{w0}\gamma_T-q)(T_b-T_{\mathrm{ocn-top}})\nonumber\\
    &=&\frac{1}{C_p\rho_{w0}\delta z}\left[\mathcal{H}_{\mathrm{ice}}-\mathcal{H}_{\mathrm{cond}}+L_fq+C_p(T_{\mathrm{ocn-top}}-T_b)q\right] \label{eq:T-tendency0}
  \end{eqnarray}
  where $\delta z=2$~km is the thickness of the ``boundary layer'' at the ocean-ice interface.

  If we allow the freezing/melting of ice and the ocean circulation to feedback onto one-another, the positive feedback between them renders it difficult to find consistent solutions. We therefore cut off this feedback loop by setting the freezing rate $q$ to that which is required to sustain the prescribed ice sheet geometry (details can be found in the next section, ice flow model), whilst allowing a heating term to balance the heat budget (Eq.\ref{eq:boundary-heat}). The amplitude of this heat imbalance can then be used to discriminate between different steady state solutions (Eq.~1 in the main text). This also simplifies the calculation of the T/S tendencies of the upper-most ocean grid. The S tendency can be directly calculated from Eq.~\ref{eq:S-tendency}, and the T tendency approximated by:
  \begin{eqnarray}
    \frac{dT_{\mathrm{ocn-top}}}{dt}&=&\frac{1}{\delta z}(\gamma_T-q)(T_{\mathrm{f,ocn-top}}-T_{\mathrm{ocn-top}}),\label{eq:T-tendency}
  \end{eqnarray}
  replacing the boundary layer freezing temperature $T_b=T_f(S_b,P)$ in Eq.~\ref{eq:T-tendency0} with $T_{\mathrm{f,ocn-top}}=T_f(S_{\mathrm{ocn-top}}, P)$, the freezing temperature determined by the upmost ocean grid salinity and pressure. The difference between the $S_b$ and $S_{\mathrm{ocn-top}}$ can be estimated by $\mathcal{F}_{\mathrm{ocn}}/(\rho_0\gamma_S)=qS_{\mathrm{ocn-top}}/(\rho_0\gamma_S)$, according to Eq.\ref{eq:S-ocn} and Eq.\ref{eq:S-tendency}, given that $|q|\lesssim 10^{-7}$~kg/$m^2$/s is orders of magnitude smaller than $\rho_0\gamma_S=0.01$~kg/m$^2$/s. Even in the saltiest scenario we consider here, $|S_b-S_{\mathrm{ocn-top}}|$ does not exceed $0.0004$~psu, and the associated freezing point change is lower than $10^{-5}$K.
  Readers interested in the formulation of a freely evolving ice-water system are referred to the method section of \textit{Kang et al. 2021}\cite{Kang-Bire-Campin-et-al-2020:differing} and \textit{Losch 2008}\cite{Losch-2008:modeling}.
  
  In addition to the above conditions on temperature and salinity, the tangential velocity is relaxed back to zero at a rate of $\gamma_M=10^{-3}$m/s at the upper and lower boundaries.

 \underline{Ice flow model}

  We prescribe $q$ using the divergence of the ice flow, assuming the ice sheet geometry is in equilibrium. We use an upside-down land ice sheet model following \textit{Ashkenazy et al. 2018} \cite{Ashkenazy-Sayag-Tziperman-2018:dynamics}. The ice flows down its thickness gradient, driven by the pressure gradient induced by the spatial variation of the ice top surface, somewhat like a second order diffusive process. At the top, the speed of the ice flow is negligible because the upper part of the shell is so cold and hence rigid; at the bottom, the vertical shear of the ice flow speed vanishes, as required by the assumption of zero tangential stress there. This is the opposite to that assumed in the land ice sheet model. In rough outline, we calculate the ice flow using the expression below obtained through repeated vertical integration of the force balance equation (the primary balance is between the vertical flow shear and the pressure gradient force), using the aforementioned boundary conditions to arrive at the following formula for ice transport $\mathcal{Q}$,
\begin{eqnarray}
  \mathcal{Q}(\phi)&=& \mathcal{Q}_0H^3(\partial_\phi H/a) \label{eq:ice-flow}\\
  \mathcal{Q}_0&=&\frac{2(\rho_0-\rho_i)g}{\eta_{\mathrm{melt}}(\rho_0/\rho_i)\log^3\left(T_f/T_s\right)}\int_{T_s}^{T_f}\int_{T_s}^{T(z)}\exp\left[-\frac{E_{a}}{R_{g} T_{f}}\left(\frac{T_{f}}{T'}-1\right)\right]\log(T')~\frac{dT'}{T'}~\frac{dT}{T}.\nonumber 
\end{eqnarray}
Here, $\phi$ denotes latitude, $a=252$~km and $g=0.113$~m/s$^2$ are the radius and surface gravity of Enceladus, $T_s$ and $T_f$ are the temperature at the ice surface and the water-ice interface (equal to local freezing point, Eq.~\ref{eq:freezing-point}), and $\rho_i=917$~kg/m$^3$ and $\rho_0$ are the ice density and the reference water density (Eq.~\ref{eq:EOS-linear}). $E_a=59.4$~kJ/mol is the activation energy for diffusion creep, $R_g=8.31$~J/K/mol is the gas constant and $\eta_{\mathrm{melt}}$ is the ice viscosity at the freezing point. The latter has considerable uncertainty ($10^{13}$-$10^{16}$~Pa$\cdot$s) \cite{Tobie-Choblet-Sotin-2003:tidally} but we choose to set $\eta_{\mathrm{melt}}=10^{14}$~Pa$\cdot$s.

In steady state, the freezing rate $q$ must equal the divergence of the ice transport thus:
\begin{equation}
    q=-\frac{1}{a\cos\phi}\frac{\partial}{\partial \phi} (Q\cos\phi).
    \label{eq:freezing-rate}
\end{equation}
As shown by the dashed curve in Fig.1b of the main text, ice melts in high latitudes and forms in low latitudes at a rate of a few kilometers every million years. A more detailed description of the ice flow model can be found in \textit{Kang and Flierl 2020}\cite{Kang-Flierl-2020:spontaneous} and \textit{Ashkenazy et al. 2018} \cite{Ashkenazy-Sayag-Tziperman-2018:dynamics}. Freezing and melting leads to changes in local salinity and thereby a buoyancy flux. At $S_0=30$~psu, the salinity-associated buoyancy flux is approximately $gq\beta_SS_0\approx 10^{-13}$~m$^2$/s$^3$, which is 3-6 orders of magnitude smaller than the buoyancy flux used by \textit{Lobo et al. 2021}\cite{Lobo-Thompson-Vance-et-al-2021:pole}.

\subsection{Model of tidal dissipation in the ice shell}
\label{sec:tidal-dissipation-model}

Enceladus's ice shell is periodically deformed by tidal forcing and the resulting strains in the ice sheet produce heat. We follow \textit{Beuthe 2019}\cite{Beuthe-2019:enceladuss} to calculate the implied dissipation rate. Instead of repeating the whole derivation here, we only briefly summarize the procedure and present the final result. Unless otherwise stated, parameters are the same as assumed in \textit{Kang \& Flierl 2020}\cite{Kang-Flierl-2020:spontaneous}.

Tidal dissipation consists of three components \cite{Beuthe-2019:enceladuss}: a membrane mode $\mathcal{H}_{\mathrm{ice}}^{\mathrm{mem}}$ due to the extension/compression and tangential shearing of the ice membrane, a mixed mode $\mathcal{H}_{\mathrm{ice}}^{mix}$ due to vertical shifting, and a bending mode $\mathcal{H}_{\mathrm{ice}}^{bend}$ induced by the vertical variation of compression/stretching. Following \textit{Beuthe 2019}\cite{Beuthe-2019:enceladuss}, we first assume the ice sheet to be completely flat. By solving the force balance equation, we obtain the auxiliary stress function $F$, which represents the horizontal displacements, and the vertical displacement $w$. The dissipation rate $\mathcal{H}_{\mathrm{ice}}^{\mathrm{flat,x}}$ (where $x=\{\mathrm{mem},\mathrm{mix},\mathrm{bend}\}$ ) can then be written as a quadratic form of $F$ and $w$. In the calculation, the ice properties are derived assuming a globally-uniform surface temperature of 60K and a melting viscosity of $5\times10^{13}$~Pa$\cdot$s. 

Ice thickness variations are accounted for by multiplying the membrane mode dissipation $\mathcal{H}_{\mathrm{ice}}^{\mathrm{flat,mem}}$, by a factor that depends on ice thickness. This makes sense because this is the only mode which is amplified in thin ice regions (see \textit{Beuthe 2019}\cite{Beuthe-2019:enceladuss}). This results in the expression:
\begin{equation}
  \label{eq:H-tide}
  \mathcal{H}_{\mathrm{ice}}=(H/H_0)^{p_\alpha}\mathcal{H}_{\mathrm{ice}}^{\mathrm{flat,mem}}+\mathcal{H}_{\mathrm{ice}}^{\mathrm{flat,mix}}+\mathcal{H}_{\mathrm{ice}}^{\mathrm{flat,bend}},
\end{equation}
where $H$ is the prescribed thickness of the ice shell as a function of latitude and $H_0$ is the global mean of $H$. Since thin ice regions deform more easily and produce more heat, $p_\alpha$ is negative. Because more heat is produced in the ice shell, the overall ice temperature rises, which, in turn, further increases the mobility of the ice and leads to more heat production (the rheology feedback).

Using reasonable parameters for Enceladus, $\mathcal{H}_{\mathrm{ice}}$ turns out to be at least an order of magnitude smaller than the heat loss rate $\mathcal{H}_{\mathrm{cond}}$. This is a universal flaw of present tidal dissipation models, and could be due to use of an over-simplified Maxwell rheology \cite{McCarthy-Cooper-2016:tidal, Renaud-Henning-2018:increased}. We therefore scale up $\mathcal{H}_{\mathrm{ice}}$ by a constant factor to obtain the desired magnitude.
The tidal heating profile corresponding to $p_\alpha=-1.5$ is the red solid curve plotted in Fig.~\ref{fig:heating-profiles-Ts}. In Fig.~4(b,e) of the main text, we show the tidal heating profile for $p_\alpha=-1$ and $p_\alpha=-2$. 

\begin{table}[hptb!]
  
  \centering
  \begin{tabular}{lll}
    \hline
    Symbol & Name & Definition/Value\\
    \hline
    \multicolumn{3}{c}{Enceladus parameters}\\
    \hline
    $a$ & radius & 252~km\\
    $\delta$ & obliquity & 27$^\circ$\\
    $H$ & global mean ice thickness & 20.8~km: ref \cite{Hemingway-Mittal-2019:enceladuss}  \\
    $D$ & global mean ocean depth& 39.2~km: ref \cite{Hemingway-Mittal-2019:enceladuss} \\
    $\Omega$ & rotation rate & 5.307$\times$10$^{-5}$~s$^{-1}$\\
    $g_0$ & surface gravity & 0.113~m/s$^2$\\
    $\bar{T_s}$ & mean surface temperature& 59K\\
    \hline
    \multicolumn{3}{c}{Europa parameters}\\
    \hline
    $a$ & radius & 1561~km\\
    $\delta$ & obliquity & 3.1$^\circ$\\
    $H$ & global mean ice thickness  &  15~km: ref \cite{Hand-Chyba-2007:empirical} \\
    $D$ & global mean ocean depth & 85~km: ref \cite{Hand-Chyba-2007:empirical} \\
    $\Omega$ & rotation rate & 2.05$\times$10$^{-5}$~s$^{-1}$\\
    $g_0$ & surface gravity & 1.315~m/s$^2$\\
    $\bar{T_s}$ & mean surface temperature & 110K \\
    \hline
    \multicolumn{3}{c}{Physical constants}\\
    \hline
    $L_f$ & fusion energy of ice & 334000~J/kg\\
    $C_p$ & heat capacity of water & 4000~J/kg/K\\
    $T_f(S,P)$ & freezing point & Eq.\ref{eq:freezing-point}\\
    $\rho_i$ & density of ice & 917~kg/m$^3$ \\
    $\rho_w$ & density of the ocean & Eq.\ref{eq:EOS-linear}\\
    $\alpha,\beta$ & thermal expansion \& saline contraction coeff. &  using Gibbs Seawater Toolbox: ref  \cite{McDougall-Barker-2011:getting} \\
    $\kappa_0$ & conductivity coeff. of ice & 651~W/m:  ref \cite{Petrenko-Whitworth-1999:physics}\\
    $p_\alpha$ & ice dissipation amplification factor & -2 $\sim$ -1 \\
    $\eta_m$ & ice viscosity at freezing point & 10$^{14}$~Ps$\cdot$s\\
    \hline
    \multicolumn{3}{c}{Default parameters in the ocean model}\\
    \hline
    $\nu_h,\ \nu_v$ & horizontal/vertical viscosity & 50~m$^2$/s\\
    $\tilde{\nu}_h,\ \tilde{\nu}_v$ & bi-harmonic hyperviscosity & 3$\times$10$^9$~m$^4$/s\\ 
    $\kappa_h,\ \kappa_v$ & horizontal/vertical diffusivity & 0.005~m$^2$/s\\
    $\kappa_\mathrm{GM}$ & Gent-McWilliams diffusivity & 0.1~m$^2$/s\\
    $(\gamma_T,\ \gamma_S,\ \gamma_M)$ & water-ice exchange coeff. for T, S \& momentum & (10$^{-5}$, 10$^{-5}$, 10$^{-3}$)~m/s\\
    $g$ & gravity in the ocean & Eq.\ref{eq:g-z}\\
    $P_0$ & reference pressure & $\rho_ig_0H=2.16\times10^6$~Pa \\
    $T_0$ & reference temperature & $T_f(S_0,P_0)$ \\
    $\rho_{w0}$ & reference density of ocean & Eq.\ref{eq:rho0} \\
    $\mathcal{H}_{\mathrm{cond}}$ & conductive heat loss through ice & Eq.\ref{eq:H-cond}, Fig.\ref{fig:heating-profiles-Ts}\\
    $\mathcal{H}_{\mathrm{ice}}$ & tidal heating produced in the ice & Eq.\ref{eq:H-tide}, Fig.\ref{fig:heating-profiles-Ts} \\
    $\mathcal{H}_{\mathrm{core}}$ & bottom heat flux powered by the core & Eq.\ref{eq:H-core}, Fig.\ref{fig:heating-profiles-Ts} \\
    $A$ & surface albedo & 0.81 \\ 
    $T_s$ & surface temperature profile & Fig.\ref{fig:heating-profiles-Ts}\\
    \hline
     \end{tabular}
  \caption{Model parameters used in our study. }
  \label{tab:parameters}
  
\end{table}

\begin{table}[hptb!]
  
  \centering
  \begin{tabular}{cccc|c}
    \hline
    Control exp & sensitivity-1 & sensitivity-2 & sensitivity-3 & Conceptual model \\
    \hline
    1.5         & 1.1           & 1.1           & 0.6           & 2.9                              \\
    2.0         & 1.2           & 0.3           & 0.2           & 2.2                              \\
    0.95        & 0.5           & 0.3           & 0.3           & 0.44                             \\
    1.2         & 0.7           & 0.4           & 0.4           & 0.8                              \\
    1.5         & 0.92          & 0.5           & 0.4           & 1.3                              \\
    1.8         & 1.0           & 0.5           & 0.4           & 1.9                              \\
    2.1         & 1.2           & 0.6           & 0.5           & 2.5                              \\
    2.3         & 1.3           & 0.6           & 0.6           & 3.2                              \\
    2.4         & 1.5           & 0.7           & 0.6           & 3.8                              \\
        \hline
     \end{tabular}
  \caption{\small{The maximum northern-hemispheric streamfunctions (in unit of $10^8$~kg/s) in numerical models and the conceptual model. The control experiment parameters can be found in Tab.~\ref{tab:parameters}. Sensitivity test 1 is the same as control except that $\kappa_{\mathrm{GM}}=0$~m$^2$/s. Sensitivity test 2 uses $\kappa_{\mathrm{GM}},\kappa_v,\kappa_h=0,10^{-3},10^{-3}$~m$^2$/s. Sensitivity test 3 uses $\kappa_{\mathrm{GM}},\kappa_v,\kappa_h=0,10^{-3},10^{-5}$~m$^2$/s. When changing diffusivity, the circulation depth also changes, so the absolute values of streamfunction are not directly comparable from one group of experiments to another. However, the general trends are very similar -- as salinity increases, the streamfunction first decreases then increases.}}
  \label{tab:streamfunction}
  
\end{table}

\section{Exploring the sensitivity of ocean model solutions to parameters}
\label{sec:sensitivity}
\subsection{Sensitivity to heat partition between the core and the shell}
\label{sec:sensitivity-heat-partition}
\begin{figure*}[htp!]
    \centering
    \includegraphics[page=9,width=0.85\textwidth]{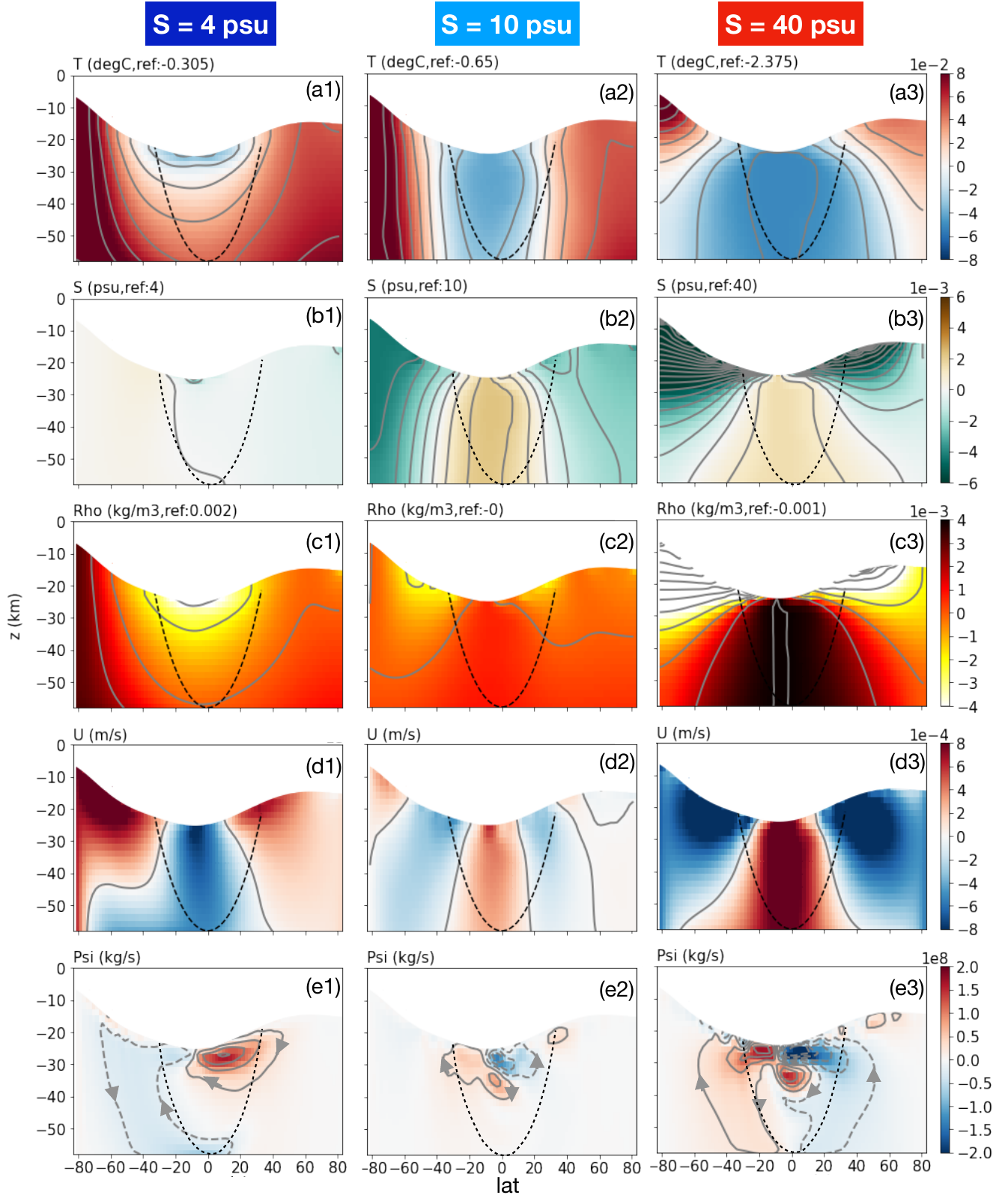}
    
    \caption{\small{As in Fig.~3 of the main text but with 20\% heat assumed to be produced in the silicate core and 80\% in the ice shell. Results are presented for three different salinities: 4~psu, 10~psu and 40~psu. Default mixing parameters are used.} }
    
    \label{fig:T-S-U-Psi-core20}
  \end{figure*}

\begin{figure*}[htp!]
    \centering
    \includegraphics[page=8,width=0.85\textwidth]{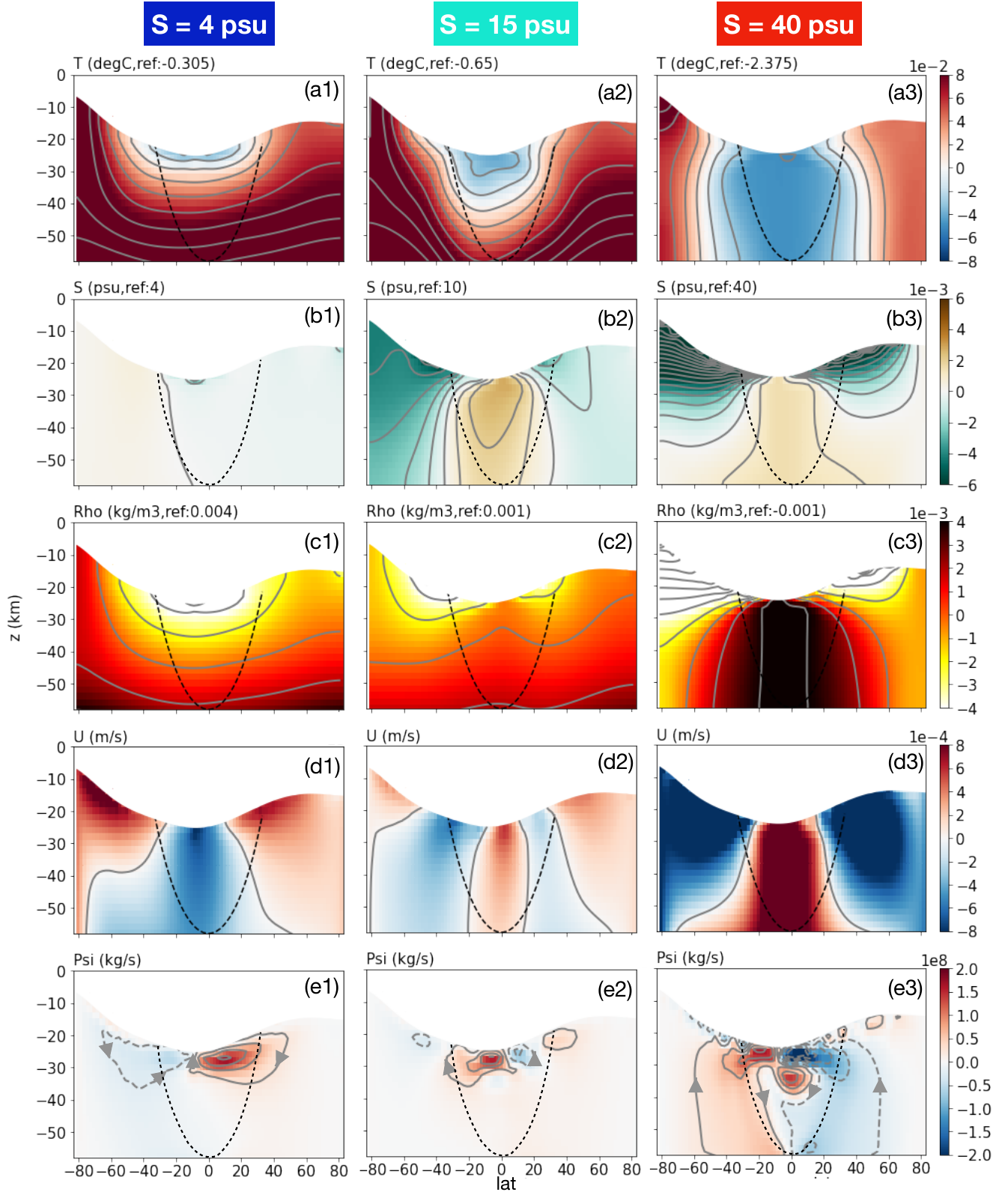}
    
    \caption{\small{As in Fig.\ref{fig:T-S-U-Psi-core20} but with 100\% heating in the core and none in the shell. Default mixing parameters are used.}}
    
    \label{fig:T-S-U-Psi-core100}
  \end{figure*}

To examine the sensitivity of ocean circulation to core-shell heat partition, we repeat the same set of simulations with first 20\% and then 100\% heat produced in the core. The equilibrium ocean solutions are presented in Fig.\ref{fig:T-S-U-Psi-core20} and Fig.\ref{fig:T-S-U-Psi-core100} for the two heat partitions. Compared to our default calculation, the shell-heating scenario shown in Fig.3 of the main text, there is no qualitative change. This is to be expected because the dominant forcing of the flow is the salinity and heat exchange between ice and ocean: the vertical temperature gradient induced by bottom heating is much smaller than the temperature gradient at the water-ice interface induced by the pressure dependence of the freezing point of water. Bottom warming induces stronger stratification if the ocean is fresher than 22~psu (when $\alpha<0$), and vice versa. As can be seen by comparing Fig.\ref{fig:T-S-U-Psi-core20} and Fig.\ref{fig:T-S-U-Psi-core100} with Fig.3, the strengthening/weakening of the stratification suppresses/enhances the vertical extent over which the overturning circulation reaches into the deep ocean. The change is most pronounced at low salinity (4~psu), because the negative thermal expansion coefficient in a fresh ocean suppresses the parameterized convection, resulting in bottom water warming up. However, even with a mean salinity of 4~psu, the response of the dynamics to these stratification changes is rather small (compare the left columns of Fig.3 and Fig.\ref{fig:T-S-U-Psi-core100} here). Note also that all experiments are run out to full equilibrium and so the bottom heat flux is transmitted upward to the water-ice interface without loss in an integral sense, but with ocean currents shaping regional contributions.

  \subsection{Sensitivity to diffusivity and viscosity}
  \label{sec:sensitivity-diffusivity}
\begin{figure*}[htp!]
    \centering
    \includegraphics[page=12,width=0.85\textwidth]{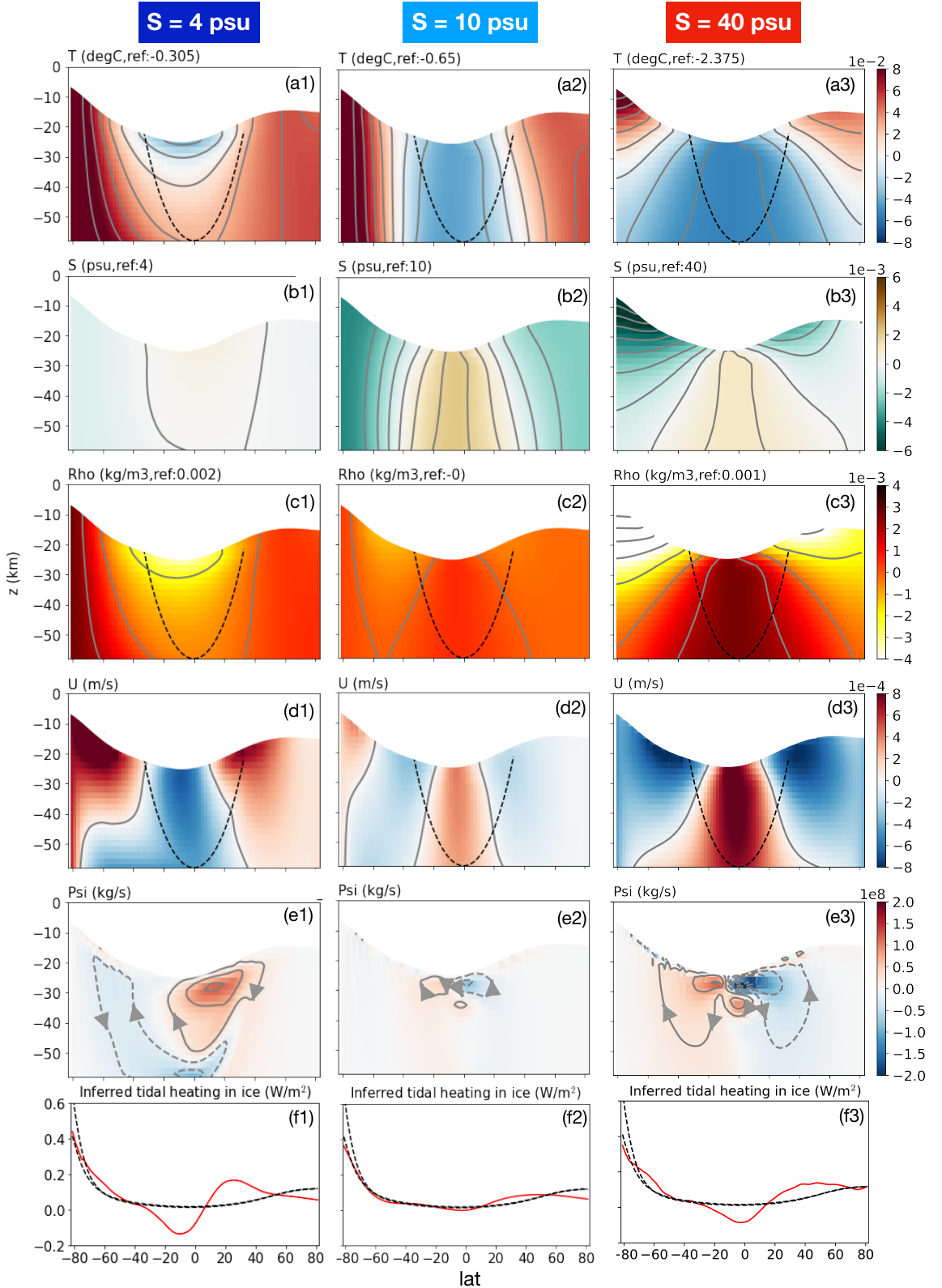}
    
    \caption{\small{The 100\% shell-heating solution with default parameters except $\kappa_{\mathrm{GM}}=0$~m$^2$/s. Row (a-e) are the same figure as in Fig.3 of the main text. Row (f) is similar to Fig.~4(b,e) of the main text and shows the inferred tidal dissipation $\hat{\mathcal{H}}_{\mathrm{ice}}$ (red solid line, calculated using Eq.~1 in the main text), compared with the dissipation rate predicted by our tidal dissipation model (black dashed lines, Eq.\ref{eq:H-tide}). } }
        \label{fig:T-S-U-Psi-noGM}
  \end{figure*}

\begin{figure*}[htp!]
    \centering
    \includegraphics[page=13,width=0.85\textwidth]{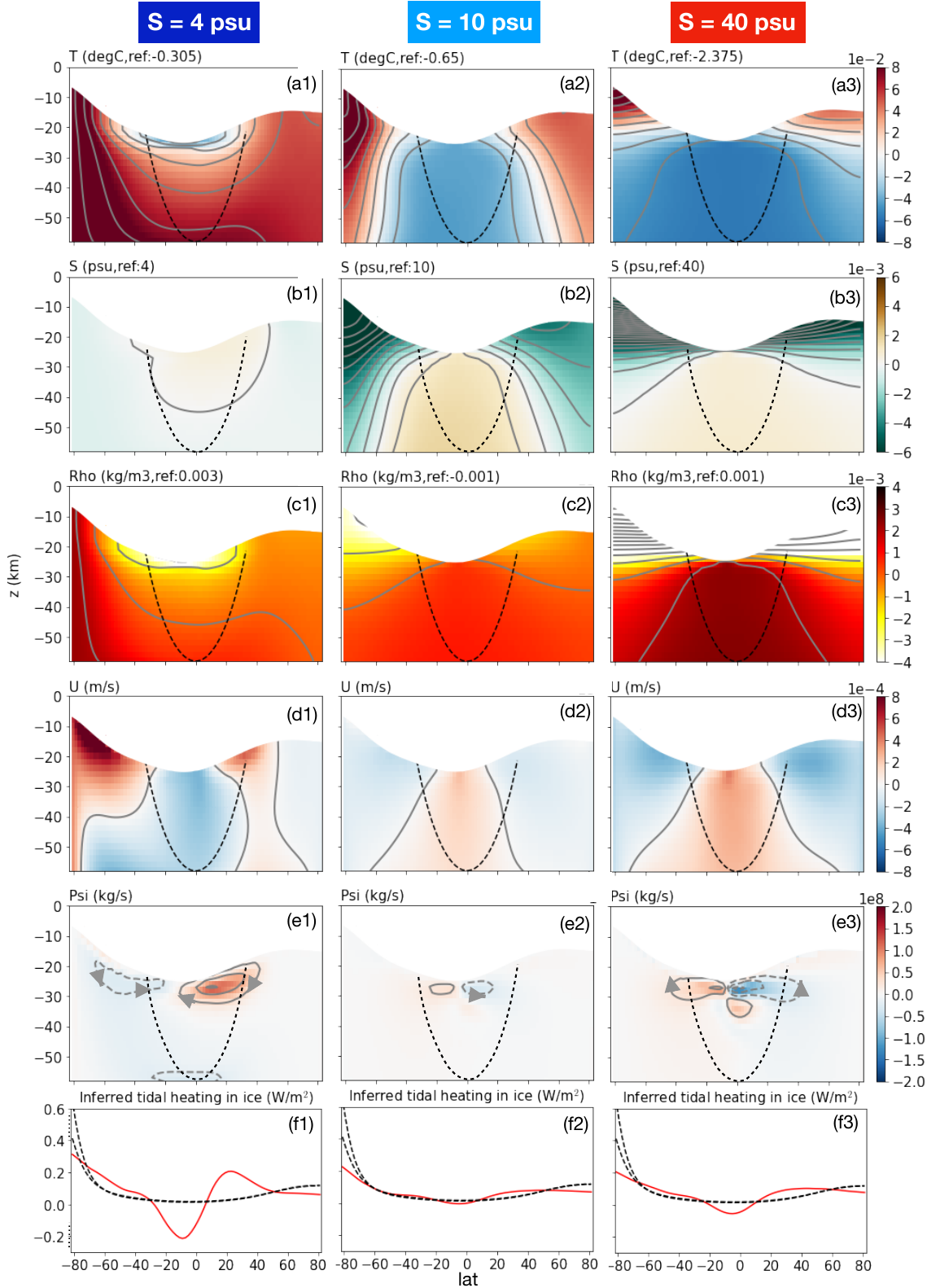}
    
    \caption{\small{The sensitivity of the 100\% shell-heating solution to lower explicit diffusivity ($\kappa_{\mathrm{GM}}=0$~m$^2$/s, $\kappa_v=\kappa_h=10^{-3}$~m$^2$/s), set out as in Fig.~S5.}}
    
    \label{fig:T-S-U-Psi-diff1e-3}
  \end{figure*}

\begin{figure*}[htp!]
    \centering
    \includegraphics[page=14,width=0.85\textwidth]{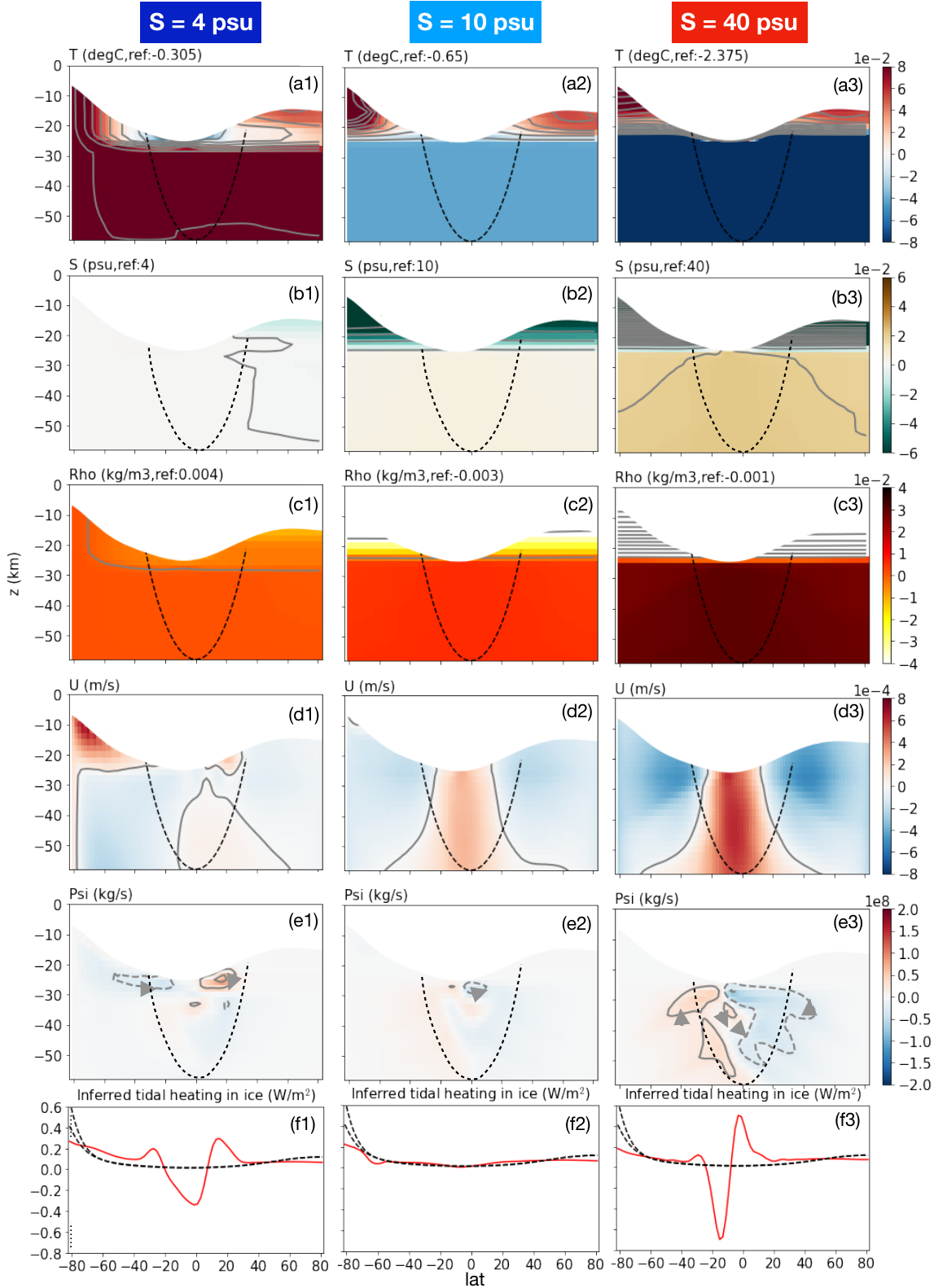}
    
    \caption{\small{The sensitivity of the 100\% shell-heating solution to even lower explicit diffusivity ($\kappa_{\mathrm{GM}}=0$~m$^2$/s, $\kappa_v=10^{-5}$~m$^2$/s, $\kappa_h=10^{-3}$~m$^2$/s), set out as in Fig.~S5. The interval between salinity and density contours are 20 times larger than in other plots.  } }
    
    \label{fig:T-S-U-Psi-diff1e-5}
  \end{figure*}

To examine the sensitivity to diffusivity, we carried out three additional sets of experiments for the shell-heating scenario using different GM and diapycnal diffusivities: one with $\kappa_{\mathrm{GM}}=0$~m$^2$/s, $\kappa_v=\kappa_h=5\times 10^{-3}$~m$^2$/s, one with $\kappa_{\mathrm{GM}}=0$~m$^2$/s, $\kappa_v=\kappa_h=10^{-3}$~m$^2$/s, and the third with $\kappa_{\mathrm{GM}}=0$~m$^2$/s, $\kappa_v=10^{-5}$~m$^2$/s, $\kappa_h=10^{-3}$~m$^2$/s. The corresponding solutions for $S_0=4,\ 10$ and $40$ psu are shown in Fig.~\ref{fig:T-S-U-Psi-noGM}, Fig.~\ref{fig:T-S-U-Psi-diff1e-3} and Fig.~\ref{fig:T-S-U-Psi-diff1e-5}, respectively. The mismatch index that measures the discrepancy between the inferred and predicted ice tidal dissipation (Eq.~2 in the main text) are plotted on Fig.~4c in the main text using triangular markers.

On changing GM and diapycnal diffusivities, the dependence of the meridional heat transport and hence the inferred tidal dissipation on salinity remains qualitatively similar to the control experiments: compare the bottom panels of Fig.~\ref{fig:T-S-U-Psi-noGM}, Fig.~\ref{fig:T-S-U-Psi-diff1e-3} and Fig.~\ref{fig:T-S-U-Psi-diff1e-5} with Fig.~4b in the main text. The mismatch between the inferred tidal dissipation $\hat{\mathcal{H}}_{\mathrm{ice}}$ and the modeled dissipation $\mathcal{H}_{\mathrm{ice}}$ is smallest when the reference salinity is in the range 10-20~psu regardless of the spread of diffusivities being used. The ocean solutions also remain qualitatively similar to the control experiments shown in Fig.~3 in the main text. Low salinity cases have sinking over the poles, driven in the main by the density gradient associated with temperature anomalies (see left panels of Fig.~\ref{fig:T-S-U-Psi-noGM}, Fig.~\ref{fig:T-S-U-Psi-diff1e-3} and Fig.~\ref{fig:T-S-U-Psi-diff1e-5}). The opposite is true for the high salinity cases (see the right panels). At intermediate salinities ($\sim$10~psu), the density gradient and overturning circulation are weak (see the middle panels), just as in the control solution (Fig.~3 in the main text). This weak circulation, in turn, leads to a weaker heat convergence toward the equator compared to the end-member cases (see Fig.~\ref{fig:T-S-U-Psi-noGM}f, Fig.~\ref{fig:T-S-U-Psi-diff1e-3}f and Fig.~\ref{fig:T-S-U-Psi-diff1e-5}f and Fig.~4b in the main text), and the resulting $\hat{\mathcal{H}}_{\mathrm{ice}}$ is more consistent with $\mathcal{H}_{\mathrm{ice}}$ (black dashed curves). This general trend is found in all diffusivity scenarios (Fig.~4c in the main text), suggesting that our main conclusions are indeed robust.
  
There are quantitative changes to our solutions, however. When GM is turned off, the overall stratification becomes weaker due to the lack of parameterized slantwise convection, the isopycnal slope becomes steeper (row-c of Fig.~\ref{fig:T-S-U-Psi-noGM} and Fig.~3 in the main text), the salinity contrast is slightly reduced (row-b of Fig.~\ref{fig:T-S-U-Psi-noGM} and Fig.~3 in the main text) and the circulation weakens. Furthermore, reducing the explicit diffusivity suppresses mixing and increases salinity/temperature variations over the globe (Fig.~\ref{fig:T-S-U-Psi-diff1e-3} and Fig.~\ref{fig:T-S-U-Psi-diff1e-5}). That leads to stronger stratification, shallower circulation and weaker meridional heat transport, as can be seen by comparing Fig.~\ref{fig:T-S-U-Psi-diff1e-3} and Fig.~\ref{fig:T-S-U-Psi-diff1e-5} against Fig.~\ref{fig:T-S-U-Psi-noGM}. These changes are particularly significant in the 40~psu case perhaps because the ocean circulation cannot be efficiently energized when the buoyancy source is located higher in the water column than the buoyancy sink \cite{Zeng-Jansen-2021:ocean}.

\begin{figure*}[htp!]
    \centering
    \includegraphics[page=17,width=0.85\textwidth]{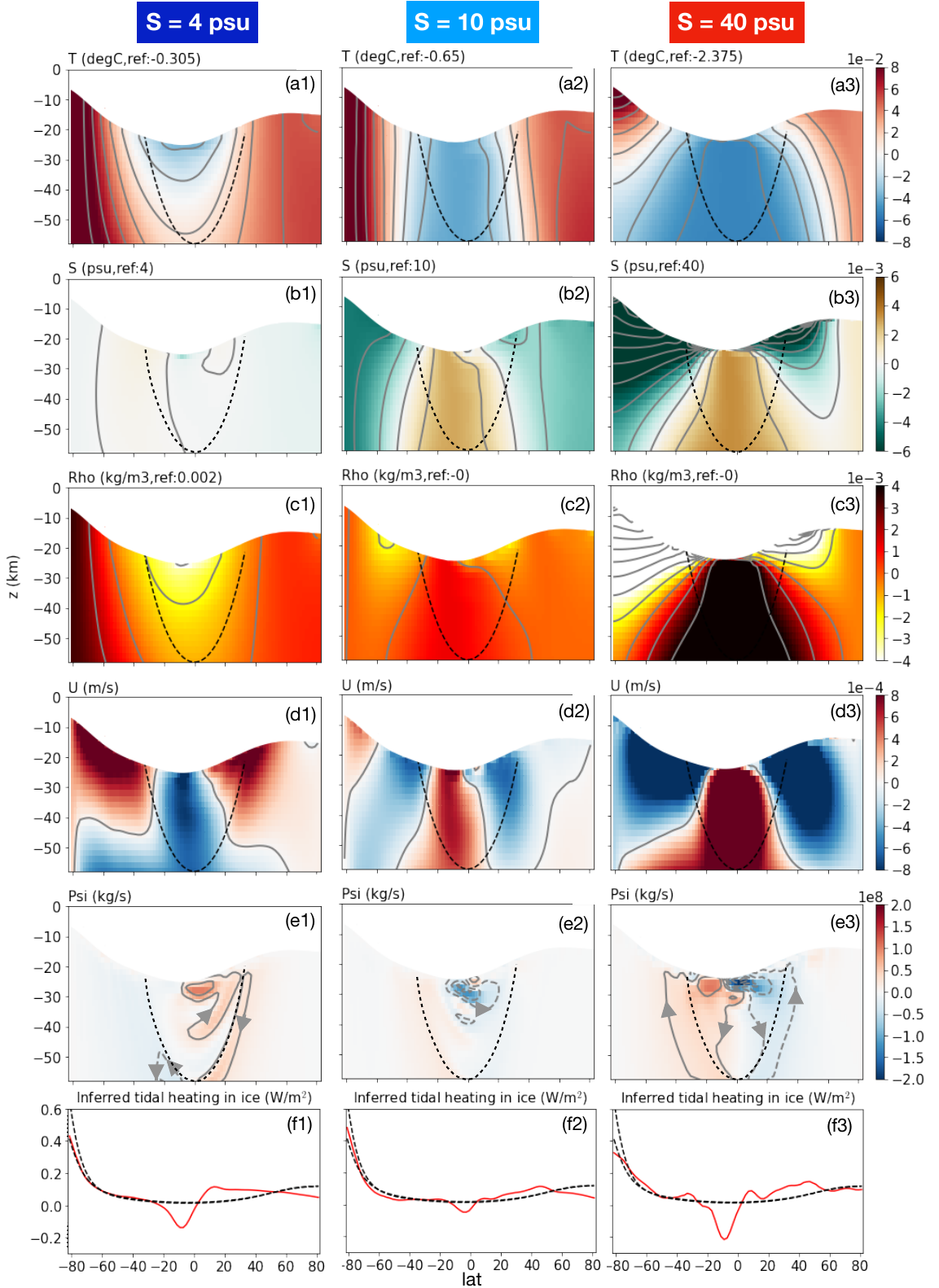}
    
    \caption{\small{The sensitivity of the shell-heating scenario solution to lower viscosity ($\nu_v=\nu_h=10$~m$^2$/s instead of $50$~m$^2$/s). Row (a-e) are the same figure as Fig.3 in the main text, showing the ocean circulation and thermodynamic state. Row (f) is similar to Fig.~4(b,e) in the main text, showing the inferred tidal dissipation $\hat{\mathcal{H}}_{\mathrm{ice}}$ (red solid line, calculated using Eq.~1 in the main text), in comparison with the dissipation rate predicted by tidal dissipation model (black dashed lines, Eq.\ref{eq:H-tide}). To remove the grid-size noise from the heat transport profile, we apply a 9-point smoothing.} }
    
    \label{fig:T-S-U-Psi-lowvisc}
  \end{figure*}

We also carried out two sets of experiments to test the sensitivity to lower viscosity. The solutions with default diffusivities but 5 times lower viscosities ($\nu_v=\nu_h=10$~m$^2$/s instead of $50$~m$^2$/s) are shown in Fig.~\ref{fig:T-S-U-Psi-lowvisc}. To reduce grid-scale noise caused by the low viscosity, the streamfunctions (row-e) and the heat transport curve (row-f) were smoothed. While the temperature, salinity, zonal flow fields remains almost identical to the control experiments (Fig.~3 in the main text), the overturning circulation follows the Taylor columns more closely. Without viscosity, the angular momentum of water conserves before it encounters the rough boundary at the top or bottom, and as a result, flow must follow the direction of rotating axis to avoid drastic change in zonal momentum, which would require strong density variation according to the thermal wind balance. The meridional heat transport decreases by only by a small portion, leading to a slightly lower mismatch index, as can be seen from Fig.~4 in the main text (dots indicate the default setup and left triangle the case with low viscosity).

\begin{figure*}[htp!]
    \centering
    \includegraphics[page=24,width=0.85\textwidth]{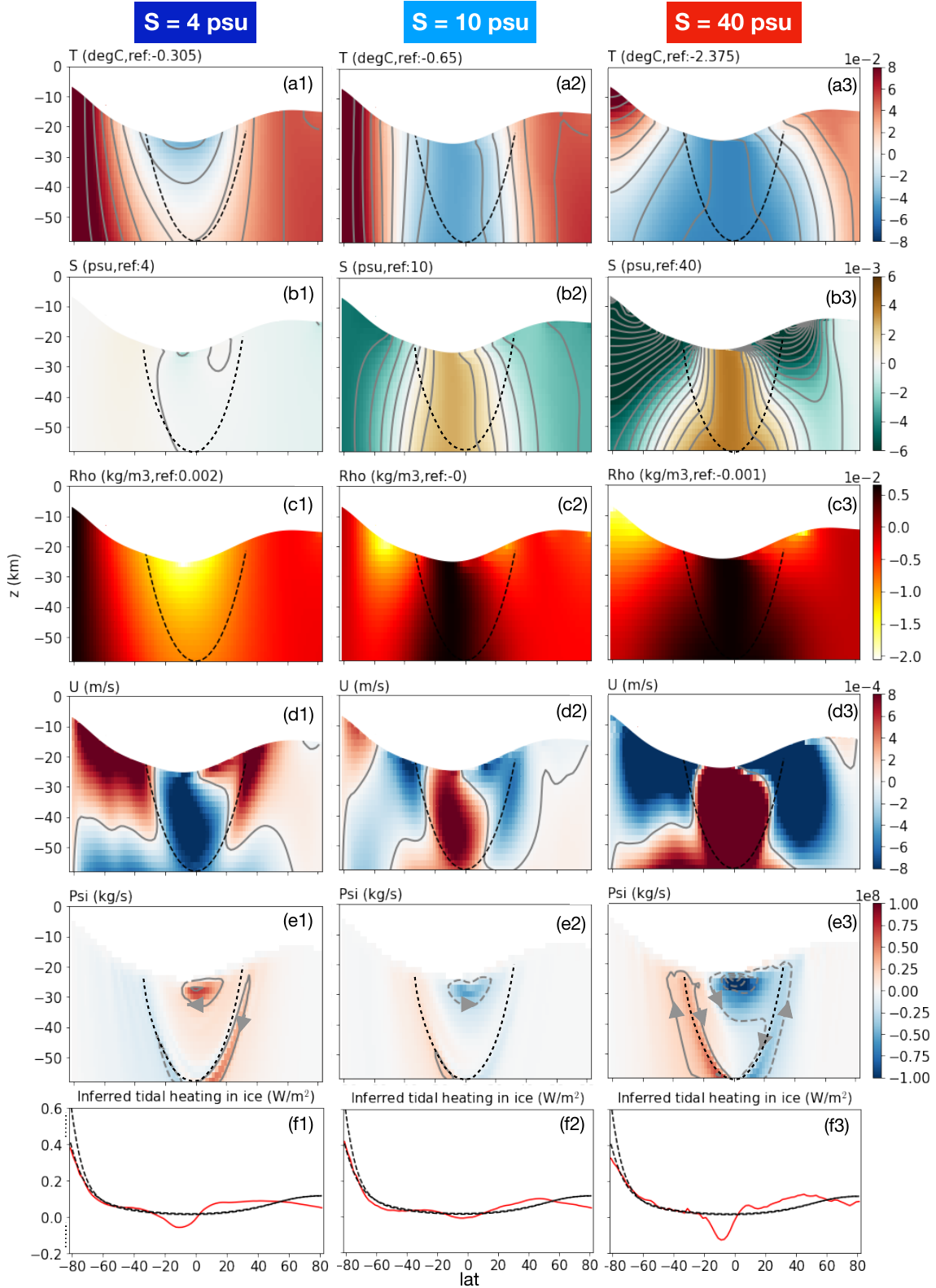}
    
    \caption{\small{The sensitivity of the shell-heating scenario solution to even lower viscosity ($\nu_v=\nu_h=1$~m$^2$/s instead of $50$~m$^2$/s). To remove the grid-size noise from the heat transport and streamfunction profiles, we apply a 9-point smoothing.} }
        \label{fig:T-S-U-Psi-llowvisc}
      \end{figure*}

\begin{figure*}[htp!]
    \centering
    \includegraphics[page=25,width=0.85\textwidth]{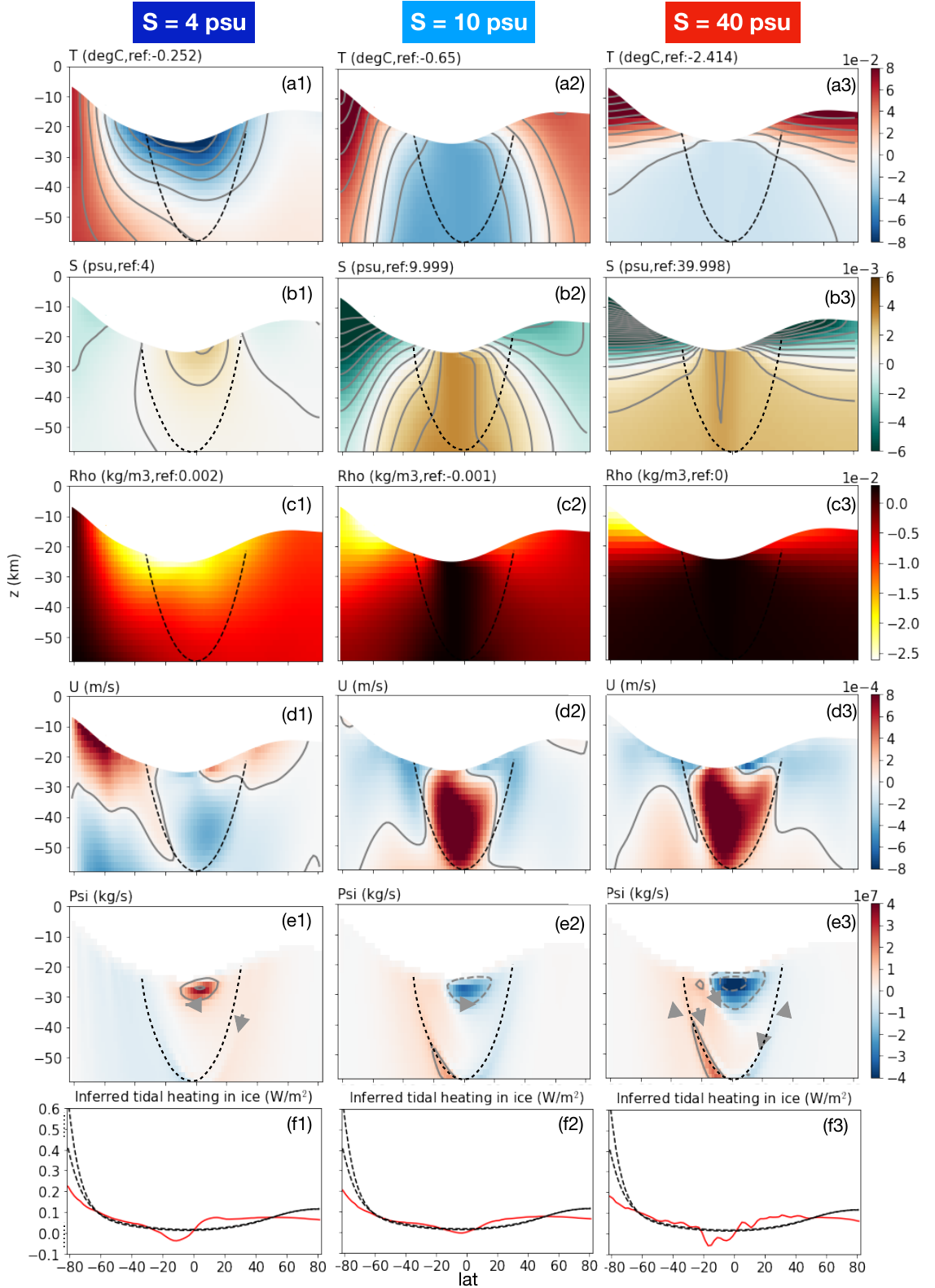}
    
    \caption{\small{Same as Fig.\ref{fig:T-S-U-Psi-llowvisc} except lower diffusivity is used ($\kappa_v=\kappa_h=10^{-3}$~m$^2$/s). To remove the grid-size noise from the heat transport and streamfunction profiles, we apply a 9-point smoothing.} }
        \label{fig:T-S-U-Psi-llowvisc-lowdiff}
      \end{figure*}

When the viscosities are further reduced to $\nu_v=\nu_h=2$~m$^2$/s, 25 times lower than the default viscosities, the vertical flow speed becomes pixelated (not shown). Although we usually avoid noises like this by increasing damping, they sometimes show up in proper solutions for terrestrial ocean when the resolution is coarse, leaving the solution's relevance somewhat uncertain. In row-e of Fig.~\ref{fig:T-S-U-Psi-llowvisc} and Fig.~\ref{fig:T-S-U-Psi-llowvisc-lowdiff}, we present the smoothed streamfunction for the 4~psu, 10~psu and 40~psu scenarios. It is noticeable that the shallow cells along the ice shell disappear, and a deeper circulation that is overridden before shows up. This deep circulation closely follows the tangent cylinder and connects the ice shell all the way to the seafloor.

To understand this change, we need to consider the momentum budget. When water moves toward (away from) the rotating axis under the influence of pressure gradient, by conserving angular momentum, the flow will tend to accelerate eastward (westward). This tendency has to be removed by either friction or viscosity, otherwise the resultant zonal jet will form the so-called geostrophic balance with the pressure gradient force, the driving force of the meridional circulation. The roughness of the water-ice interface and the seafloor allows water to move along these two boundaries meridionally. In the interior, without boundary friction, flow tends to be aligned with the direction of the rotating axis to avoid zonal acceleration, unless viscosity can transport the gained zonal momentum toward a rough boundary nearby, so that momentum can be dissipated. In the viscous layer, the dominant momentum balance should be
\begin{equation}
  \label{eq:dominant-balance-boundary}
 fV\sim\nu_v\partial_z\partial_zU\sim \nu_v\frac{\Delta U}{\delta^2},
\end{equation}
where $V$ and $U$ denote flow speeds in north/south direction and east/west direction, respectively, $\Delta U$ denotes the zonal speed difference between upper branch and lower branch of the shallow cell, and $\delta$ denotes the depth of the return flow. Assuming that $\Delta U$ is 1-2 order of magnitude greater than $V$, the depth of return flow should be around $\delta=\sqrt{50\nu_v/f}\sim 5$~km ($\nu_v=50$~m$^2$/s, $\Delta U/V=50$ is used here), two times the model's vertical resolution (2~km). That is why the shallow cell is resolved in the default setup. As viscosity decreases, the lower branch needs to move closer to the ice shell to feel the viscous force. With $\nu_v=1$~m$^2$/s, $\delta$ decreases to $0.7$~km, and therefore, the $2$~km vertical resolution can no longer resolve the cell. 

How deep the shallow circulation can extend in reality depends on the vertical viscosity. On Enceladus, the ice shell wobbles back and forth by roughly 100~m due to its eccentricity on the timescale of rotation period \cite{Lemasquerier-Grannan-Vidal-et-al-2017:libration, Rekier-Trinh-Triana-et-al-2019:internal}. This by itself will leads to a viscosity around $1$~m$^2$/s. Besides, shear instability may arise when the flow shear is greater than the stratification, $U_z\sim \Delta U/\delta >2N$ ($N$ is the Brunt-Vasala frequency). In a weakly stratified environment like Enceladus, the above instability condition should be easy to satisfy and the resultant eddies may further enhance the viscosity. 

Although the circulation looks very different in experiments with and without the shallow cells, the heat transport doesn't seem to vary by a lot. This can be seen by comparing Fig.~\ref{fig:T-S-U-Psi-llowvisc}f and Fig.~\ref{fig:T-S-U-Psi-llowvisc-lowdiff}f with Fig.~3f in the main text and Fig.~\ref{fig:T-S-U-Psi-diff1e-3}f here, respectively. Since the salinity flux from the ice shell is fixed in all experiments, when ocean circulates slowly, salinity anomaly will accumulate. This will drive stronger circulation in a salty ocean, keeping the heat transport roughly the same.

\begin{figure*}[htp!]
    \centering
    \includegraphics[page=15,width=0.85\textwidth]{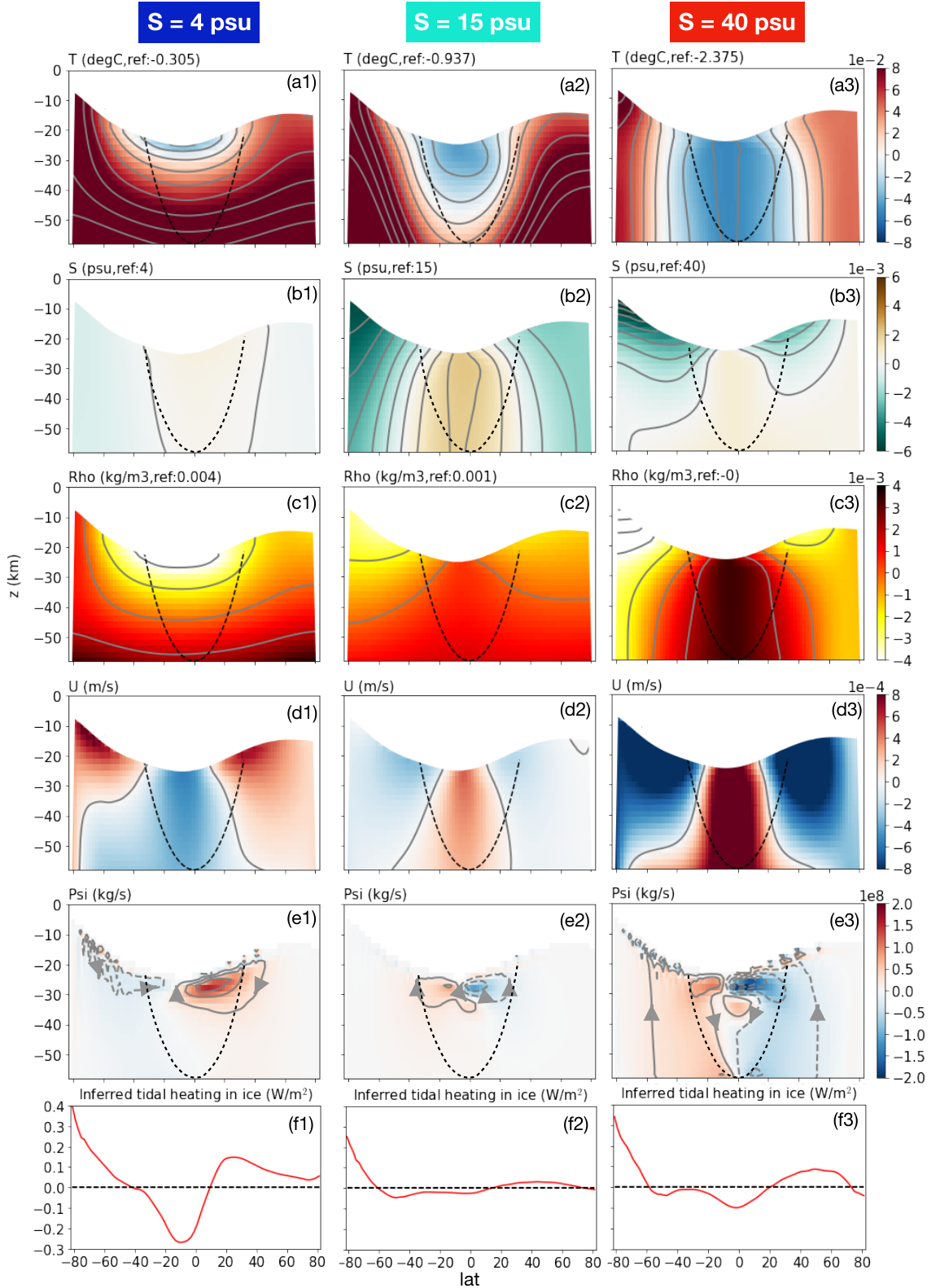}
    
    \caption{\small{The sensitivity of the 100\% core-heating solution to zero GM diffusivity ($\kappa_{\mathrm{GM}}=0$~m$^2$/s, $\kappa_v=\kappa_h=5\times10^{-3}$~m$^2$/s), set out as in Fig.~S5.  } }
    
    \label{fig:T-S-U-Psi-core100-noGM}
  \end{figure*}
\begin{figure*}[htp!]
    \centering
    \includegraphics[page=16,width=0.85\textwidth]{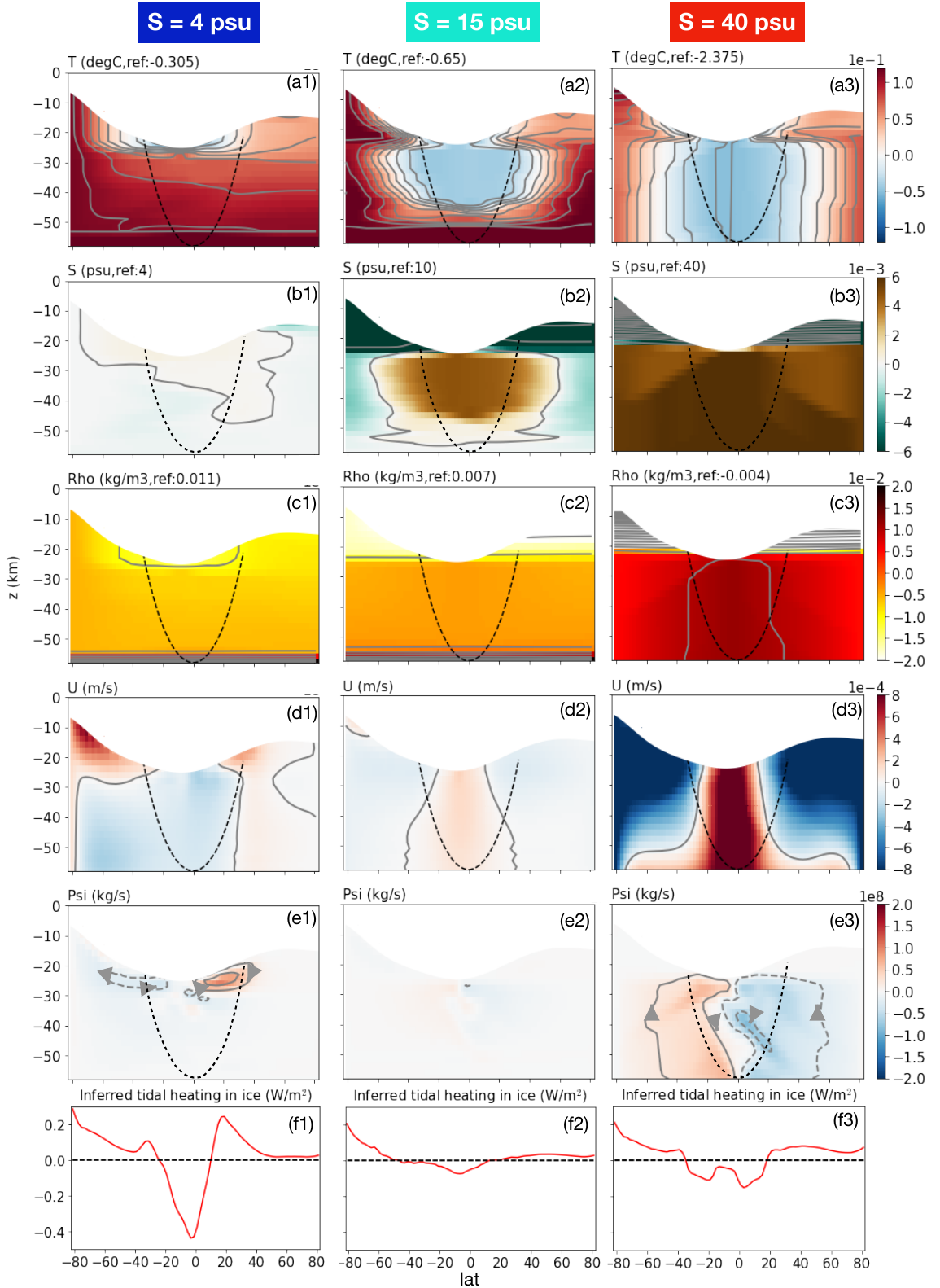}
    
    \caption{\small{The sensitivity of the 100\% core-heating solution to low diffusivities ($\kappa_{\mathrm{GM}}=0$~m$^2$/s, $\kappa_v=10^{-5}$~m$^2$/s, $\kappa_h=10^{-3}$~m$^2$/s), set out as in Fig.~S5. Note that the interval between two contour lines for salinity and density are 5 times greater than Fig.~\ref{fig:T-S-U-Psi-core100-noGM} ($10^{-2}$~psu, $10^{-2}$~kg/m$^3$).} }
    
    \label{fig:T-S-U-Psi-core100-diff1e-5}
  \end{figure*}

We also carried out sensitivity tests for the core-heating scenario as shown in Fig.~\ref{fig:T-S-U-Psi-core100-noGM} with $\kappa_\mathrm{GM}=0$ and in Fig.~\ref{fig:T-S-U-Psi-core100-diff1e-5} and low horizontal and vertical diffusivities ($\kappa_h=10^{-3}$~m$^2$/s, $\kappa_v=10^{-5}$~m$^2$/s). Again, the solutions are very similar to those with a GM parameterization and higher diffusivities (Fig.~\ref{fig:T-S-U-Psi-core100}), except for weaker stratification, weaker isopycnal slope, salinity contrast and circulation, relative to the shell-heating scenario with default parameters.

  \subsection{Sensitivity to assumed ice viscosity}
  \label{sec:sensitivity-ice-visc}

\begin{figure*}[htp!]
    \centering
    \includegraphics[page=10,width=0.85\textwidth]{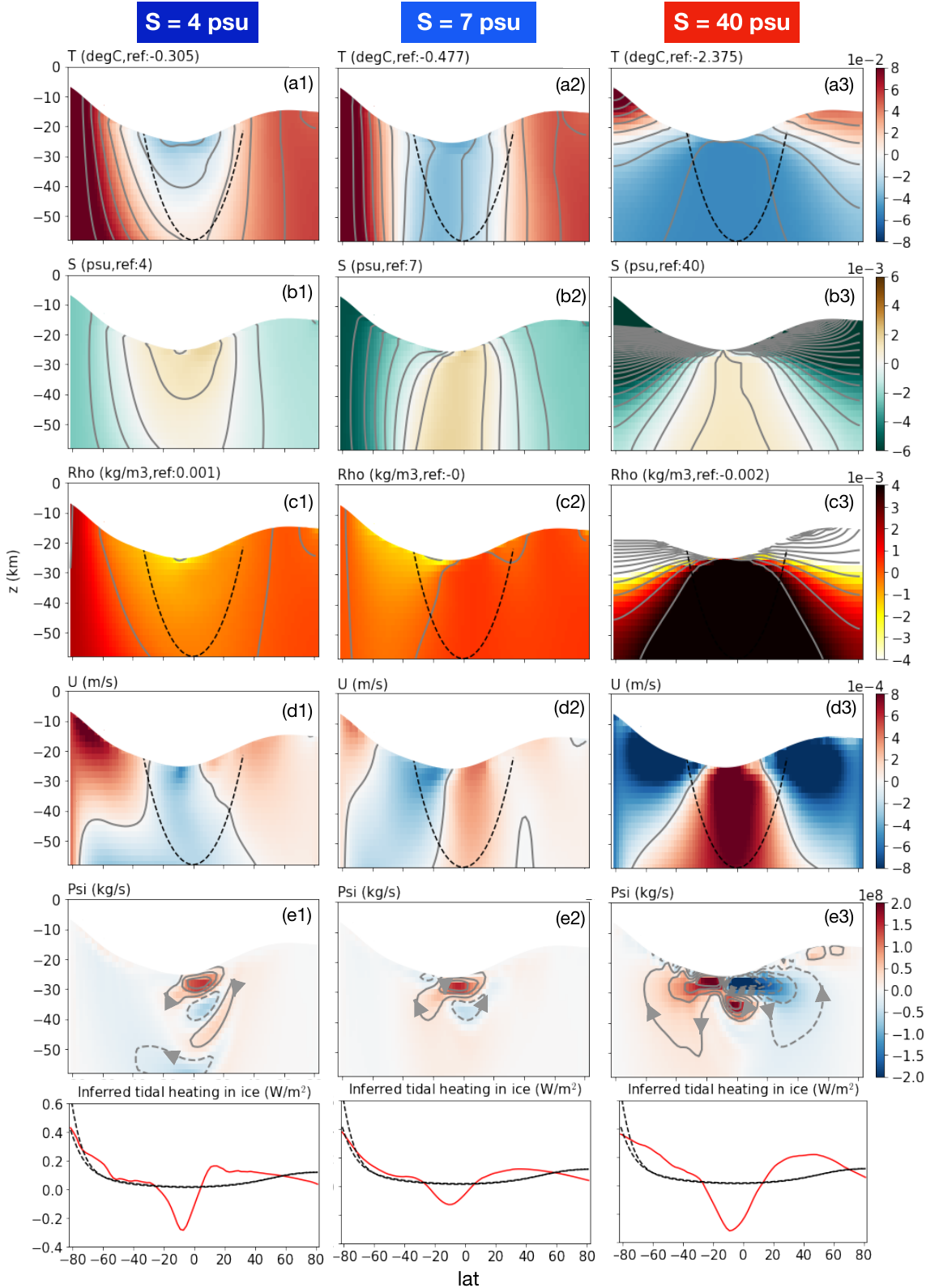}
    
    \caption{\small{The sensitivity of the 100\% shell-heating scenario solution to lower ice viscosity ($\eta_m=2\times10^{13}$~Pa$\cdot$s instead of $10^{14}$~Pa$\cdot$s), set out as in Fig.~S5. Default ocean mixing parameters are used.} }
    
    \label{fig:T-S-U-Psi-lowicev}
  \end{figure*}

  The viscosity of the ice shell controls ice speeds (Eq.~\ref{eq:ice-flow}), and thereby the freezing/melting rate needed to maintain the observed ice geometry. However, due to our limited understanding of ice rheology, the uncertainties associated with the melting point ice viscosity $\eta_m$ remain. To examine sensitivity we carried out an experiment with $\eta_m$ set to $2\times10^{13}$~Pa$\cdot$s, 5 times lower than the default value. Solutions for $S_0=4,\ 10,\ 40$~psu are presented in Fig.~\ref{fig:T-S-U-Psi-lowicev}. Decreasing the ice viscosity leads to a stronger salinity flux between the ocean and ice (Eq.~\ref{eq:S-tendency}) and stronger salinity variations. This can be clearly seen by comparing Fig.~\ref{fig:T-S-U-Psi-lowicev}b with Fig.~3b of the main text. Since the overall salinity gradient increases, the density gradient also increases (Fig.~\ref{fig:T-S-U-Psi-lowicev}c), and this in turn drives stronger circulation (Fig.~\ref{fig:T-S-U-Psi-lowicev}e). In addition to these change, increasing ice mobility lowers the transitional salinity as shown by plus sign symbols in Fig.~4c of the main text. That is because a more negative thermal expansion coefficient is required to cancel the salinity-induced density anomaly and achieve a minimum density gradient, indeed just as suggested by our conceptual model. The opposite is true with increased ice viscosity. Because the salinity flux between the ocean and ice decreases, the overall salinity gradient decreases, and that make cancellation between the temperature- and salinity-driven circulation occur at higher salinity (15~psu instead of 10~psu, see minus sign symbols in Fig.~4c of the main text).
  
  \begin{figure*}[htp!]
    \centering
    \includegraphics[page=10,width=0.85\textwidth]{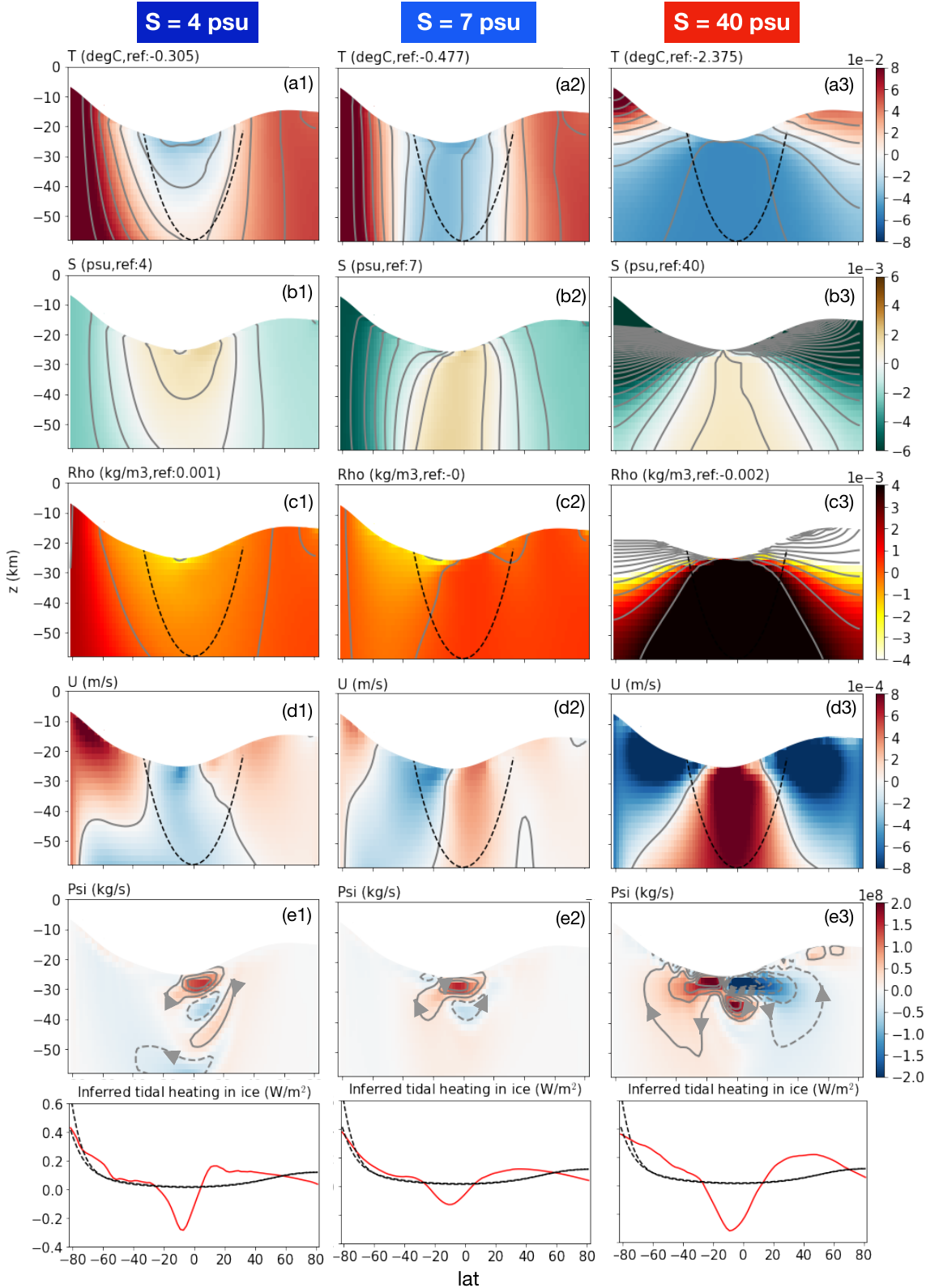}
    
    \caption{\small{The sensitivity of the 100\% shell-heating scenario solution to higher ice viscosity ($\eta_m=5\times10^{14}$~Pa$\cdot$s instead of $10^{14}$~Pa$\cdot$s), set out as in Fig.~S5. Default ocean mixing parameters are used.} }
    
    \label{fig:T-S-U-Psi-highicev}
  \end{figure*}

\subsection{Sensitivity to higher resolution and 3D dynamics}
\label{sec:sensitivity-resolution}

\begin{figure*}[htp!]
    \centering
    \includegraphics[page=19,width=0.85\textwidth]{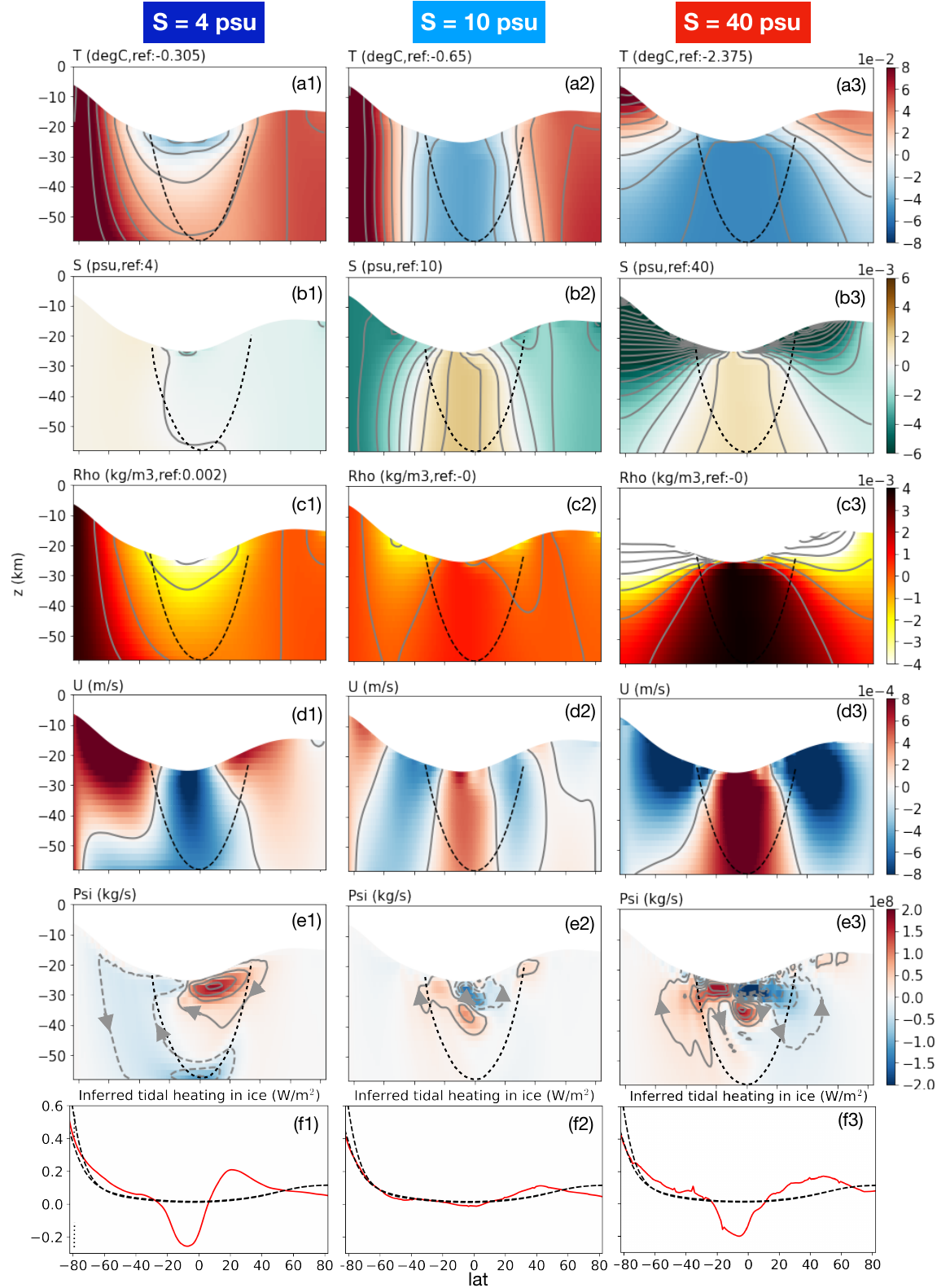}
    
    \caption{\small{The sensitivity of the 100\% shell-heating scenario solution to higher resolution (1~degree instead of 2~degree), set out as in Fig.~S5. Default parameters are used.} }
    
    \label{fig:T-S-U-Psi-highresol}
  \end{figure*}

\begin{figure*}[htp!]
    \centering
    \includegraphics[page=20,width=0.85\textwidth]{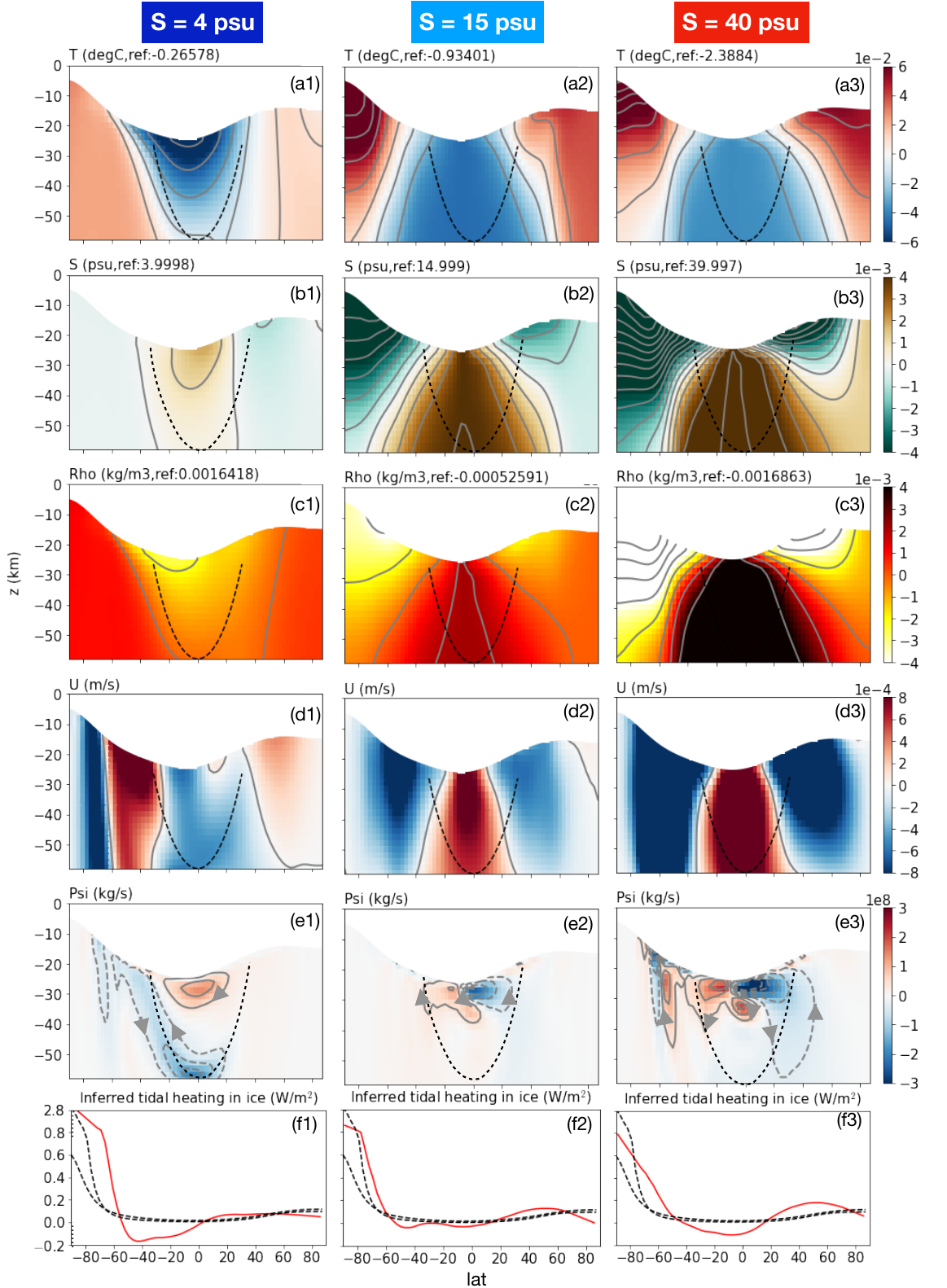}
    
    \caption{\small{The sensitivity of the 100\% shell-heating scenario solution to 3D configuration. Physical parameters are the same as Fig.~\ref{fig:T-S-U-Psi-lowicev}, except that cube sphere (cs32) grid is used. The dashed curves show the ice dissipation predicted by tidal model assuming $p_\alpha=-1.5$ and $p_\alpha=-3$, respectively. } }
    
    \label{fig:T-S-U-Psi-csc}
  \end{figure*}

We also carried out sensitivity tests at higher spatial resolution and assuming 3D rather than 2D dynamics. Doubling the resolution does not change the solution in any significant way (compare Fig.~\ref{fig:T-S-U-Psi-highresol} and Fig.3 in the main text), indicating that our default solutions are robust to increasing resolution. Adding a third dimension allows zonal wave structures to form, which transport additional heat especially when the salinity is low (see Fig.~\ref{fig:T-S-U-Psi-csc}-f1). However, the best-match heat budget again occurs at intermediate salinities (see Fig.~\ref{fig:T-S-U-Psi-csc} row-f). The detailed wave dynamics is not the focus of the present study and is left for further work.
  
  \subsection{Experiments under hemispherically-symmetric forcing}
  \label{sec:sensitivity-symmetric}

\begin{figure*}[htp!]
    \centering
    \includegraphics[page=18,width=0.85\textwidth]{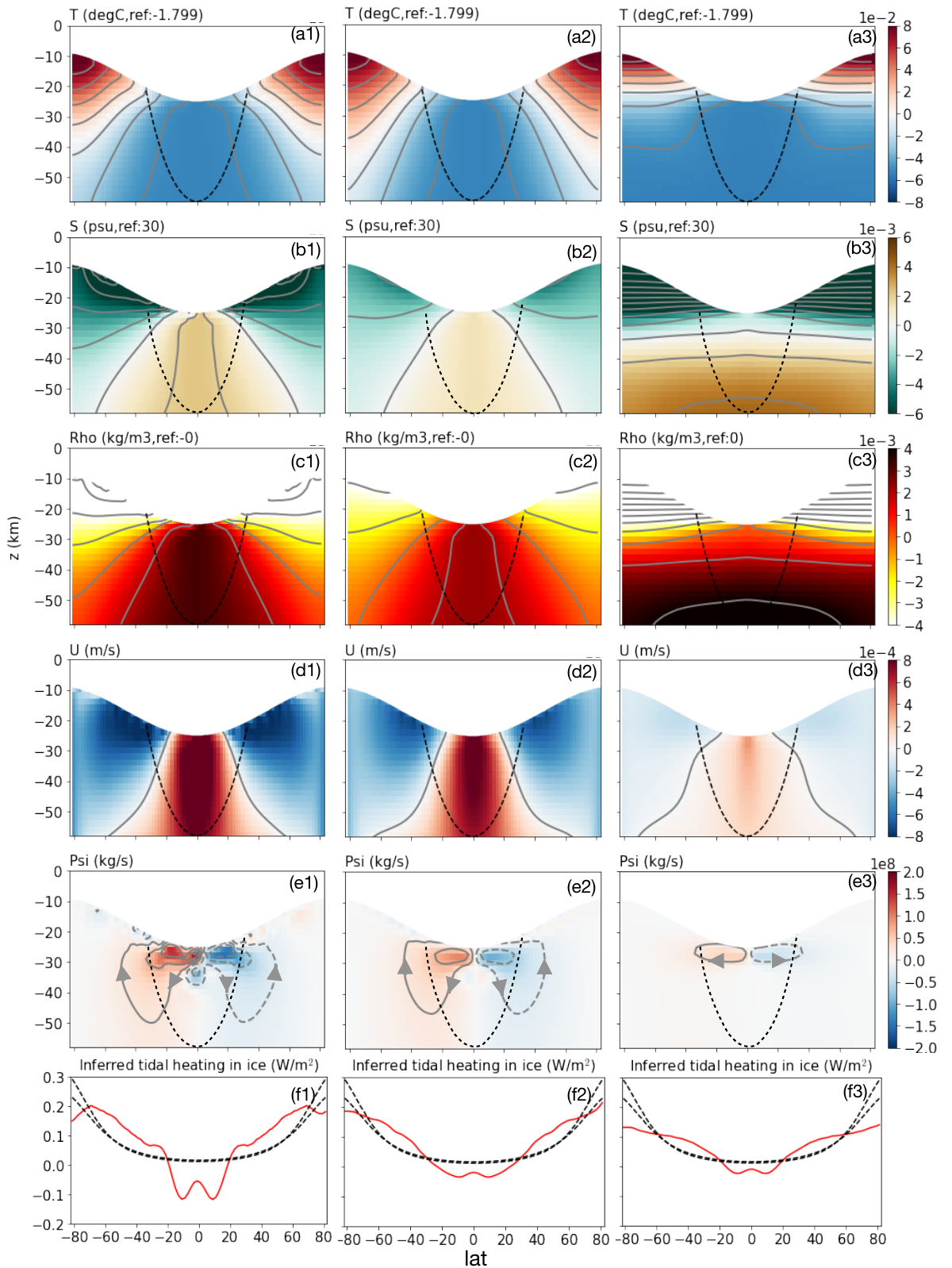}
    
    \caption{\small{Solutions under hemispherically-symmetric forcing with the mean salinity set to 30~psu. The left column shows results for default diffusivities ($\kappa_{\mathrm{GM}}=0.1$~m$^2$/s, $\kappa_v=\kappa_h=5\times 10^{-3}$~m$^2$/s), the middle column sets the GM diffusivity to zero ($\kappa_{\mathrm{GM}}=0$~m$^2$/s, $\kappa_v=\kappa_h=5\times 10^{-3}$~m$^2$/s), and the right column reduces ocean mixing coefficients  ($\kappa_{\mathrm{GM}}=0$~m$^2$/s, $\kappa_v=\kappa_h=10^{-3}$~m$^2$/s).} }
    
    \label{fig:T-S-U-Psi-symmetric}
  \end{figure*}

  In further tests we carried out experiments under hemispherically symmetric ice geometry and forcing allowing for comparison with theoretical model results reported by \textit{Lobo et al. 2021} \cite{Lobo-Thompson-Vance-et-al-2021:pole}. The symmetric ice geometry used here is constructed by averaging the default ice thickness profile shown in Fig.~1b between the two hemispheres. The freezing/melting rate forcing is calculated as before (Eq~\ref{eq:ice-flow}) and the freezing point temperature imposed as a function of pressure. Symmetric ocean solutions are shown on the left column of Fig.~\ref{fig:T-S-U-Psi-symmetric} together with inferred tidal dissipation rates $\hat{\mathcal{H}}_{\mathrm{ice}}$ with default diffusivities ($\kappa_{\mathrm{GM}}=0.1$~m$^2$/s, $\kappa_v=\kappa_h=5\times 10^{-3}$~m$^2$/s). Tracer distributions, circulation and zonal flow are all symmetric about the equator due to the symmetry in the forcing. The mean salinity is set to 30~psu to avoid anomalous expansion effects, and we assume all heat is produced in the ice shell. This broadly mimics the setup used by \textit{Lobo et al. 2021} \cite{Lobo-Thompson-Vance-et-al-2021:pole}.

  As found by \textit{Lobo et al. 2021} \cite{Lobo-Thompson-Vance-et-al-2021:pole}, the near-surface isopycnals tilt downward in polar regions (Fig.~\ref{fig:T-S-U-Psi-symmetric}-c1), driving meridional circulation confined to the near-surface layer (Fig.~\ref{fig:T-S-U-Psi-symmetric}-e1). As in \textit{Lobo et al. 2021} \cite{Lobo-Thompson-Vance-et-al-2021:pole}, we explored the sensitivity of our solution to mixing coefficients. The middle and right columns of Fig.~\ref{fig:T-S-U-Psi-symmetric} present solutions obtained using two different diapycnal diffusivities with $\kappa_{\mathrm{GM}}=0$~m$^2$/s). By turning off the GM parameterization we again observe the circulation and stratification becoming weaker (Fig.~\ref{fig:T-S-U-Psi-symmetric}-e2 and Fig.~\ref{fig:T-S-U-Psi-symmetric}-c2). Furthermore, when lower diapycnal diffusivity is used, the stratification strengthens (Fig.~\ref{fig:T-S-U-Psi-symmetric}-c3) and the circulation becomes weaker and shallower (Fig.~\ref{fig:T-S-U-Psi-symmetric}-e3).

  Finally, it should be noted that the GM diffusivity used here, $\kappa_{\mathrm{GM}}=0.1$~m$^2$/s, is estimated based on our arguments that lateral mixing scales are a few kilometers and the eddy flow speed a few millimeters per second (see Eq~\ref{eq:kappa-GM}). This is several orders of magnitude smaller than the range explored by \textit{Lobo et al. 2021} \cite{Lobo-Thompson-Vance-et-al-2021:pole}, which was presumably motivated by those typical of Earth's ocean. Furthermore, in order to sustain the observed ice geometry \cite{Hemingway-Mittal-2019:enceladuss}, we argue that the freezing/melting rate is a few kilometers per million years. This can be converted to a buoyancy flux by multiplying the haline contract coefficient $\beta_S$, ocean mean salinity $S_0$ and gravity $g$. Substituting parameters appropriate to Enceladus yields a buoyancy flux of the order of $10^{-13}$~m$^2$/s$^3$. This is a full 3-6 orders of magnitude smaller than that used in \textit{Lobo et al. 2021} \cite{Lobo-Thompson-Vance-et-al-2021:pole}. Due to these differences, we prefer not to make quantitative comparison with \textit{Lobo et al. 2021} \cite{Lobo-Thompson-Vance-et-al-2021:pole}.

 \bibliography{export}
 \bibliographystyle{Science}